\documentclass[a4paper,11pt]{article}
\pdfoutput=1

\usepackage{jcappub}
\usepackage[T1]{fontenc}
\usepackage{multirow}
\usepackage{bm}
\usepackage{mathtools, booktabs}
\usepackage[version=4]{mhchem}
\usepackage[normalem]{ulem}
\usepackage{physics, mathrsfs, comment}
\usepackage{tikz-cd}
\newcommand{\versor}[1]{\mathbf{\hat{#1}}}
\usepackage{xcolor}
\compress

\title{\boldmath CMB Lensing Trispectrum as a Probe of Parity Violation in LSS}

\author[a]{Alessandro Greco,}
\author[a]{Zachary Slepian,}
\author[b]{Jiamin Hou,}
\author[c,d,*]{and Alex Krolewski}

\affiliation[a]{Department of Astronomy, University of Florida,\\211 Bryant Space Science Center, Gainesville, FL 32611, USA}
\affiliation[b]{Max-Planck-Institut für Extraterrestrische Physik,\\Postfach 1312, Giessenbachstrasse 1, 85748, Garching bei München, Germany}
\affiliation[c]{Waterloo Centre for Astrophysics, University of Waterloo, 200 University Ave W, Waterloo, ON N2L 3G1, Canada}
\affiliation[d]{Department of Physics and Astronomy, University of Waterloo, 200 University Ave W, Waterloo, ON N2L 3G1, Canada}
\affiliation[*]{CITA National Fellow}

\emailAdd{alessandro.greco@ufl.edu}
\emailAdd{zslepian@ufl.edu}
\emailAdd{jiamin.hou@mpe.mpg.de}
\emailAdd{akrolews@uwaterloo.ca}

\abstract{We show that the Cosmic Microwave Background (CMB) lensing trispectrum is sensitive to parity violation in Large-Scale Structure (LSS). We obtain a compact expression for the reduced lensing trispectrum that is general for any input matter trispectrum. We then present as an example a simple parity-violating toy model for the latter, and explicitly compute the parity-odd lensing trispectrum, including an estimate of the Signal-to-Noise Ratio (SNR). This work serves as a proof of principle, demonstrating how future studies of more physically motivated models can be conducted. It also provides an intuitive physical explanation of why CMB lensing is sensitive to parity. Our work is the first to point out that secondary CMB anisotropies can be used to probe parity in LSS, and will be important in enabling upcoming experiments such as Simons Observatory and CMB-S4 to contribute maximal power on parity violation.}

\begin{document}
\maketitle
\flushbottom

\section{\label{sec:introduction}Introduction}

Any Lorentz-invariant quantum field theory must be invariant under the combination of charge conjugation (C), parity (P), and time reversal (T) \cite{schwinger1951theory}. However, CP symmetry, the combination of charge conjugation and parity, is not a strict requirement for all quantum field theories. The Standard Model violates CP symmetry in the weak sector \cite{lee1956question}, but this violation alone is insufficient to explain baryogenesis, the observed excess of baryons over anti-baryons. The Sakharov conditions for baryogenesis \cite{sakharov1998violation} demand CP violation, but no significant CP violation has been observed outside of the weak sector, and the CP violation in the weak sector is too small to account for the baryon asymmetry \cite{fukugita1986barygenesis}.

Since the CP violation in the Standard Model is insufficient to explain the Baryogenesis, it requires a new 
CP-violating mechanism. While CP violation does not necessarily imply P violation by iteself, new CP violation offer a more comprehensive explanation of the observed matter-antimatter asymmetry.

\subsection{Parity Violation in Large-Scale Structure}

The search for CP violation in physics, particularly in the context of baryogenesis, motivates the exploration of parity violation in cosmology as a potential signature of new physics. Several cosmological observables have been proposed as probes of fundamental parity violation \cite{Lue:1998mq, Komatsu:2022nvu, Philcox:2023uor, Shim:2024tue, Zhu:2024wme, Hou:2024udn}. Recently, \cite{cahn2023test} showed that the galaxy 4-Point Correlation Function (4PCF) is the lowest-order statistic able to probe 3D parity, using an isotropic basis developed in \cite{cahn2023isotropic}.\footnote{Earlier work suggested a parity test using position-dependent power spectra \cite{Jeong:2012df}, though this method has not yet been applied to data. Additionally, \cite{shiraishi2016parity} noted that the trispectrum of CMB temperature anisotropies is sensitive to parity.} This method was applied in \cite{hou2023measurement} to detect parity violation in 3D LSS using samples from the Data Release 12
(DR12) of the Baryon Oscillation Spectroscopic Survey (BOSS) within the Sloan Digital Sky Survey (SDSS), finding $3.1\sigma$ in the smaller-volume, lower-redshift LOWZ sample ($\bar{z}\lesssim 0.4$, 288,000 Luminous Red Galaxies, LRGs) and $7.1\sigma$ in the larger, higher-redshift CMASS sample ($\bar{z} = 0.57$, 777,202 LRGs). Subsequently, \cite{philcox2022probing} found $3\sigma$ evidence in CMASS, though that analysis was limited by suboptimal covariance weighting and coarser binning of the tetrahedron side lengths.

Since the BOSS results, several other works have studied 3D LSS to probe parity on data. \cite{krolewski2024evidenceparityviolationboss} split the CMASS sample into angular patches and cross-correlated them, reducing the sensitivity to covariance misestimation (generalizing a North-South correlation test originally presented in \cite{hou2023measurement}). That paper demonstrates that most of the parity-odd signal in BOSS data arises from a mismatch between the variance of the mocks and the data. It leaves open the possibility of parity violation at $\sim2\sigma$ when combining BOSS data from different hemispheres.

At higher redshift, \cite{adari2024searching} investigated parity in the SDSS DR16 Lyman-$\alpha$ forest, but found no evidence for parity violation. They faced several challenges with covariance estimation, including the use of a robust but suboptimal covariance matrix derived in bootstrapping, which produced spurious detections. These issues limited the strength of their conclusions. Meanwhile, \cite{Cabass:2022oap} compared inflationary parity-violating models to the BOSS 4PCF and found no strong detections. However, this result does not necessarily indicate systematics, but rather suggests that the explored models are unlikely to describe the data. Indeed, our forecasts indicate these models would remain undetectable even with DESI.\footnote{A publication on this forecast is in preparation.}

Recent studies have proposed other explanations for the BOSS results. \cite{Inomata:2024ald} investigated whether lensing by a chiral gravitational-wave background could produce a signal. By analyzing the Uchuu mock catalogues, \cite{philcox2024could} proposed that the BOSS signal arises from sample variance rather than a physical origin. However, this analysis likely overestimated the covariance due to known limitations in mocks generated by replicating a smaller underlying box.\footnote{Appendix B of \cite{Ereza:2023zmz} demonstrates that even after rescaling the volume, the covariance of the 2-Point Correlation Function (2PCF) is overestimated by 10--15\%. By scaling arguments, this would imply an even larger overestimation of the 4PCF covariance, order order of 20--30\%, as claimed in \cite{philcox2024could}.} Machine learning has also been applied to parity violation in LSS \cite{Taylor:2023deh, Hewson:2024rnb, Craigie:2024bhk}.

\subsection{Turning to the Cosmic Microwave Background}
Recent work \cite{philcox2023cmb, philcox2023testing}, has argued that, if inflationary in origin, parity violation in the matter field should also manifest in the primary anisotropies of CMB temperature and polarization. This argument may not be correct on the scales relevant for the 3D LSS detection and is taken up more carefully in a work in preparation. The main reason the primary CMB is not sensitive to parity save on very large scales, is that it is a thin, spherical shell. Thus, primary CMB temperature anisotropies in this limit represent the correlation of four density points that are all roughly co-planar. In particular, their separation relative to the distance from the last scattering is small enough that they are in the ``flat sky'' limit. Hence any parity-violating signature is strongly suppressed since it should require one direction vector to be perpendicular to the other two. 

However, the case of CMB lensing importantly differs from the primary CMB. While both involve integrated quantities, some of the shells probed by CMB lensing are relatively close to us. At a given angular scale, the line-of-sight distance to the relevant shells is small compared to the typical side length of tetrahedra formed by points on the sky, which in principle allows sensitivity to non-coplanar configurations. However, in the absence of redshift information, projection effects tend to wash out any parity-violating signal, as parity-even and parity-odd configurations become observationally indistinguishable. The key difference between primary CMB anisotropies and CMB lensing lies instead in the physical scales probed: the broader kernel in the lensing case enhances sensitivity to smaller and more widely separated scales, where parity-odd signals are less suppressed. Thus, what sets CMB lensing apart in our analysis is the role played by the observer-to-shell distance.

With these ideas in mind, in this work we focus on the CMB lensing trispectrum, showing that a parity-odd LSS 4PCF can yield a parity-odd CMB lensing trispectrum as well. Several works have addressed the information content of the CMB lensing power spectrum
\cite{seljak1995gravitational, hu2000weak, bartelmann2001weak, challinor2005lensed, lewis2006weak}. The study of CMB lensing has naturally extended to higher-order correlation functions, including the bispectrum (three-point correlation) \cite{bohm2016bias, Namikawa:2016jff, namikawa2019cmb, kalaja2023reconstructed, Namikawa:2018erh}. This is the first work to consider the CMB lensing trispectrum
(four-point correlation). Previous works have shown that parity violation can be detected in the CMB temperature field and the trispectrum (Hu and Okamoto, among others) \cite{hu2001angular, okamoto2002angular}. This leaves open the prospect that in principle one can have a rotation- and translation-invariant parity-odd CMB lensing trispectrum, as we indeed find here. 

This paper is organized as follows. In Section \ref{sec:setup} we define the lensing potential due to LSS and compute the CMB lensing power spectrum, both to set notation and to ensure our formalism is sensible. In Section~\ref{sec:bispectrum}, we review the bispectrum calculation done in \cite{bohm2016bias} and redo it. In Section~\ref{sec:trispectrum} we then present the trispectrum calculation with a particular emphasis on looking for parity violation. In Section~\ref{sec:phenomenology}, we offer a phenomenological explanation of why the parity-odd angular trispectrum is not suppressed in the CMB lensing case. We estimate the signal-to-noise ratio for the parity-odd CMB lensing trispectrum in Section~\ref{sec:SNR}. We conclude in Section~\ref{sec:conclusions} with a summary of our main findings and suggestions for future work. Appendix~\ref{app:recovering_bohm} presents a consistency check between the treatment adopted in this paper and the results of \cite{bohm2016bias}. We also offer a rigorous proof in Appendix~\ref{app:parity_on_sphere} that parity is not equivalent to a rotation on a curved spherical surface. In Appendix ~\ref{app:triple_product}, we estimated the geometric suppression of parity-odd terms in the CMB angular trispectrum by analyzing the scalar triple product of density vectors on the last-scattering surface.

\section{\label{sec:setup}Theoretical Setup}
In this section, we lay out the conventions and formalism adopted throughout this paper, along with key approximations that facilitate our analysis. To illustrate these concepts in a concrete context, we will compute the angular power spectrum of the CMB lensing potential. This calculation serves as a foundational example, demonstrating the practical application of our methods and setting the stage for the more complex analyses that follow.

\subsection{\label{sub_sec:lensing_potential}CMB Lensing Potential}
\begin{figure}
\centering
\includegraphics[width = \linewidth]{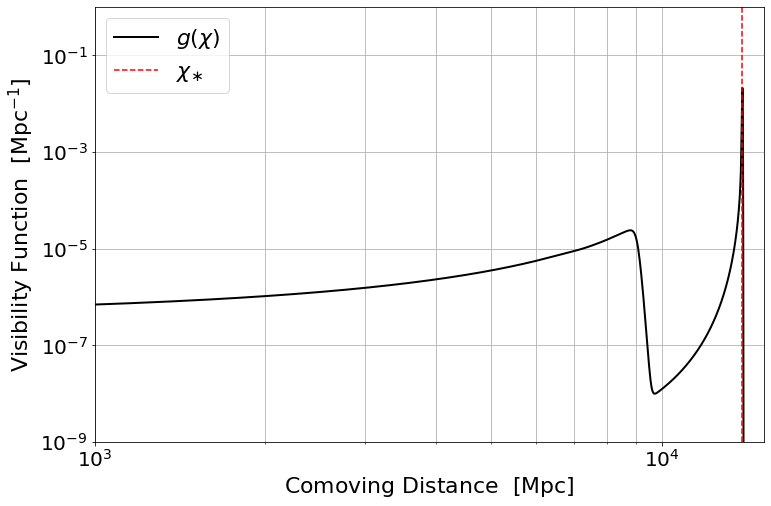}
\caption{The visibility function $g(\chi)$ plotted versus comoving distance, evaluated with \texttt{CLASS} \cite{lesgourgues2011cosmic} using parameters from  \cite{aghanim2020planck}. It represents the Poissonian probability that a photon is last scattered at a certain $\chi$, and it peaks at recombination, $\chi=\chi_{\ast}$, denoted by the dashed vertical line. Another peak at $z=10$ marks the epoch of reionization. The accuracy of the sudden recombination approximation is evident, especially considering the logarithmic scale.}
\label{fig:visibility_function}
\end{figure}

The incoming direction of CMB photons travelling through the gravitational potential produced by large-scale structure is deflected by $\grad{\phi}$, with $\phi$ the CMB lensing potential \cite{kaiser1998weak}. In a spatially flat Universe, it is \cite{hu2000weak}
\begin{equation}
\label{eqn:lensing_potential}
\phi(\versor{n})=-2\int_{0}^{\infty}\frac{\mathrm{d}\chi^{\prime}}{\chi^{\prime}}\,g(\chi^{\prime})\int_{0}^{\chi^{\prime}}\frac{\mathrm{d}\chi}{\chi}\left(\chi^{\prime}-\chi\right)\psi(\chi\,\versor{n},\eta_0-\chi),
\end{equation}
where $\chi\equiv\eta_0-\eta$ is the comoving distance, $\eta$ is the conformal time (so that $\eta_0$ is the conformal age of the Universe), $\psi$ is the gravitational potential at a point $\chi\versor{n}$ in space and a time $\eta=\eta_0-\chi$, $\versor{n}$ is the direction along the line of sight, and $g(\chi)$ is the photon visibility function. As shown in Fig.~\ref{fig:visibility_function}, since $g(\chi)$ peaks at the epoch of recombination, it is reasonable to approximate the photon visibility function as a Dirac delta function centred at $\chi_{\ast}$. i.e. the comoving distance at that time:
\begin{equation}
\label{eqn:dirac}
g(\chi^{\prime})\simeq\delta^{[1]}_{\mathrm{D}}(\chi^{\prime}-\chi_{\ast}).
\end{equation}
This approximation neglects reionization, which also produces photons that contribute to the CMB. With the approximated visibility function, i.e. Eq.~\eqref{eqn:dirac}, we then have Eq.~\eqref{eqn:lensing_potential} as
\begin{equation}
\label{eqn:lensing_potential_approximated}
\phi(\versor{n})=\int_{0}^{\chi_{\ast}}\mathrm{d}\chi\,W(\chi)\psi(\chi\,\versor{n},\eta_0-\chi),
\end{equation}
where the CMB lensing efficiency is defined as 
\begin{equation}
\label{eqn:lensing_efficiency}
W(\chi)\equiv-2\,\frac{\left(\chi_{\ast}-\chi\right)}{\chi_{\ast}\chi}.
\end{equation}
Writing the gravitational potential$\psi$ as an inverse Fourier transform\footnote{We use a tilde to denote any Fourier-transformed quantity.}, Eq.~\eqref{eqn:lensing_potential_approximated} becomes
\begin{equation}
\label{eqn:lensing_potential_fourier}
\phi(\versor{n})=\int_{0}^{\chi_{\ast}}\mathrm{d}\chi\,W(\chi)\int\frac{\mathrm{d}^3\mathbf{k}}{(2\pi)^3}\,\tilde{\psi}(\mathbf{k},\eta_0-\chi)e^{-i\mathbf{k}\cdot\chi\,\versor{n}}.
\end{equation}

To proceed further, it is convenient to express the Newtonian potential $\psi$ in terms of the matter density contrast $\delta_{\mathrm{m}}$. This can be done by using the time-time component of the Einstein field equations:
\begin{equation}
\label{eqn:einstein_equation}
R^{\,0}_{\ 0}-\frac{1}{2}R+\Lambda=-\frac{\rho}{m_{\mathrm{Pl}}^2},
\end{equation}
where $R_{\mu\nu}$ is the Ricci tensor, $R$ is the Ricci scalar, $\Lambda$ is the cosmological constant, $\rho=\rho(\eta,\mathbf{x})$ is the energy density of the Universe, with $\mathbf{x}$ a vector of the comoving spatial coordinates, $m_{\mathrm{Pl}}\equiv1/\sqrt{8\pi G}$ is the reduced Planck mass, and $G$ is Newton's constant. We work at linear order in perturbation theory, and in conformal Newtonian gauge:
\begin{align}
\label{eqn:newtonian_gauge}
g_{\mu\nu}(\mathbf{x},\eta)&=a^2(\eta)\begin{pmatrix}
-\left[1+2\psi(\mathbf{x},\eta)\right] & 0 \\
0 & \left[1+2\varphi(\mathbf{x},\eta)\right]\delta_{ij}^{\mathrm{K}}
\end{pmatrix},\\
\label{eqn:density_contrast}
\rho(\mathbf{x},\eta)&=\overline{\rho}(\eta)\left[1+\delta(\mathbf{x},\eta)\right],
\end{align}
with $g_{\mu\nu}$ and $\overline{\rho}$ respectively the physical metric tensor and the background density of the Universe, $\delta_{ij}^{\mathrm{K}}$ the Kronecker delta, and $a$ the scale factor. We choose this gauge because it easily links the gravitational potentials $\psi$ and $\phi$ to the total density contrast $\delta$, whose matter component is $\delta_{\mathrm{m}}$. By evaluating Eq.~\eqref{eqn:einstein_equation} for the quantities defined in Eq.~\eqref{eqn:newtonian_gauge} at first order in perturbation theory, we obtain the relativistic Poisson equation:
\begin{equation}
\label{eqn:relativistic_poisson_equation}
3\mathcal{H}\frac{\mathrm{d}\varphi}{\mathrm{d}\eta}-3\mathcal{H}^2\psi-\nabla^2\varphi=\frac{a^2\overline{\rho}}{2m_{\mathrm{Pl}}^2}\delta,
\end{equation}
where $\mathcal{H}\equiv a^{-1}\mathrm{d}a/\mathrm{d}\eta$ is the conformal Hubble parameter. We now take the Fourier transform of this equation, finding
\begin{equation}
\label{eqn_relativistic_poisson_equation_fourier}
3\mathcal{H}\frac{\mathrm{d}\tilde{\varphi}}{\mathrm{d}\eta}-3\mathcal{H}^2\tilde{\psi}+k^2\tilde{\varphi}=\frac{a^2\overline{\rho}}{2m_{\mathrm{Pl}}^2}\tilde{\delta}.
\end{equation}
We solve this equation in the matter-dominated epoch,\footnote{In the matter-dominated epoch $\overline{\rho}\,\tilde{\delta}$ can be approximated as just its matter component $\overline{\rho}_{\mathrm{m}}\tilde{\delta}_{\mathrm{m}}$. This happens because 1) background matter dominates over background radiation and 2) perturbations of photons and neutrinos are suppressed because of free-streaming. } that is when
\begin{equation}
\label{eqn:matter_dominated_epoch}
\overline{\rho}(\eta)\simeq\frac{3H_0^2m_{\mathrm{Pl}}^2\Omega_{\mathrm{m},0}}{a^3(\eta)},\qquad\qquad
\tilde{\delta}(\mathbf{k},\eta)\simeq\tilde{\delta}_{\mathrm{m}}(\mathbf{k},\eta),
\end{equation}
so that $a\propto\eta^2$, whence $\mathcal{H}\propto\eta^{-1}$. We then consider very small scales,  i.e. deep inside the particle horizon, which implies that the dominant term on the left-hand side of Eq.~\eqref{eqn_relativistic_poisson_equation_fourier} is proportional to $k^2$:
\begin{equation}
\label{eqn:relativistic_poisson_small_scales}
k\eta\gg1\implies k^2\tilde{\varphi}(\mathbf{k},\eta)\simeq\frac{3}{2}\frac{\Omega_{\mathrm{m},0}H_0^2}{a(\eta)}\tilde{\delta}_{\mathrm{m}}(\eta,\mathbf{k}),
\end{equation}
where $H_0$ and $\Omega_{\mathrm{m},0}$ are the Hubble parameter and the matter density parameter, respectively, both evaluated at present. The relation between the two scalar perturbations $\tilde{\psi}$ and $\tilde{\phi}$ can be read off from the scalar sector of the space-space component of Eq.~\eqref{eqn:einstein_equation}, whose Fourier transform at leading order in conformal Newtonian gauge is 
\begin{equation}
2m_{\mathrm{Pl}}^2k^2\left(\tilde{\varphi}+\tilde{\psi}\right)=3a^2\hat{k}_i\hat{k}^{j}\pi^{i}_{\ j},
\end{equation}
where $\pi^{i}_{\ j}$ is the anisotropic stress. This stress exists for
relativistic species such as photons and neutrinos because of the quadrupole moments of their distributions (as detailed in e.g. \cite{ma1995cosmological}). Since the energy density of both photons and neutrinos is negligible during the matter-dominated epoch, we may write that $\tilde{\varphi}\simeq-\tilde{\psi}$, so that Eq.~\eqref{eqn:relativistic_poisson_small_scales} reduces to
\begin{equation}
\label{eqn:from_psi_to_delta}
\tilde{\psi}(\mathbf{k},\eta)\simeq-\frac{\gamma(\eta)}{k^2}\,\tilde{\delta}_{\mathrm{m}}(\mathbf{k},\eta),
\end{equation}
where we have defined
\begin{equation}
\label{eqn:gamma_definition}
\gamma(\eta)\equiv\frac{3}{2}\frac{\Omega_{\mathrm{m},0}H_0^2}{a(\eta)}.
\end{equation}
As previously noted, Eq.~\eqref{eqn:from_psi_to_delta} is valid only during the matter-dominated epoch and for small scales. However, since the majority of the region of integration in Eq.~\eqref{eqn:lensing_potential_fourier} falls in the matter-dominated epoch, it is a reasonable approximation to write
\begin{equation}
\label{eqn:lensing_potential_matter}
\phi(\versor{n})=-\int_{0}^{\chi_{\ast}}\mathrm{d}\chi\,W(\chi)\gamma(\eta_0-\chi)\int\frac{\mathrm{d}^3\mathbf{k}}{(2\pi)^3}\,\frac{\tilde{\delta}_{\mathrm{m}}(\mathbf{k},\eta_0-\chi)}{k^2}\,e^{-i\mathbf{k}\cdot\chi\,\versor{n}}.
\end{equation}
At the end of this section, we will compare the results found with the approximations ere to the exact results from using \texttt{CLASS} \cite{lesgourgues2011cosmic}.

To proceed further, we must now evaluate the harmonic transform of the CMB lensing potential, which is
\begin{equation}
\label{eqn:lensing_potential_harmonic}
\phi_{\ell m}\equiv\int\mathrm{d}^2\versor{n}\,Y_{\ell m}^*(\versor{n})\phi(\versor{n});
\end{equation}
the reason we are using spin-$0$ spherical harmonics to decomposition $\phi(\versor{n})$ is that the lensing potential is a scalar quantity. We use the plane-wave expansion (as detailed in e.g. \cite{mehrem2011plane}),
\begin{equation}
\label{eqn:plane_wave_expansion}
e^{-i\mathbf{k}\cdot\versor{n}\chi}=4\pi\sum_{L=0}^{\infty}\sum_{M=-L}^{L}(-i)^{L}j_{L}(k\chi)Y_{LM}^*(\versor{k})Y_{LM}(\versor{n}),
\end{equation}
where $j_{L}(k\chi)$ is the spherical Bessel Function (sBF) of order $L$. By substituting Eq.~\eqref{eqn:plane_wave_expansion} in Eq.~\eqref{eqn:lensing_potential_matter} and then inserting what results from this into Eq.~\eqref{eqn:lensing_potential_harmonic}, we find
\begin{equation}
\label{eqn:lensing_potential_harmonic_fourier}
\phi_{\ell m}=-4\pi(-i)^{\ell}\int\frac{\mathrm{d}^3\mathbf{k}}{(2\pi)^3}\,Y_{\ell m}^*(\versor{k})\int_{0}^{\chi_{\ast}}\frac{\mathrm{d}\chi}{k^2}\,W(\chi)\gamma(\chi)j_{\ell}(k\chi)\tilde{\delta}_{\mathrm{m}}(\mathbf{k},\eta_0-\chi),
\end{equation}
where we have used the orthogonality of the spherical harmonics \cite{varshalovich1988quantum},
\begin{equation}
\label{eqn:orthogonality}
\int\mathrm{d}^2\versor{n}\,Y_{\ell m}^*(\versor{n})Y_{LM}(\versor{n})=\delta_{\ell L}^{\mathrm{K}}\delta_{mM}^{\mathrm{K}}.
\end{equation}

\subsection{\label{sub_sec:power_spectrum}CMB Lensing Power Spectrum}
We now compute the angular 2-point correlation function, 
\begin{equation}
\begin{split}
\label{eqn:2PCF}
&\expval{\phi_{\ell m}^*\phi_{\ell^{\prime}m^{\prime}}}=\\
&\qquad=(4\pi)^2i^{\ell-\ell^{\prime}}\int\frac{\mathrm{d}^3\mathbf{k}}{(2\pi)^3}\,Y_{\ell m}(\versor{k})\int\frac{\mathrm{d}^3\mathbf{k}^{\prime}}{(2\pi)^3}\,Y^{*}_{\ell^{\prime}m^{\prime}}(\versor{k}^{\prime})\int_{0}^{\chi_{\ast}}\frac{\mathrm{d}\chi}{k^2}\,W(\chi)\gamma(\chi)j_{\ell}(k\chi)\\
&\qquad\qquad\qquad\qquad\times\int_0^{\chi_{\ast}}\frac{\mathrm{d}\chi^{\prime}}{k^{\prime2}}\,W(\chi^{\prime})\gamma(\chi^{\prime})j_{\ell^{\prime}}(k^{\prime}\chi^{\prime})\expval{\tilde{\delta}^*_{\mathrm{m}}(\mathbf{k},\eta_0-\chi)\tilde{\delta}_{\mathrm{m}}(\mathbf{k}^{\prime},\eta_0-\chi^{\prime})}.
\end{split}
\end{equation}
In the last line, we recognize the matter power spectrum, which, by neglecting non-linear terms, can be written as\footnote{This can be understood by recalling Eq.~\eqref{eqn:gamma_definition}:
\begin{equation}
\tilde{\delta}_{\mathrm{m}}(\mathbf{k},\eta_0-\chi)\simeq\frac{2k^2a(\chi)}{3\Omega_{\mathrm{m},0}H_0^2}\tilde {\varphi}(\mathbf{k},\eta_0-\chi),
\end{equation}
and by using that $\tilde{\varphi}$ can be rewritten in terms of the inflationary comoving curvature perturbation $\tilde{\mathcal{R}}$ as
\begin{equation}
\label{eqn:from_phi_to_R}
\tilde{\varphi}(\mathbf{k},\eta_0-\chi)=\frac{3}{5}\frac{D(\chi)}{a(\chi)}\mathcal{T}_{\delta}(k)\tilde{\mathcal{R}}(\mathbf{k}),
\end{equation}
where $D(\chi)$ and $\mathcal{T}_{\delta}(k)$ are defined below Eq.~\eqref{eqn:matter_power_spectrum}.} \cite{eisenstein1998baryonic, eisenstein1999power, borges2008evolution, slepian2016simple}
\begin{equation}
\begin{split}
\label{eqn:matter_power_spectrum}
&\expval{\tilde{\delta}^*_{\mathrm{m}}(\mathbf{k},\eta_0-\chi)\tilde{\delta}_{\mathrm{m}}(\mathbf{k}^{\prime},\eta_0-\chi^{\prime})}=\\
&\qquad\qquad\qquad\qquad\qquad=D(\chi)D(\chi
^{\prime})\left[\frac{2k^2D_0\mathcal{T}_{\delta}(k)}{5\Omega_{\mathrm{m},0}H_0^2}\right]^2(2\pi)^3\delta_{\mathrm{D}}^{[3]}(\mathbf{k}-\mathbf{k}^{\prime})P_{\mathcal{R}}(k),
\end{split}
\end{equation}
where $D(\chi)$ is the linear growth factor, which describes the wavelength-independent growth of the matter density perturbations, $\mathcal{T}_{\delta}(k)$ is the matter transfer function encoding the dependence on $k$, and $P_{\mathcal{R}}(k)$ is the primordial power spectrum of the comoving curvature perturbation $\tilde{\mathcal{R}}$. 

Therefore, Eq.~\eqref{eqn:2PCF} becomes
\begin{equation}
\begin{split}
\label{eqn:2PCF_power_spectrum}
\expval{\phi_{\ell m}^*\phi_{\ell^{\prime}m^{\prime}}}&=\frac{8\,\delta^{\mathrm{K}}_{\ell\ell^{\prime}}\delta^{\mathrm{K}}_{mm^{\prime}}}{25\pi\,\Omega_{\mathrm{m},0}^2H_0^4}\int_0^{\infty}\mathrm{d}k\,k^2\mathcal{T}^2_{\delta}(k)P_{\mathcal{R}}(k)\\
&\quad\times\int_{0}^{\chi_{\ast}}\mathrm{d}\chi\,W(\chi)\gamma(\chi)j_{\ell}(k\chi)D(\chi)\int_0^{\chi_{\ast}}\mathrm{d}\chi^{\prime}\,W(\chi^{\prime})\gamma(\chi^{\prime})j_{\ell}(k\chi^{\prime})D(\chi^{\prime}),
\end{split}
\end{equation}
where we have used Eq.~\eqref{eqn:orthogonality} to perform the angular integration over $\versor{k}$. To simplify the expression above, we now invoke the Limber approximation \cite{limber1953analysis, lemos2017effect},
\begin{equation}
\label{eqn:limber_approximation}
j_{\ell}(k\chi)\simeq\sqrt{\frac{\pi}{2\ell+1}}\,\frac{1}{k}\,\delta_{\mathrm{D}}^{[1]}\left(k-\frac{\ell}{\chi}\right),
\end{equation}
valid for large multipoles ($\ell\ge10)$, leading in this case to the approximation that $\chi\simeq\chi^{\prime}\simeq\ell/k$. Therefore, we see that for large $\ell$,
\begin{equation}
\label{eqn:2PCF_limber}
\boxed{C_{\ell}^{\phi\phi}\simeq\left(-\frac{5}{2}\Omega_{\mathrm{m},0}H_0^2\right)^{-2}\int_{0}^{\chi_{\ast}}\frac{\mathrm{d}\chi}{\chi^3}\,P_{\mathcal{R}}\left(\frac{\ell}{\chi}\right)\left[D(\chi)W(\chi)\gamma(\chi)\mathcal{T}_{\delta}\left(\frac{\ell}{\chi}\right)\right]^2}
\end{equation}
where we have used
\begin{equation}
\label{eqn:angular_power_spectrum}
\expval{\phi_{\ell m}^*\phi_{\ell^{\prime}m^{\prime}}}=C_{\ell}^{\phi\phi}\delta^{\mathrm{K}}_{\ell\ell^{\prime}}\delta^{\mathrm{K}}_{m m^{\prime}}
\end{equation}
this relationship defines $C_{\ell}^{\phi\phi}$ under the assumption of statistical isotropy. 
Eq.~\eqref{eqn:2PCF_limber} holds for any statically isotropic primordial power spectrum and linear matter transfer function, and except for the usage of the Limber approximation, it is thus completely general.

\subsubsection{Specifying the Primordial Power Spectrum}
\begin{figure}
\centering
\includegraphics[width = \linewidth]{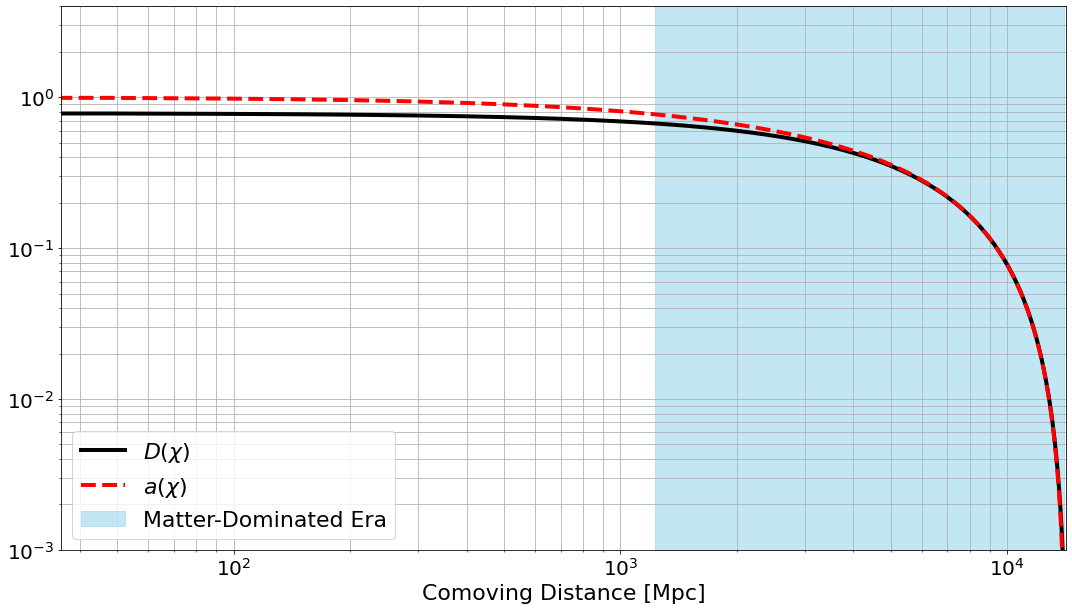}
\caption{The linear growth factor $D$ (black solid line) and the scale factor $a$ (red dashed line) as functions of the comoving distance $\chi$. We have performed the numerical calculation using \texttt{CLASS} \cite{lesgourgues2011cosmic} for the fiducial parameters of the $\Lambda$CDM model \cite{aghanim2020planck}. During the matter-dominated epoch, the linear growth factor is approximately proportional to the scale factor.}
\label{fig:growth_factor}
\end{figure}

To proceed further, we have to specify the primordial power spectrum, which is usually parameterized as\footnote{We have assumed adiabatic initial conditions.}
\begin{equation}
\label{eqn:primordial_power_spectrum}
P_{\mathcal{R}}(k)=\frac{2\pi^2}{k^3}A_s(k_p)\left(\frac{k}{k_p}\right)^{n_s(k_p)-1}
\end{equation}
with $A_s$ and $n_s$ respectively the scalar amplitude and the scalar spectral tilt at the pivot scale $k_p$. We now evaluate the integral expression in Eq.~\eqref{eqn:2PCF_limber}; for this purpose, we substitute the explicit functional forms of $W(\chi)$, $\gamma(\chi)$ and $D(\chi)$. We can exploit the fact that $D\sim D_0a$, as shown in Fig.~\ref{fig:growth_factor}.
Thanks to this simplification the dependence on $\chi$ encoded in $D^2(\chi)$ perfectly cancels with that in $\gamma^2(\chi)$, leaving\footnote{The $D_0\equiv D(\chi=0)$ factor arises from the normalization of the matter transfer function. In our paper, we normalize $\mathcal{T}_{\delta}$ to unity at large scales (as done e.g. in \cite{eisenstein1999power}), resulting in $D_0\simeq0.78$. Conversely,  \texttt{CLASS} \cite{lesgourgues2011cosmic} normalizes $D(\chi)$ to unity at present (i.e. $D_0=1$), while $\mathcal{T}_{\delta}$ is not normalized to unity at large scales.}
\begin{equation}
\label{eqn:2PCF_limber_n_s_1}
\ell^3C_{\ell}^{\phi\phi}=2\pi^2A_s(k_p)\left(\frac{6}{5}D_0\right)^2\int_{0}^{\chi_{\ast}}\frac{\mathrm{d}\chi}{\chi}\,\left(1-\frac{\chi}{\chi_{\ast}}\right)^2\mathcal{T}^{\,2}_{\delta}\left(\frac{\ell}{\chi}\right),
\end{equation}
where we have substituted Eq.~\eqref{eqn:lensing_efficiency} and taken $n_s\simeq1$ for $k_p=0.05\,\mathrm{Mpc}^{-1}$ \cite{akrami2020planck_inflation}. 

\subsubsection{Specifying the Matter Transfer Function}
\begin{figure}
\centering
\includegraphics[width = \linewidth]{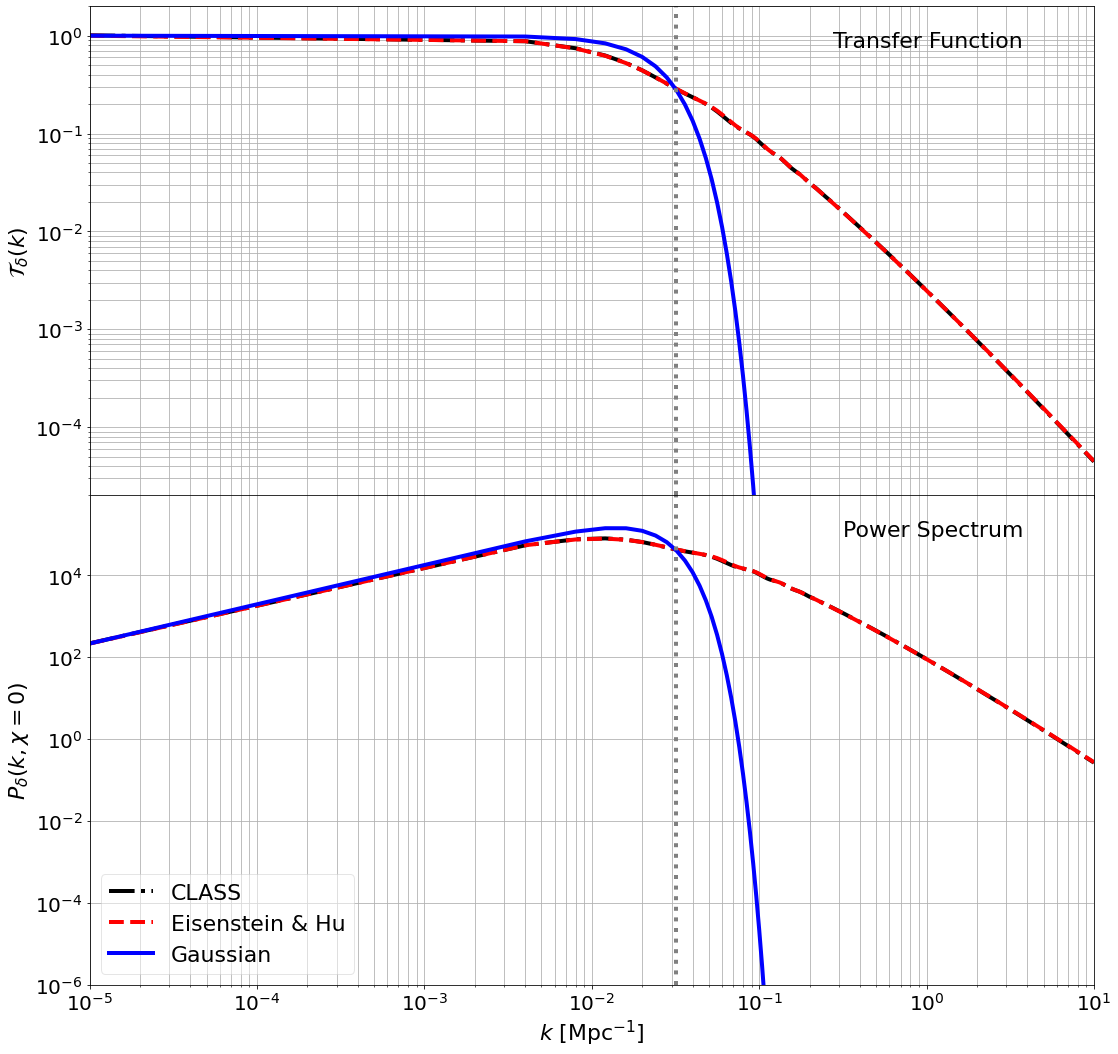}
\caption{Matter power spectrum $P_{\delta}$ evaluated today ($\chi=0$), and the associated transfer function. The black dash-dotted lines represent the result obtained with \texttt{CLASS} for the fiducial parameters of the $\Lambda$CDM model from \cite{aghanim2020planck}, while the dashed red curves have been obtained by numerically evaluating the Eisenstein \& Hu (EH) transfer function \cite{eisenstein1998baryonic} with \texttt{CosmoloPy} \cite{kramer2020cosmolopy}. The blue solid lines in the upper and lower panels display, respectively, Eq.~\eqref{eqn:matter_power_spectrum} and Eq.~\eqref{eqn:gaussian_transfer_function}, for $\sigma=0.02\,\mathrm{Mpc}^{-1}$, tuned to match the exact result as much as possible. The vertical dotted grey lines indicate when the mismatch between the Gaussian transfer function and the EH one starts to be larger than 10\%. The EH and CLASS curves match almost perfectly.}
\label{fig:matter}
\end{figure}

To perform the integration over $\chi$, the last ingredient is $\mathcal{T}_{\delta}(k)$. The rigorous way to obtain it would be to fully solve the system of Einstein-Boltzmann equations, but we can avoid this procedure by using a fit, such as Eisenstein \& Hu (EH) or Bardeen-Bond-Kaiser-Szalay (BBKS). \cite{eisenstein1998baryonic, eisenstein1999power, bardeen1986statistics}. Here, we also test a Gaussian matter transfer function that effectively mimics the quadratic fall-off observed at scales crossing the horizon approximately at the time of matter-radiation equality:\footnote{Indeed, a rough approximation for the overall matter transfer function shape is $1/[1+(k/k_{\mathrm{eq}})^2]$, where $k_{\mathrm{eq}}$ is the wavenumber corresponding to a scale which enters the horizon at the equivalence epoch. The Gaussian matches this at leading order in a Taylor expansion for $k\ll \sigma\sim k_{\mathrm{eq}}\sim10^{-2}$.}
\begin{equation}
\label{eqn:gaussian_transfer_function}
\mathcal{T}_{\delta}(k)\simeq\exp\left(-\frac{k^2}{2\sigma^2}\right).
\end{equation}
The comparison between the different matter transfer functions and the impact of using each in the matter power spectrum is shown in Fig.~\ref{fig:matter}.
Although our Gaussian ansatz for the matter transfer function seems not to be accurate at small scales, we will soon see that it produces a very good approximation for the CMB lensing power spectrum. Furthermore, it permits us to obtain a closed-form result for the latter.

\begin{figure}
\centering
\includegraphics[width = \linewidth]{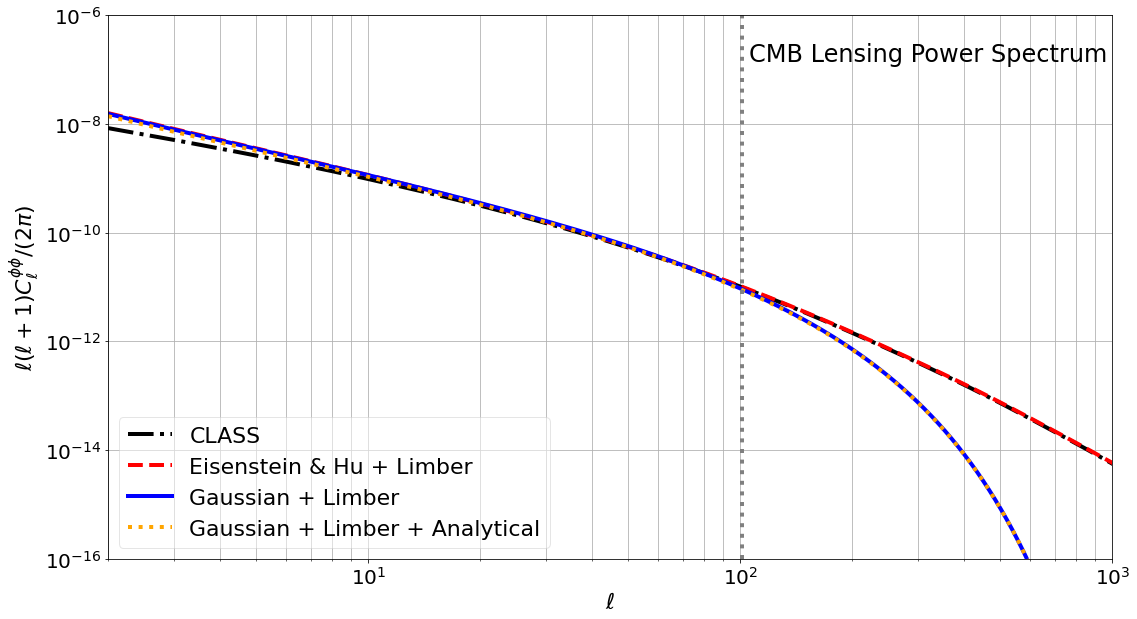}
\caption{Angular power spectrum of the CMB lensing potential. The black dash-dotted curve is the $C_{\ell}$ fully obtained with \texttt{CLASS} without relying on any approximation, while the red dashed curve stems from using the Limber approximation and adopting the EH matter transfer function given Eq.~\eqref{eqn:2PCF_limber_n_s_1}. The blue solid line is from numerical integration of Eq.~\eqref{eqn:2PCF_limber_gaussian}, and the orange dotted line is nothing but the closed-form analytic result of Eq.~\eqref{eqn:2PCF_expn}, for $A_s=2.1\times10^{-9}$, $\Omega_{\mathrm{m},0}=0.3$, $H_0=2.3\times10^{-4}\,\mathrm{Mpc}^{-1}$, $\chi_{\ast}=1.4\times10^{4}\,\mathrm{Mpc}$ and $\sigma=0.02\,\mathrm{Mpc}^{-1}$. The vertical dotted grey line indicates when the mismatch between the angular power spectrum evaluated by using the Gaussian transfer function, and that evaluated using the EH transfer function, starts to exceed 10\%.}
\label{fig:lensing}
\end{figure}

By substituting Eq.~\eqref{eqn:gaussian_transfer_function} in Eq.~\eqref{eqn:2PCF_limber_n_s_1}, and changing variable to $y\equiv\chi_{\ast}/\chi$, we find
\begin{equation}
\label{eqn:2PCF_limber_gaussian}
\ell^3C_{\ell}^{\phi\phi}=2\pi^2A_s(k_p)\,\alpha^2\int_{1}^{\infty}\frac{\mathrm{d}y}{y^3}\,\left(1-y\right)^2e^{-2\beta^2\ell^2y^2},
\end{equation}
where we have defined
\begin{equation}
\label{eqn:alpha_and_beta}
\alpha\equiv-\frac{6}{5}D_0,\qquad\qquad\qquad\beta\equiv\frac{1}{\sqrt{2}\,\sigma\chi_{\ast}}.
\end{equation}
As we will see, we define $\alpha$ in this way because it will enter with different exponents in the higher-order correlation functions (so its sign does matter). The integral on the right-hand side of Eq.~\eqref{eqn:2PCF_limber_gaussian} can be performed in terms of the generalized exponential integral \cite{abramowitz1968handbook}:
\begin{equation}
\label{eqn:generalized_exponential_integral}
\mathrm{E}_n(x)\equiv\int_1^{\infty}\frac{\mathrm{d}t}{t^n}e^{-xt},
\end{equation}
yielding
\begin{equation}
\label{eqn:2PCF_expn}
\ell^3C_{\ell}^{\phi\phi}=A_s(k_p)\pi^2\,\alpha^2\left[\mathrm{E}_{2}\left(\xi_2\right)-2\mathrm{E}_{3/2}\left(\xi_2\right)+\mathrm{E}_{1}\left(\xi_2\right)\right],
\end{equation}
where we have defined $\xi_2\equiv2\ell^2\beta^2$. The structure of Eq.~\eqref{eqn:2PCF_expn} follows from expanding the integrand using the binomial expansion. As detailed e.g. in \href{https://dlmf.nist.gov/8.20}{Section~8.20 of} \cite{NIST:DLMF}, for $x$ approaching infinity, 
\begin{equation}
\label{eqn:exponential_asymptotics}
\mathrm{E}_n(x\to\infty)\sim\frac{e^{-x}}{x}+\sum_{s=1}^{\infty}\frac{(-1)^s}{x^s}\frac{\Gamma(n+s)}{\Gamma(n)},
\end{equation}
where $\Gamma$ is the Euler gamma function. Hence, at leading order $\mathrm{E}_n(x)$ becomes independent of $n$, so that all the terms on the right-hand side of Eq.~\eqref{eqn:2PCF_expn} perfectly cancel out, yielding an identically zero result. This is consistent with the behaviour we see in Fig.~\ref{fig:lensing}, and the same feature will also be present in the lensing bispectrum and trispectrum. We have numerically evaluated Eq.~\eqref{eqn:2PCF_limber_gaussian} and we have compared it
both with the exact \texttt{CLASS} result and with the result from using EH transfer function with the Limber approximation applied. This comparison is shown in Fig.~\ref{fig:lensing}.

The results of this section are not new, but our closed-form expressions with the Gaussian toy transfer function are. Here, we have tested the approximations used to simplify our calculations, namely the Limber approximation, the assumption \( D \simeq D_0 a \), and the Gaussian transfer function. While not critical for the 2-point correlation function, these approximations will be valuable for the bispectrum and trispectrum, given computational demands of these latter. Fig.~\ref{fig:lensing} shows that the power spectrum of the CMB lensing potential is accurate for $\ell\lesssim100$ with the Gaussian transfer function.
 
\section{\label{sec:bispectrum}CMB Lensing Bispectrum}
In this section, we revisit the lensing bispectrum calculation of \cite{bohm2016bias}. While we do not introduce any new methods, we explicitly present several intermediate steps to clarify the derivation. We do so since some of these steps will need to change when we move to the lensing trispectrum. Hence establishing them first will enable more clearly displaying what changes when we consider the trispectrum. Furthermore, the conventions already introduced in Section~\ref{sec:setup} are slightly different from those in \cite{bohm2016bias}, and so it is good to have a self-consistent derivation of the lensing bispectrum and trispectrum, to better enable comparison. We also add a bit more extended explanation relative to \cite{bohm2016bias}, to build physical understanding that will usefully guide the trispectrum calculation. 

We begin with the angular 3-point correlation of the lensing potential $\phi$ as a function of the direction vector $\versor{n}$ to each of three CMB points from us, as the observer. As we implicitly assumed in Section~\ref{sec:setup}, the observer is at the origin of coordinates. By recalling Eq.~\eqref{eqn:lensing_potential_harmonic_fourier}, we then move to Fourier space, finding
\begin{equation}
\label{eqn:3PCF}
\begin{split}
\expval{\phi_{\ell_1m_1}\phi_{\ell_2m_2}\phi_{\ell_3m_3}}&=-\int\frac{\mathrm{d}^3\mathbf{k}_1}{(2\pi)^3}\int\frac{\mathrm{d}^3\mathbf{k}_2}{(2\pi)^3}\int\frac{\mathrm{d}^3\mathbf{k}_3}{(2\pi)^3}\int_{0}^{\chi_{\ast}}\frac{\mathrm{d}\chi_1}{k^2_1}\int_{0}^{\chi_{\ast}}\frac{\mathrm{d}\chi_2}{k^2_2}\int_{0}^{\chi_{\ast}}\frac{\mathrm{d}\chi_3}{k^2_3}\,\\
&\qquad\quad\times\prod_{n=1}^3\left[4\pi(-i)^{\ell_n}Y_{\ell_nm_n}^*(\versor{k}_n)W(\chi_n)\gamma(\chi_n)j_{\ell_n}(k_n\chi_n)\right]\\
&\qquad\qquad\quad\times\expval{\tilde{\delta}_{\mathrm{m}}(\mathbf{k}_1,\eta_0-\chi_1)\tilde{\delta}_{\mathrm{m}}(\mathbf{k}_2,\eta_0-\chi_2)\tilde{\delta}_{\mathrm{m}}(\mathbf{k}_3,\eta_0-\chi_3)}
\end{split}
\end{equation}
We note that when there is a constant factor inside the $\Pi_{n=1}^N$ operator (such as the $4\pi$), it scales each term in the product by that constant, resulting in an overall $(4\pi)^N$ factor.

The quantity in the last line of the right-hand side of Eq.~\eqref{eqn:3PCF} is the matter bispectrum, which in analogy with the matter power spectrum, can be factorized into its primordial source, i.e. the inflationary three-point correlation function, and its deterministic component associated with the matter evolution, which latter is a  product of linear growth factors and matter transfer functions (as detailed in e.g. \cite{shiraishi2011cmb}):
\begin{equation}
\label{eqn:matter_bispectrum}
\begin{split}
&\expval{\tilde{\delta}_{\mathrm{m}}(\mathbf{k}_1,\eta_0-\chi_1)\tilde{\delta}_{\mathrm{m}}(\mathbf{k}_2,\eta_0-\chi_2)\tilde{\delta}_{\mathrm{m}}(\mathbf{k}_3,\eta_0-\chi_3)}=\\
&\qquad\qquad\qquad=\prod_{n=1}^3\left[\frac{2}{5}\frac{D(\chi_n)k_n^2}{\Omega_{\mathrm{m},0}H_0^2}\mathcal{T}_{\delta}(k_n)\right](2\pi)^3\delta_{\mathrm{D}}^{[3]}\left(\mathbf{k}_1+\mathbf{k}_2+\mathbf{k}_3\right)B_{\mathcal{R}}(k_1,k_2,k_3),
\end{split}
\end{equation}
where $B_{\mathcal{R}}(k_1,k_2,k_3)$ is the primordial bispectrum of the comoving curvature perturbation. The bispectrum's amplitude is predicted to be tiny in the standard single-field slow-roll models of inflation \cite{bartolo2004non, akrami2020planck_nongaussianity}. 
Importantly, in Eq.~\eqref{eqn:matter_bispectrum} the matter bispectrum's spatial wave-vector dependence can be parameterized in terms of just the lengths $k_n$. This is a crucial simplification that will not be available to us in the trispectrum case. For the matter bispectrum, the whole dependence on $\versor{k}_n$ is just encoded in the Dirac delta ensuring homogeneity. As we will show very soon, this is the reason why the CMB lensing angular bispectrum is not sensitive to parity violation. We now rewrite the Dirac delta in terms of its definition in Fourier space, and use again the plane-wave expansion introduced in Eq.~\eqref{eqn:plane_wave_expansion}:
\begin{equation}
\label{eqn:dirac_delta_plane_waves}
\begin{split}
\delta_{\mathrm{D}}^{[3]}\left(\mathbf{k}_1+\mathbf{k}_2+\mathbf{k}_3\right)&=\int\frac{\mathrm{d}^3\mathbf{x}}{(2\pi)^3}\,\exp\left[i\left(\mathbf{k}_1+\mathbf{k}_2+\mathbf{k}_3\right)\cdot\mathbf{x}\right]\\
&=8\sum_{L_aM_a}\int_0^{\infty}\mathrm{d}x\,x^2\prod_{n=1}^3\left[i^{L_n}j_{L_n}(k_nx)Y_{L_nM_n}(\versor{k}_n)\right]\\
&\qquad\qquad\qquad\qquad\qquad\quad\times\int\mathrm{d}^2\versor{x}\,Y_{L_1M_1}^*(\versor{x})Y_{L_2M_2}^*(\versor{x})Y_{L_3M_3}^*(\versor{x}),
\end{split}
\end{equation}
with $a=1,2,3$. The last line is the Gaunt integral, defined as \cite{varshalovich1988quantum}:
\begin{equation}
\label{eqn:gaunt_integral}
\mathcal{G}_{L_1L_2L_3}^{M_1M_2M_3}\equiv\sqrt{\frac{(2L_1+1)(2L_2+1)(2L_3+1)}{4\pi}}\begin{pmatrix}
L_1 & L_2 & L_3 \\
0 & 0 & 0
\end{pmatrix}
\begin{pmatrix}
L_1 & L_2 & L_3 \\
M_1 & M_2 & M_3
\end{pmatrix}
\end{equation}
The quantities resembling matrices are Wigner 3-$j$ symbols, related to the Clebsch–Gordan coefficients, and they obey the following selection rules (as detailed e.g. in \href{https://dlmf.nist.gov/34.2}{Section~34.2 of} \cite{NIST:DLMF}):
\begin{equation}
\label{eqn:selection_rules}
|L_1-L_2|\le L_3\le L_1+L_2,\qquad\qquad M_1+M_2+M_3=0.
\end{equation}
In particular, when the elements of the second row of a Wigner 3-$j$ symbol are all zeros, then $L_1+L_2+L_3$ must be even. Therefore, Eq.~\eqref{eqn:3PCF} reduces to
\begin{equation}
\label{eqn:3PCF_Gaunt}
\begin{split}
&\expval{\phi_{\ell_1m_1}\phi_{\ell_2m_2}\phi_{\ell_3m_3}}=-(-i)^{\ell_1+\ell_2+\ell_3}\sum_{L_1M_1}\sum_{L_2M_2}\sum_{L_3M_3}i^{L_1+L_2+L_3}\mathcal{G}_{L_1L_2L_3}^{M_1M_2M_3}\\
&\quad\times\int\mathrm{d}^3\mathbf{k}_1\int\mathrm{d}^3\mathbf{k}_2\int\mathrm{d}^3\mathbf{k}_3\,B_{\mathcal{R}}(k_1,k_2,k_3)\int_{0}^{\chi_{\ast}}\mathrm{d}\chi_1\int_{0}^{\chi_{\ast}}\mathrm{d}\chi_2\int_{0}^{\chi_{\ast}}\mathrm{d}\chi_3\int_{0}^{\infty}\mathrm{d}x\,x^2\\
&\qquad\times\prod_{n=1}^3\left[\frac{4D_0W(\chi_n)\gamma(\chi_n)}{5\pi\Omega_{\mathrm{m},0}H_0^2}D(\chi_n)j_{\ell_n}(k_n\chi_n)j_{L_n}(k_nx)\mathcal{T}_{\delta}(k_n)Y_{\ell_nm_n}^*(\versor{k}_n)Y_{L_nM_n}(\versor{k}_n)\right].\\
\end{split}
\end{equation}
By recalling Eq.~\eqref{eqn:orthogonality}, we can easily evaluate the integrals over $\versor{k}_n$, so that we obtain
\begin{equation}
\label{eqn:3PCF_only_ells}
\begin{split}
&\expval{\phi_{\ell_1m_1}\phi_{\ell_2m_2}\phi_{\ell_3m_3}}=-\mathcal{G}_{\ell_1\ell_2\ell_3}^{m_1m_2m_3}\int_0^{\infty}\mathrm{d}x\,x^2\int_0^{\infty}\mathrm{d}k_1\int_0^{\infty}\mathrm{d}k_2\int_0^{\infty}\mathrm{d}k_3\,B_{\mathcal{R}}(k_1,k_2,k_3)\\
&\times\int_{0}^{\chi_{\ast}}\mathrm{d}\chi_1\int_{0}^{\chi_{\ast}}\mathrm{d}\chi_2\int_{0}^{\chi_{\ast}}\mathrm{d}\chi_3\prod_{n=1}^3\left[\frac{4D_0W(\chi_n)\gamma(\chi_n)}{5\pi\Omega_{\mathrm{m},0}H_0^2D^{-1}(\chi_n)}j_{\ell_n}(k_n\chi_n)j_{\ell_n}(k_nx)k^2_n\mathcal{T}_{\delta}(k_n)\right].
\end{split}
\end{equation}

Proceeding to the next step, we find that while the argument presented in \cite{bohm2016bias} appears flawed, the final result is nonetheless correct. Specifically, \cite{bohm2016bias} suggests that the closure relation of spherical Bessel functions must be used to show that \(\chi_1 \simeq \chi_2 \simeq \chi_3\). It does not appear to us that this is the correct logic. It could be that \cite{bohm2016bias} meant that we should try to perform the $k_{n}$ integrals. However, these integrals also involve the bispectrum, which explicitly depends on $k_{n}$; thus they do not give us the closure relation. Attempting to perform integration over each \(k_n\) while enforcing this condition leads to a problem. The integrals also involve the bispectrum, which explicitly depends on \(k_n\). Similarly, attempting to perform the \(\chi_n\) integrals does not yield pairs of spherical Bessel functions, and the same issue arises when considering the \(r\) integral. Furthermore, applying the closure relation for spherical Bessel functions would leave residual SBFs in the expression, undermining the justification for employing the Limber approximation. Hence, we adopt again the Limber approximation introduced in Eq.~\eqref{eqn:limber_approximation} for each of the six spherical Bessel functions on the right-hand side of Eq.~\eqref{eqn:3PCF_only_ells}, leading to
\begin{equation}
\label{eqn:limber_on_bispectrum}
\chi_n\simeq\frac{\ell_n}{k_n}\simeq x\equiv\chi\quad\text{for $\,n=1,2,3$},
\end{equation}
so that we obtain that for large $\ell$
\begin{equation}
\label{eqn:3PCF_limber}
\begin{split}
&\expval{\phi_{\ell_1m_1}\phi_{\ell_2m_2}\phi_{\ell_3m_3}}\simeq\left(\frac{5}{2}\Omega_{\mathrm{m},0}H_0^2\right)^{-3}\mathcal{G}_{\ell_1\ell_2\ell_3}^{m_1m_2m_3}\\
&\qquad\qquad\qquad\quad\times\int_{0}^{\chi_{\ast}}\frac{\mathrm{d}\chi}{\chi^4}D^3(\chi)W^3(\chi)\,\mathcal{T}_{\delta}\left(\frac{\ell_1}{\chi}\right)\mathcal{T}_{\delta}\left(\frac{\ell_2}{\chi}\right)\mathcal{T}_{\delta}\left(\frac{\ell_3}{\chi}\right)B_{\mathcal{R}}\left(\frac{\ell_1}{\chi},\frac{\ell_2}{\chi},\frac{\ell_3}{\chi}\right).
\end{split}
\end{equation}
We notice that Eq.~\eqref{eqn:3PCF_limber} is the same result as \cite{bohm2016bias}, despite the slightly different approach (as shown in Appendix~\ref{app:recovering_bohm}). We see that the 3-point angular correlation function of the CMB lensing potential can only be parity-even, because of the Gaunt integral defined in Eq.~\eqref{eqn:gaunt_integral}, which requires that the sum $\ell_1 + \ell_2 + \ell_3$ is even. While the geometry of the angular bispectrum is encoded in the Gaunt integral, the physical quantity is the so called reduced bispectrum \cite{Spergel:1999xn}, defined as $b_{\ell_1\ell_2\ell_3}^{\phi\phi\phi}\equiv\expval{\phi_{\ell_1m_1}\phi_{\ell_2m_2}\phi_{\ell_3m_3}}\left(\mathcal{G}_{\ell_1\ell_2\ell_3}^{m_1m_2m_3}\right)^{-1}$, so that we find
\begin{equation}
\label{eqn:reduced_bispectrum}
\boxed{\begin{aligned}
b_{\ell_1\ell_2\ell_3}^{\phi\phi\phi}&=\left(-\frac{5}{2}\Omega_{\mathrm{m},0}H_0\right)^{-3}\\
&\qquad\qquad\times\int_{0}^{\chi_{\ast}}\frac{\mathrm{d}\chi}{\chi^4}\prod_{n=1}^3\left[D(\chi)W(\chi)\gamma(\chi)\,\mathcal{T}_{\delta}\left(\frac{\ell_n}{\chi}\right)\right]B_{\mathcal{R}}\left(\frac{\ell_1}{\chi},\frac{\ell_2}{\chi},\frac{\ell_3}{\chi}\right),
\end{aligned}}
\end{equation}
valid if $\ell_1+\ell_2+\ell_3$ is even \cite{komatsu2001acoustic}, as in our case. Eq.~\eqref{eqn:2PCF_limber} holds for any statically isotropic primordial bispectrum and linear matter transfer function, and except for the usage of the Limber approximation, it is thus completely general.

\subsection{Specifying the Primordial Bispectrum}
We assume a local shape for the primordial bispectrum \cite{acquaviva2002second, maldacena2003non}, as
\begin{equation}
\label{eqn:primordial_bispectrum}
B_{\mathcal{R}}(k_1,k_2,k_3)=\left[2\pi^2A_s(k_p)\right]^2f_{\text{NL}}^{\text{loc}}\left(\frac{k_1^3+k_2^3+k_3^3}{k_1^3k_2^3k_3^3}\right),
\end{equation}
conceptually aligned with the scale-invariant power spectrum defined in Eq.~\eqref{eqn:primordial_power_spectrum}, $f_{\text{NL}}^{\text{loc}}$ encodes the amount of primordial non-Gaussianity \cite{bartolo2004non}. We thus evaluate Eq.~\eqref{eqn:reduced_bispectrum}, by substituting the definition of $W(\chi)$ and $\gamma(\chi)$:
\begin{equation}
\label{eqn:3PCF_numerical}
\begin{split}
b_{\ell_1\ell_2\ell_3}^{\phi\phi\phi}&\simeq-\left(\frac{6}{5}D_0\right)^3\left[2\pi^2A_s(k_p)\right]^2f_{\text{NL}}^{\text{loc}}\\
&\qquad\times\left(\frac{\ell_1^3+\ell_2^3+\ell_3^3}{\ell_1^3\ell_2^3\ell_3^3}\right)\int_{1}^{\infty}\frac{\mathrm{d}y}{y^4}\left(1-y\right)^3\,\mathcal{T}_{\delta}\left(\frac{\ell_1y}{\chi_\ast}\right)\mathcal{T}_{\delta}\left(\frac{\ell_2y}{\chi_\ast}\right)\mathcal{T}_{\delta}\left(\frac{\ell_3y}{\chi_\ast}\right),
\end{split}
\end{equation}
where we have used again $D\simeq D_0a$, and performed the change of variable $y\equiv\chi_{\ast}/\chi$.

\subsection{Specifying the Matter Transfer Function}
\begin{figure}
\centering
\includegraphics[width =\linewidth]{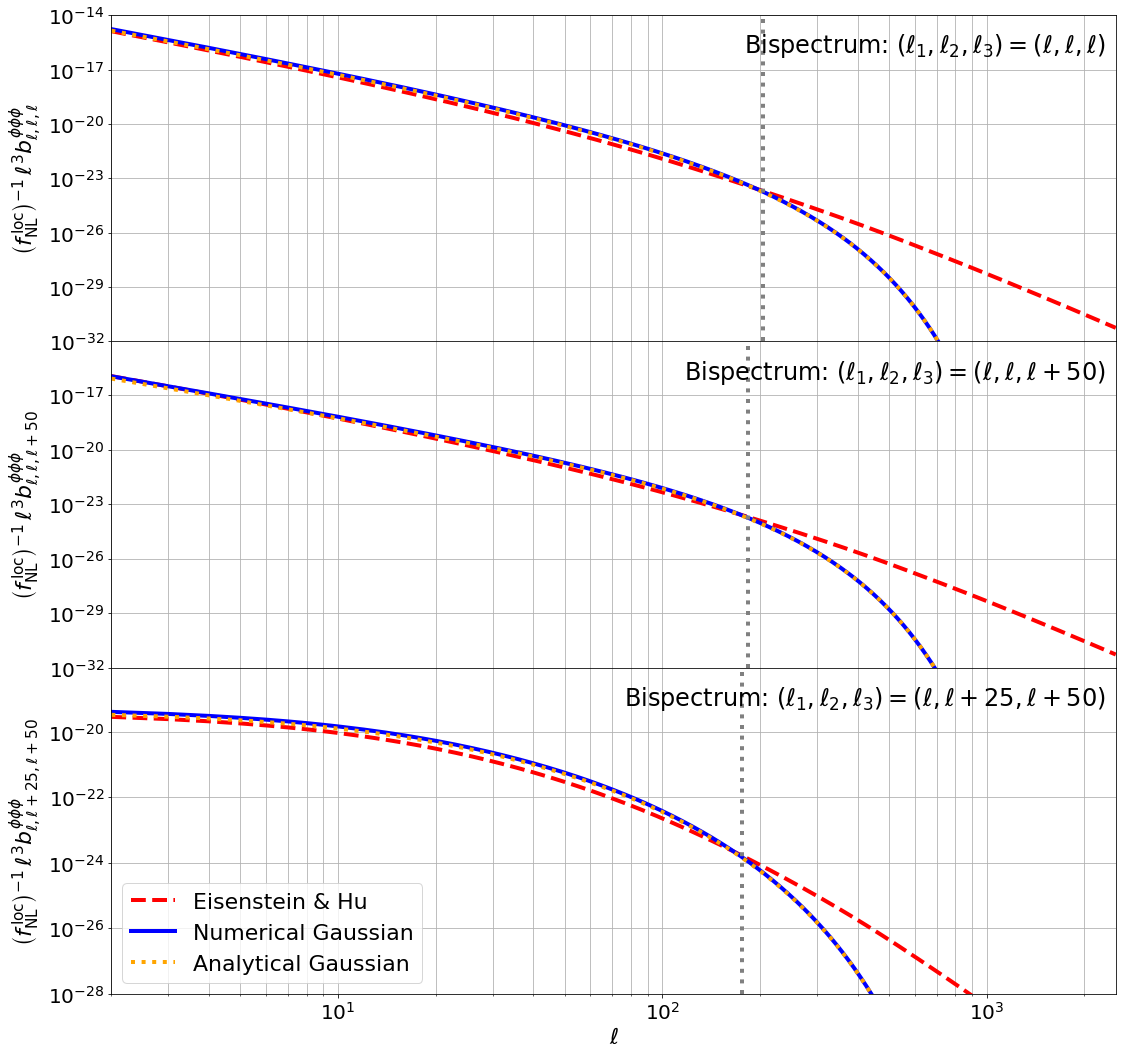}
\caption{The reduced bispectrum given Eq.~\eqref{eqn:3PCF_numerical} as a function of $\ell$ in three different configurations: the equilateral (upper), an example of isosceles (middle), and an example of scalene (bottom). The red dashed curve has been obtained by using the EH matter transfer function. The blue solid line is from numerical integration of Eq.~\eqref{eqn:gaussian_numerical}, and the orange dotted line is from Eq.~\eqref{eqn:3PCF_expn}. The vertical dotted grey lines indicate when the mismatch between the bispectrum evaluated by using the Gaussian transfer function and that evaluated with the EH transfer function starts to be larger than 10\%.}
\label{fig:bispectrum_1D_configs}
\end{figure}

\begin{figure}
\centering
\includegraphics[width =\linewidth]{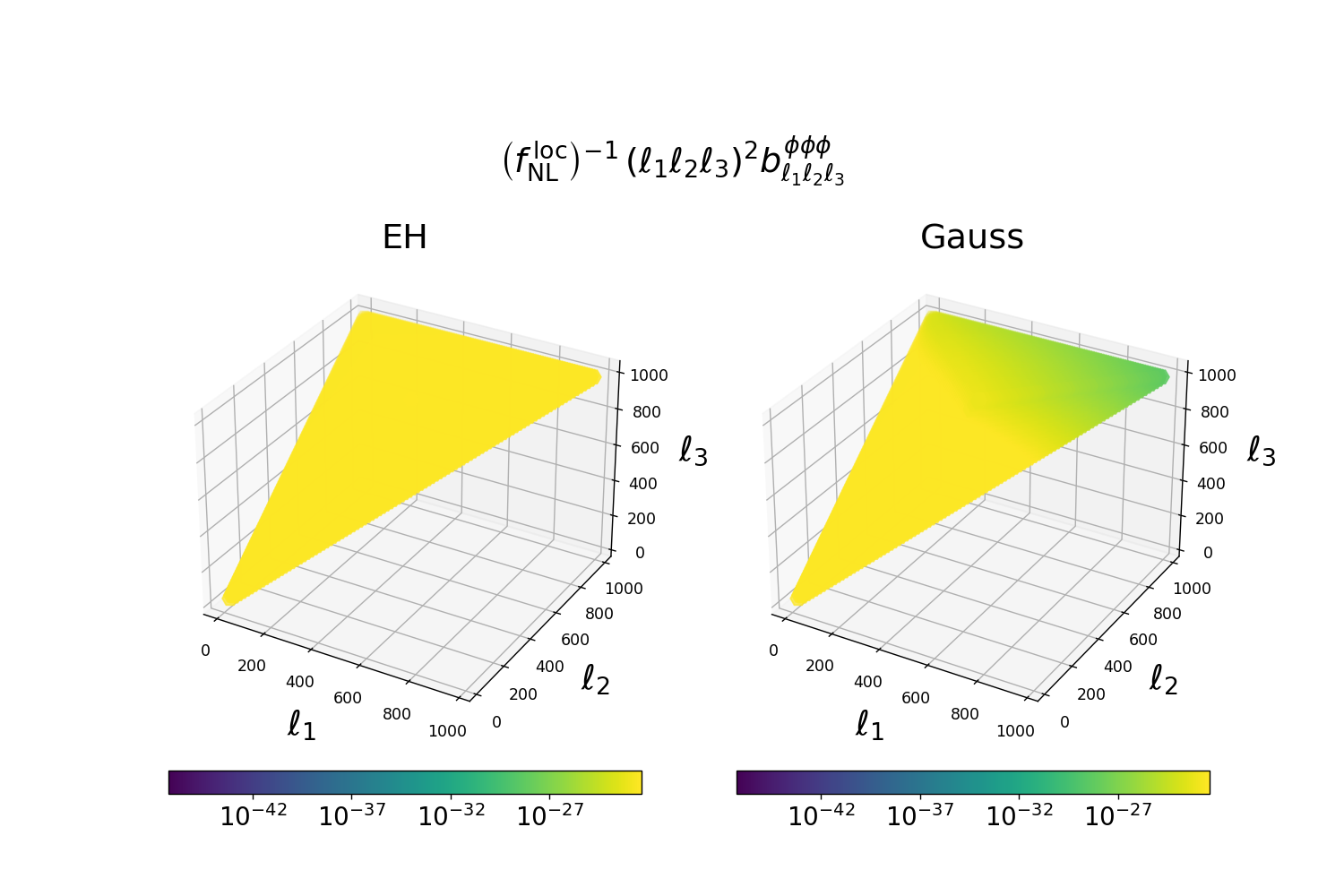}
\caption{The reduced bispectrum given Eq.~\eqref{eqn:3PCF_numerical} as a function of $\ell_1$, $\ell_2$, $\ell_3$. We have restricted our analysis to $\ell_1<\ell_2<\ell_3$, and we chosen $\ell$'s satisfying the Gaunt integral's selection rules, and $\ell_1+\ell_2+\ell_3=$ even.}
\label{fig:bispectrum_grid}
\end{figure}

We now use the Gaussian toy model for the matter transfer function we already introduced in Section~\ref{sub_sec:power_spectrum}:
\begin{equation}
\label{eqn:gaussian_numerical}
\begin{split}
b_{\ell_1\ell_2\ell_3}^{\phi\phi\phi}&\simeq\left[2\pi^2A_s(k_p)\right]^2f_{\text{NL}}^{\text{loc}}\left(\frac{\ell_1^3+\ell_2^3+\ell_3^3}{\ell_1^3\ell_2^3\ell_3^3}\right)\alpha^3\int_{1}^{\infty}\frac{\mathrm{d}y}{y^4}\left(1-y\right)^3e^{-\beta^2(\ell_1^2+\ell_2^2+\ell_3^2)y^2},
\end{split}
\end{equation}
where we have substituted the definitions of $\alpha$ and $\beta$, given in Eq.~\eqref{eqn:alpha_and_beta}. We notice that the structure of this expression greatly resembles that of Eq.~\eqref{eqn:2PCF_limber_gaussian}. Thus we can again use Eq.~\eqref{eqn:generalized_exponential_integral} to perform the integral over $y$, finding
\begin{equation}
\label{eqn:3PCF_expn}
\begin{split}
&\left[\pi^2A_s(k_p)\right]^{-2}\left(\frac{\ell_1^3\ell_2^3\ell_3^3}{\ell_1^3+\ell_2^3+\ell_3^3}\right)\frac{b_{\ell_1\ell_2\ell_3}^{\phi\phi\phi}}{2\alpha^3f_{\text{NL}}^{\text{loc}}}=\\
&\qquad\qquad\qquad\qquad\qquad\qquad\qquad=\left[\mathrm{E}_{5/2}(\xi_3)-3\mathrm{E}_{2}(\xi_3)+3\mathrm{E}_{3/2}(\xi_3)-\mathrm{E}_{1}(\xi_3)\right],
\end{split}
\end{equation}
where $\xi_3\equiv\beta^2(\ell_1^2+\ell_2^2+\ell_3^2)$. We numerically evaluated Eqs.~\eqref{eqn:3PCF_expn} and \eqref{eqn:3PCF_limber} to compare the bispectrum from a Gaussian transfer function with that from the EH fit of the exact transfer function.
The results of the comparison are shown in Figs.~\ref{fig:bispectrum_1D_configs}--\ref{fig:bispectrum_grid}.
As in the power spectrum case, we here notice that the mismatch between the two bispectra (EH versus Gaussian transfer functions) is below 10\% for $\ell\lesssim100$, while it worsens for larger multiples.

\section{\label{sec:trispectrum}CMB Lensing Trispectrum} 
Following the same approach as in the previous sections, we now compute the CMB lensing angular trispectrum. A disconnected trispectrum arises naturally even in the absence of primordial non-Gaussianity. It can be expressed as a sum of products of power spectra. due to the contribution of disconnected terms:
\begin{equation}
\label{eqn:full_4PCF}
\begin{split}
\expval{\phi_{\ell_1m_1}\phi_{\ell_2m_2}\phi_{\ell_3m_3}\phi_{\ell_4m_4}}&=\expval{\phi_{\ell_1m_1}\phi_{\ell_2m_2}}\expval{\phi_{\ell_3m_3}\phi_{\ell_4m_4}}+\expval{\phi_{\ell_1m_1}\phi_{\ell_3m_3}}\expval{\phi_{\ell_2m_2}\phi_{\ell_4m_4}}\\
&\,\,\,\,\,+\expval{\phi_{\ell_1m_1}\phi_{\ell_4m_4}}\expval{\phi_{\ell_2m_2}\phi_{\ell_3m_3}}+\expval{\phi_{\ell_1m_1}\phi_{\ell_2m_2}\phi_{\ell_3m_3}\phi_{\ell_4m_4}}_{\text{c}}.
\end{split}
\end{equation}
We then also a have a connected contribution, which we indicate with a subscript $c$. We have used Isserlis' theorem \cite{isserlis1918formula}, also known as Wick's theorem, to write down explicitly the Gaussian disconnected piece. It is then trivial to see that any kind of parity-violating signature can only arise in the non-Gaussian term in Eq.~\eqref{eqn:full_4PCF}, since the disconnected contributions must to have $\ell_1+\ell_2+\ell_3+\ell_4=$ even to an even number by virtue of Eq.~\eqref{eqn:angular_power_spectrum}. Therefore, we now focus on the purely non-Gaussian (connected) piece, which is
\begin{equation}
\label{eqn:4PCF}
\begin{split}
&\expval{\phi_{\ell_1m_1}\phi_{\ell_2m_2}\phi_{\ell_3m_3}\phi_{\ell_4m_4}}_{\text{c}}=\int\frac{\mathrm{d}^3\mathbf{k}_1}{(2\pi)^3}\int\frac{\mathrm{d}^3\mathbf{k}_2}{(2\pi)^3}\int\frac{\mathrm{d}^3\mathbf{k}_3}{(2\pi)^3}\int\frac{\mathrm{d}^3\mathbf{k}_4}{(2\pi)^3}\int_{0}^{\chi_{\ast}}\frac{\mathrm{d}\chi_1}{k^2_1}\int_{0}^{\chi_{\ast}}\frac{\mathrm{d}\chi_2}{k^2_2}\\
&\qquad\qquad\qquad\quad\times\int_{0}^{\chi_{\ast}}\frac{\mathrm{d}\chi_3}{k^2_3}\int_{0}^{\chi_{\ast}}\frac{\mathrm{d}\chi_4}{k^2_4}\prod_{n=1}^4\left[4\pi(-i)^{\ell_n}Y_{\ell_nm_n}^*(\versor{k}_n)W(\chi_n)\gamma(\chi_n)j_{\ell_n}(k_n\chi_n)\right]\\
&\qquad\qquad\qquad\quad\times\expval{\tilde{\delta}_{\mathrm{m}}(\mathbf{k}_1,\eta_0-\chi_1)\tilde{\delta}_{\mathrm{m}}(\mathbf{k}_2,\eta_0-\chi_2)\tilde{\delta}_{\mathrm{m}}(\mathbf{k}_3,\eta_0-\chi_3)\tilde{\delta}_{\mathrm{m}}(\mathbf{k}_4,\eta_0-\chi_4)}_{\text{c}}.
\end{split}
\end{equation}
In contrast to the bispectrum, we cannot rewrite the matter trispectrum so that the dependence of the matter trispectrum on the $\versor{k}$ is just encoded in the Dirac delta. The connected matter trispectrum is
\begin{equation}
\label{eqn:matter_trispectrum}
\begin{split}
&\expval{\tilde{\delta}_{\mathrm{m}}(\mathbf{k}_1,\eta_0-\chi_1)\tilde{\delta}_{\mathrm{m}}(\mathbf{k}_2,\eta_0-\chi_2)\tilde{\delta}_{\mathrm{m}}(\mathbf{k}_3,\eta_0-\chi_3)\tilde{\delta}_{\mathrm{m}}(\mathbf{k}_4,\eta_0-\chi_4)}_{\text{c}}=\\
&\qquad\quad=\prod_{n=1}^4\left[\frac{2}{5}\frac{D(\chi_n)k_n^2}{\Omega_{\mathrm{m},0}H_0^2}\mathcal{T}_{\delta}(k_n)\right](2\pi)^3\delta_{\mathrm{D}}^{[3]}\left(\mathbf{k}_1+\mathbf{k}_2+\mathbf{k}_3+\mathbf{k}_4\right)T_{\mathcal{R}}(\mathbf{k}_1,\mathbf{k}_2,\mathbf{k}_3,\mathbf{k}_4).
\end{split}
\end{equation}
This is why the trispectrum is the lowest-order statistic able to probe cosmological parity violation in the scalar sector --- statistical isotropy does not automatically erase the dependence on the $\versor{k}$ in it \cite{shiraishi2016parity, cahn2023isotropic, cahn2023test}. We rewrite the primordial trispectrum using the isotropic basis functions \cite{cahn2023isotropic} as 
\begin{equation}
\label{eqn:isotropic_basis}
T_{\mathcal{R}}(\mathbf{k}_1,\mathbf{k}_2,\mathbf{k}_3,\mathbf{k}_4)=\sum_{\ell_1^{\prime}\ell_2^{\prime}\ell_3^{\prime}}T^{\mathcal{R}}_{\ell_1^{\prime}\ell_2^{\prime}\ell_3^{\prime}}(k_1,k_2,k_3,k_4)\sum_{m_1^{\prime}m_2^{\prime}m_3^{\prime}}\begin{pmatrix}
\ell_1^{\prime} & \ell_2^{\prime} & \ell_3^{\prime} \\
m_1^{\prime} & m_2^{\prime} & m_3^{\prime}
\end{pmatrix}\prod_{n=1}^3Y_{\ell_n^{\prime}m_n^{\prime}}(\versor{k}_n).
\end{equation}
Again, the reason we can adopt this treatment is that the lensing potential is a scalar quantity, and so its angular dependence on directions on the sky can be described using spin-zero spherical harmonics \cite{newman1966note}, with the Wigner 3-$j$ symbol automatically ensuring statistical isotropy, as a consequence of the Wigner-Eckhart theorem \cite{wigner1927einige, eckart1930application}.

Now, as we did in Section~\ref{sec:bispectrum}, it is convenient to rewrite the Dirac delta by using Eq.~\eqref{eqn:dirac_delta_plane_waves}:
\begin{equation}
\label{eqn:dirac_delta_plane_waves_trispectrum}
\begin{split}
\delta_{\mathrm{D}}^{[3]}\left(\mathbf{k}_1+\mathbf{k}_2+\mathbf{k}_3+\mathbf{k}_4\right)&=32\pi\sum_{L_aM_a}\int_0^{\infty}\mathrm{d}x\,x^2\prod_{n=1}^4\left[ i^{L_n}j_{L_n}(k_nx)Y_{L_nM_n}(\versor{k}_n)\right]\\
&\qquad\qquad\quad\times\int\mathrm{d}^2\versor{x}\,Y_{L_1M_1}^*(\versor{x})Y_{L_2M_2}^*(\versor{x})Y_{L_3M_3}^*(\versor{x})Y_{L_4M_4}^*(\versor{x}),
\end{split}
\end{equation}
where $a=1,2,3,4$. To perform the angular integration over $\versor{x}$, we can rewrite the product of two spherical harmonics by using the composition of angular momenta:\cite{varshalovich1988quantum},
\begin{equation}
\label{eqn:composition_angular_momentum}
Y_{L_3M_3}^*(\versor{x})Y_{L_4M_4}^*(\versor{x})=\sum
_{L^{\prime}M^{\prime}}\mathcal{G}_{L_3L_4L^{\prime}}^{M_3M_4M^{\prime}}Y_{L^{\prime}M^{\prime}}(\versor{x}),
\end{equation}
so that after recalling Eq.~\eqref{eqn:gaunt_integral}, the last line of the right-hand side of Eq.~\eqref{eqn:dirac_delta_plane_waves_trispectrum} becomes
\begin{equation}
\label{eqn:gaunt_integral_trispectrum}
\begin{split}
\mathcal{G}_{L_1L_2L_3L_4}^{M_1M_2M_3M_4}&\equiv\int\mathrm{d}^2\versor{x}\,Y_{L_1M_1}^*(\versor{x})Y_{L_2M_2}^*(\versor{x})Y_{L_3M_3}^*(\versor{x})Y_{L_4M_4}^*(\versor{x})\\
&=\sum
_{L^{\prime}M^{\prime}}(-1)^{M^{\prime}}\mathcal{G}_{L_1L_2L^{\prime}}^{M_1M_2-M^{\prime}}\mathcal{G}_{L_3L_4L^{\prime}}^{M_3M_4M^{\prime}}.
\end{split}
\end{equation}
Now using Eq.~\eqref{eqn:gaunt_integral_trispectrum} in Eq.~\eqref{eqn:dirac_delta_plane_waves_trispectrum}, and then substituting the result in Eq.~\eqref{eqn:matter_trispectrum} and using Eq.~\eqref{eqn:isotropic_basis}, we obtain
\begin{equation}
\label{eqn:matter_trispectrum_isotropic_basis}
\begin{split}
&\expval{\tilde{\delta}_{\mathrm{m}}(\mathbf{k}_1,\eta_0-\chi_1)\tilde{\delta}_{\mathrm{m}}(\mathbf{k}_2,\eta_0-\chi_2)\tilde{\delta}_{\mathrm{m}}(\mathbf{k}_3,\eta_0-\chi_3)\tilde{\delta}_{\mathrm{m}}(\mathbf{k}_4,\eta_0-\chi_4)}_{\text{c}}=\\
&\quad=(2\pi)^3\sum_{L_1M_1}\sum_{L_2M_2}\sum_{L_3M_3}\sum_{L_4M_4}\mathcal{G}_{L_1L_2L_3L_4}^{M_1M_2M_3M_4}\sum_{\ell_1^{\prime}m_1^{\prime}}\sum_{\ell_2^{\prime}m_2^{\prime}}\sum_{\ell_3^{\prime}m_3^{\prime}}\\
&\qquad\times32\pi\int_0^{\infty}\mathrm{d}x\,x^2\prod_{n=1}^3\left[ \frac{2}{5}\frac{D(\chi_n)k_n^2}{\Omega_{\mathrm{m},0}H_0^2}\,i^{L_n}j_{L_n}(k_n\,x)\mathcal{T}_{\delta}(k_n)Y_{L_nM_n}(\versor{k}_n)Y_{\ell_n^{\prime}m_n^{\prime}}(\versor{k}_n)\right]\\
&\qquad\times\frac{2}{5}\frac{D(\chi_4)k_4^2}{\Omega_{\mathrm{m},0}H_0^2}i^{L_4}j_{L_4}(k_4x)\mathcal{T}_{\delta}(k_4)Y_{L_4M_4}(\versor{k}_4)T^{\mathcal{R}}_{\ell_1^{\prime}\ell_2^{\prime}\ell_3^{\prime}}(k_1,k_2,k_3,k_4)\begin{pmatrix}
\ell_1^{\prime} & \ell_2^{\prime} & \ell_3^{\prime} \\
m_1^{\prime} & m_2^{\prime} & m_3^{\prime}
\end{pmatrix},
\end{split}
\end{equation}
so that Eq.~\eqref{eqn:4PCF} becomes
\begin{equation}
\label{eqn:4PCF_isotropic_basis}
\begin{split}
&\expval{\phi_{\ell_1m_1}\phi_{\ell_2m_2}\phi_{\ell_3m_3}\phi_{\ell_4m_4}}_{\text{c}}=\left(\frac{4D_0}{5\pi\Omega_{\mathrm{m},0}H_0^2}\right)^4\sum_{L_1M_1}\sum_{L_2M_2}\sum_{L_3M_3}\mathcal{G}_{L_1L_2L_3\ell_4}^{M_1M_2M_3m_4}\sum_{\ell_1^{\prime}m_1^{\prime}}\sum_{\ell_2^{\prime}m_2^{\prime}}\sum_{\ell_3^{\prime}m_3^{\prime}}\\
&\times\int_0^{\infty}\mathrm{d}k_1\int_0^{\infty}\mathrm{d}k_2\int_0^{\infty}\mathrm{d}k_3\int_0^{\infty}\mathrm{d}k_4\int_{0}^{\chi_{\ast}}\mathrm{d}\chi_1\int_{0}^{\chi_{\ast}}\mathrm{d}\chi_2\int_{0}^{\chi_{\ast}}\mathrm{d}\chi_3\int_{0}^{\chi_{\ast}}\mathrm{d}\chi_4\int_0^{\infty}\mathrm{d}x\,x^2\\
&\times\prod_{n=1}^3\left[(-1)^{m_n}i^{L_n}(-i)^{\ell_n}W(\chi_n)\gamma(\chi_n)D(\chi_n)j_{\ell_n}(k_n\chi_n)j_{L_n}(k_nx)k_n^2\mathcal{T}_{\delta}(k_n)\mathcal{G}_{L_n\ell_n^{\prime}\ell_n}^{M_nm_n^{\prime}-m_n}\right]\\
&\times W(\chi_4)\gamma(\chi_4)D(\chi_4)j_{\ell_4}(k_4\chi_4)j_{\ell_4}(k_4x)k_4^2\mathcal{T}_{\delta}(k_4)T^{\mathcal{R}}_{\ell_1^{\prime}\ell_2^{\prime}\ell_3^{\prime}}(k_1,k_2,k_3,k_4)\begin{pmatrix}
\ell_1^{\prime} & \ell_2^{\prime} & \ell_3^{\prime} \\
m_1^{\prime} & m_2^{\prime} & m_3^{\prime}
\end{pmatrix},
\end{split}
\end{equation}
where we have used Eq.~\eqref{eqn:orthogonality} and Eq.~\eqref{eqn:gaunt_integral}. To proceed further, we now adopt the Limber approximation for each of the eight spherical Bessel functions. As in the previous sections, using Eq.~\eqref{eqn:limber_approximation}, we find 
\begin{equation}
\label{eqn:limber_on_trispectrum}
\begin{cases}
k_n\simeq\ell_n/\chi_n & \qquad\text{for $\,n=1,2,3,4$},\\
k_n\simeq L_n/x & \qquad\text{for $\,n=1,2,3$},\\
k_4\simeq\ell_4/x. & \\
\end{cases}
\end{equation}
However, we notice that the 3-$j$ symbols inside the Gaunt integrals link $L_n,$ $\ell^{\prime}_n$, and $\ell_n$, because of Eq.~\eqref{eqn:selection_rules}, i.e.
\begin{equation}
\label{eqn:triangular_relations}
|L_n-\ell_n^{\prime}|\le \ell_n\le L_n-\ell_n^{\prime}\qquad\text{for $\,n=1,2,3$}.
\end{equation}
Hence, as long as $\ell_{n}\gg\ell_{n}^{\prime}$, then we can approximate that $L_n \approx \ell_n$. We also note that the Limber approximation requires high $\ell\gtrsim20$ to be accurate. Thus, as long as $\ell_n^{\prime}\ll L_n$, we may approximate that $L_n\simeq\ell_n$, so that Eq.~\eqref{eqn:limber_on_trispectrum} reduces to
\begin{equation}
\label{eqn:limber_on_trispectrum_squeezed}
k_n\simeq\frac{\ell_n}{\chi_n}\qquad\implies\qquad\chi_n=x\equiv\chi\qquad\text{for $\,n=1,2,3,4$}.
\end{equation}
With small $\ell_n^{\prime}$, the fact that $L_n \simeq \ell_n$ will, in turn, also ensure that the $L_n$ are high enough to allow the use of the Limber approximation. Therefore, we may recast Eq.~\eqref{eqn:4PCF_isotropic_basis} as
\begin{equation}
\label{eqn:4PCF_limber}
\begin{split}
&\expval{\phi_{\ell_1m_1}\phi_{\ell_2m_2}\phi_{\ell_3m_3}\phi_{\ell_4m_4}}_{\text{c}}\approx(-1)^{m_1+m_2+m_3}\sum_{L_1M_1}\sum_{L_2M_2}\sum_{L_3M_3}\mathcal{G}_{L_1L_2L_3\ell_4}^{M_1M_2M_3m_4}\\
&\qquad\qquad\qquad\qquad\times\sum_{\ell_1^{\prime}m_1^{\prime}}\mathcal{G}_{L_1\ell_1^{\prime}\ell_1}^{M_1m_1^{\prime}-m_1}\sum_{\ell_2^{\prime}m_2^{\prime}}\mathcal{G}_{L_2\ell_2^{\prime}\ell_2}^{M_2m_2^{\prime}-m_2}\sum_{\ell_3^{\prime}m_3^{\prime}}\mathcal{G}_{L_3\ell_3^{\prime}\ell_3}^{M_3m_3^{\prime}-m_3}\begin{pmatrix}
\ell_1^{\prime} & \ell_2^{\prime} & \ell_3^{\prime} \\
m_1^{\prime} & m_2^{\prime} & m_3^{\prime}
\end{pmatrix}\\
&\times\left(-\frac{5}{2} \Omega_{\mathrm{m,0}}H_0^2\right)^{-4}\int_0^{\chi_{\ast}}\frac{\mathrm{d}\chi}{\chi^6}\prod_{n=1}^4\left[D(\chi)W(\chi)\gamma(\chi)\mathcal{T}_{\delta}\left(\frac{\ell_n}{\chi}\right)\right]T^{\mathcal{R}}_{\ell_1^{\prime}\ell_2^{\prime}\ell_3^{\prime}}\left(\frac{\ell_1}{\chi},\frac{\ell_2}{\chi},\frac{\ell_3}{\chi},\frac{\ell_4}{\chi}\right).
\end{split}
\end{equation}

\subsection{Reduced Angular Trispectrum}
Although we are searching for a parity-odd signal, we still assume statistical isotropy, meaning there is no preferred $z$-axis. Now, $m$ represents the projection of the total angular momentum onto the $z$-axis. Since the CMB correlations should be invariant under rotation about the $z$-axis, the sum of the $m_i$ must be zero. Moreover, if we rotate the four direction vectors simultaneously, observing a different region of the sky, the CMB fluctuations should remain \textit{statistically} identical. This reflects the invariance under 3D rotations around the observer, statistical isotropty, corresponding to 2D translations on the spherical CMB shell where the temperature fluctuations reside.

This general 3D rotational invariance allows us to apply the Wigner-Eckart theorem \cite{wigner1927einige, eckart1930application}, implying that CMB correlations are rotation-invariant tensors, expressible in terms of products of 3-$j$ symbols. This interpretation aligns with Eq.~(15) of \cite{hu2001angular} and is consistent with \cite{shiraishi2014signatures}. Consequently, the angular trispectrum of the CMB lensing potential can be expressed in terms of a reduced trispectrum $Q^{\ell_1\ell_2}_{\ell_3\ell_4}(L)$, defined via
\begin{equation}
\label{eqn:intermediate_sums}
\expval{\phi_{\ell_1m_1}\phi_{\ell_2m_2}\phi_{\ell_3m_3}\phi_{\ell_4m_4}} = \sum_{LM} 
\begin{pmatrix}
\ell_1 & \ell_2 & L \\
m_1 & m_2 & M
\end{pmatrix}\begin{pmatrix}
L & \ell_3 & \ell_4 \\
-M & m_3 & m_4
\end{pmatrix}Q^{\ell_1 \ell_2}_{\ell_3 \ell_4}(L), 
\end{equation}
where we needed two 3-$j$ symbols to deal with $m_1$, $m_2$, $m_3$, $m_4$. We introduced intermediates $L$ and $M$ to enable constructing such a pair (as a pair of 3-$j$ requires six angular momenta). However, we then sum over these intermediates since we only need to have four independent $\ell_i$ and $m_i$ in the end. The reduced trispectrum $Q^{\ell_1 \ell_2}_{\ell_3 \ell_4}(L)$ may be extracted by using the orthogonality of the 3-$j$ symbols \cite{varshalovich1988quantum}, i.e.
\begin{equation}
\label{eqn:wigner_orthogonality}
\left(2 j_3+1\right) \sum_{s_1 s_2}\begin{pmatrix}
j_1 & j_2 & j_3 \\
s_1 & s_2 & s_3
\end{pmatrix}\begin{pmatrix}
j_1 & j_2 & j_3^{\prime} \\
s_1 & s_2 & s_3^{\prime}
\end{pmatrix}=\delta_{j_3^{\prime}j_3}^{\mathrm{K}} \delta_{s_3^{\prime}s_3}^{\mathrm{K}},
\end{equation}
so that we have the the reduced trispectrum as
\begin{equation}
\label{eqn:reduced_trispectrum}
\begin{split}
Q_{\ell_3\ell_4}^{\ell_1\ell_2}(L)&=(2L+1)\sum_{M}(-1)^{M}\\
&\quad\times\sum_{m_1m_2m_3m_4}\begin{pmatrix}
\ell_1 & \ell_2 & L \\
m_1 & m_2 & M
\end{pmatrix}\begin{pmatrix}
\ell_3 & \ell_4 & L \\
m_3 & m_4 & -M
\end{pmatrix}\expval{\phi
_{\ell_1m_1}\phi_{\ell_2m_2}\phi_{\ell_3m_3}\phi_{\ell_4m_4}}.
\end{split}
\end{equation}
We notice that this definition involves the full angular 4-point correlation function, not just the non-Gaussian contribution. However, since our interest will be in the parity-odd case, if we assume that $\ell_1+\ell_2+\ell_3+\ell_4=\text{odd}$, then all the disconnected terms identically vanish, being just the products of angular power spectra, and hence parity-even. For this reason we may now simply substitute our expression for the CMB lensing trispectrum into Eq.~\eqref{eqn:reduced_trispectrum}, finding 
\begin{equation}
\label{eqn:reduced_trispectrum_monster}
\begin{split}
&Q_{\ell_3\ell_4}^{\ell_1\ell_2}(L)=(2L+1)\sum_{L_1L_2L_3}\sum
_{L^{\prime}}\mathcal{F}_{L_1L_2L^{\prime}}\mathcal{F}_{L_3\ell_4L^{\prime}}\sum_{\ell_1^{\prime}\ell_2^{\prime}\ell_3^{\prime}}\mathcal{F}_{L_1\ell_1^{\prime}\ell_1}\mathcal{F}_{L_2\ell_2^{\prime}\ell_2}\mathcal{F}_{L_3\ell_3^{\prime}\ell_3}\,\mathcal{I}_{\ell_1\ell_2\ell_3\ell_4}^{\,\ell_1^{\prime}\ell_2^{\prime}\ell_3^{\prime}}\\
&\,\qquad\qquad\qquad\times\sum_{M}(-1)^{M}\sum_{m_1m_2m_3m_4}(-1)^{m_1+m_2+m_3}\begin{pmatrix}
\ell_1 & \ell_2 & L \\
m_1 & m_2 & M
\end{pmatrix}\begin{pmatrix}
\ell_3 & \ell_4 & L \\
m_3 & m_4 & -M
\end{pmatrix}\\
&\qquad\qquad\qquad\qquad\qquad\qquad\quad\times\sum_{M^{\prime}}(-1)^{M^{\prime}}\sum_{M_1M_2M_3}\begin{pmatrix}
L_1 & L_2 & L^{\prime} \\
M_1 & M_2 & -M^{\prime}
\end{pmatrix}\begin{pmatrix}
L_3 & \ell_4 & L^{\prime} \\
M_3 & m_4 & M^{\prime}
\end{pmatrix}\\
&\qquad\quad\times\sum_{m_1^{\prime}m_2^{\prime}m_3^{\prime}}\begin{pmatrix}
L_1 & \ell_1^{\prime} & \ell_1 \\
M_1 & m_1^{\prime} & - m_1
\end{pmatrix}\begin{pmatrix}
L_2 & \ell_2^{\prime} & \ell_2 \\
M_2 & m_2^{\prime} & - m_2
\end{pmatrix}\begin{pmatrix}
L_3 & \ell_3^{\prime} & \ell_3 \\
M_3 & m_3^{\prime} & - m_3
\end{pmatrix}\begin{pmatrix}
\ell_1^{\prime} & \ell_2^{\prime} & \ell_3^{\prime} \\
m_1^{\prime} & m_2^{\prime} & m_3^{\prime}
\end{pmatrix},
\end{split}
\end{equation}
where we have defined 
\begin{equation}
\label{eqn:definition_of_F}
\mathcal{F}_{j_1j_2j_3}\equiv\mathcal{G}_{j_1j_2j_3}^{s_1s_2s_3}\begin{pmatrix}
	j_1 & j_2 & j_3 \\
	s_1 & s_2 & s_3
\end{pmatrix}^{-1}=\sqrt{\frac{(2j_1+1)(2j_2+1)(2j_3+1)}{4\pi}}\begin{pmatrix}
j_1 & j_2 & j_3 \\
0 & 0 & 0
\end{pmatrix}.
\end{equation}
To perform the sum over the $m$, we use the \textsc{Python} code \texttt{Reduce3j} \cite{xiang2021program}, which allows us to express everything in terms of the Wigner 3\,-$j$, 6\,-$j$ and 9\,-$j$ symbols:
\begin{equation}
\label{eqn:reduced_trispectrum_compact}
\boxed{\begin{aligned}
Q_{\ell_3\ell_4}^{\ell_1\ell_2}(L)&=(-1)^{\ell_4}(2L+1)\sum_{L_1L_2L_3}\sum
_{L^{\prime}}\mathcal{F}_{L_1L_2L^{\prime}}\mathcal{F}_{L_3\ell_4L^{\prime}}\\
&\quad\times\sum_{\ell_1^{\prime}\ell_2^{\prime}\ell_3^{\prime}}(-1)^{\ell_1^{\prime}+\ell_2^{\prime}}\mathcal{F}_{L_1\ell_1^{\prime}\ell_1}\mathcal{F}_{L_2\ell_2^{\prime}\ell_2}\mathcal{F}_{L_3\ell_3^{\prime}\ell_3}\left\{\begin{matrix}
\ell_4 & L_3 & L^{\prime} \\
\ell_3^{\prime} & L & \ell_3
\end{matrix}\right\}\left\{\begin{matrix}
L & L^{\prime} & \ell_3^{\prime} \\
\ell_2 & L_2 & \ell_2^{\prime} \\
\ell_1 & L_1 & \ell_1^{\prime}
\end{matrix}\right\}\mathcal{I}_{\ell_1\ell_2\ell_3\ell_4}^{\,\ell_1^{\prime}\ell_2^{\prime}\ell_3^{\prime}},
\end{aligned}}
\end{equation}
$\mathcal{F}$ is defined in Eq.~\eqref{eqn:definition_of_F}, while
\begin{equation}
\label{eqn:integral_trispectrum_general}
\boxed{\begin{aligned}
\mathcal{I}_{\ell_1\ell_2\ell_3\ell_4}^{\ell_1^{\prime}\ell_2^{\prime}\ell_3^{\prime}}&\equiv\left(-\frac{5}{2} \Omega_{\mathrm{m,0}}H_0^2\right)^{-4}\\
&\qquad\qquad\times\int_0^{\chi_{\ast}}\frac{\mathrm{d}\chi}{\chi^6}\prod_{n=1}^4\left[D(\chi)W(\chi)\gamma(\chi)\mathcal{T}_{\delta}\left(\frac{\ell_n}{\chi}\right)\right]T^{\mathcal{R}}_{\ell_1^{\prime}\ell_2^{\prime}\ell_3^{\prime}}\left(\frac{\ell_1}{\chi},\frac{\ell_2}{\chi},\frac{\ell_3}{\chi},\frac{\ell_4}{\chi}\right).
\end{aligned}}
\end{equation} 
Eq.~\eqref{eqn:reduced_trispectrum_compact} is a general expression for the reduced trispectrum $Q_{\ell_3\ell_4}^{\ell_1\ell_2}(L)$, and is a main result of this paper. Analogously, Eq.~\eqref{eqn:integral_trispectrum_general} holds for any primordial trispectrum and linear matter transfer function, and except for the usage of the Limber approximation, it is thus completely general. Moreover, Eq.~\eqref{eqn:reduced_trispectrum_compact} allows us to read off parity constraints from the Wigner 3\,-$j$ symbols inside the $\mathcal{F}$ coefficients, each of which demands that the sum of its total angular momenta is even. Using $\mathsf{P}(n)$ to denote the parity of a given natural number $n$, we must have
\begin{equation}
\label{eqn:parity_of_multipoles}
\begin{cases}
&\mathsf{P}(\ell_i+ L_i)\,\,\,=\mathsf{P}(\ell_i^{\prime})\qquad\qquad\qquad\qquad\text{for $\,n=1,2,3$},\\
&\mathsf{P}(L_1 + L_2) =\mathsf{P}(L)=\mathsf{P}(L_3 + L_4),
\end{cases}
\end{equation}
from looking at the 3\,-$j$ symbols in Eq.~\eqref{eqn:gaunt_integral_trispectrum} and Eq.~\eqref{eqn:4PCF_limber}. Summing the two constraint equations, we find
\begin{equation}
\label{eqn:parity_of_esternal_multipoles}
\mathsf{P}(L_1 + L_2 + L_3 + \ell_1 + \ell_2+\ell_3)=\mathsf{P}(\ell_1^{\prime} +\ell_2^{\prime} +\ell_3^{\prime}).
\end{equation}
Since $\mathsf{P}(L_1+L_2)=\mathsf{P}(L_3+L_4)$, we also know that $\mathsf{P}(L_1 + L_2 + L_3 + L_4)$ is even, so that
\begin{equation}
\label{eqn:parity_of_internal_multipoles}
\mathsf{P}(L_1+L_2+L_3)=\mathsf{P}(L_4)=\mathsf{P}(\ell_4)\implies
\boxed{\mathsf{P}( \ell_1 + \ell_2 + \ell_3 + \ell_4 )=\mathsf{P}(\ell_1' + \ell_2' + \ell_3').}
\end{equation}
Hence, the presence of a parity-odd primordial scalar trispectrum automatically implies a parity-odd angular CMB lensing trispectrum.

\subsection{Specifying the Primordial Parity-Odd Trispectrum}
Following \cite{jamieson2024parity}, we write the primordial trispectrum as\footnote{The factor of $\left(\frac{5}{3}\right)^4$ arises because we consider the primordial trispectrum of $\tilde{\mathcal{R}}$, rather than $\tilde{\varphi}$, which are related to each other via Eq.~\eqref{eqn:from_phi_to_R}.}
\begin{equation}
\label{eqn:POP_trispectrum}
T_{\mathcal{R}}(\mathbf{k}_1,\mathbf{k}_2,\mathbf{k}_3,\mathbf{k}_4)=\left(\frac{5}{3}\right)^4\left[T_{+}(\mathbf{k}_1,\mathbf{k}_2,\mathbf{k}_3,\mathbf{k}_4)+iT_{-}(\mathbf{k}_1,\mathbf{k}_2,\mathbf{k}_3,\mathbf{k}_4)\right], 
\end{equation}
where the parity-even component of the trispectrum is its real part, $T_{+}$, while the parity-odd component is its imaginary part, $T_{-}$. We notice that the four wave-vectors are not independent of each other because of statistical homogeneity: $\mathbf{k}_4=-\mathbf{k}_1-\mathbf{k}_2-\mathbf{k}_3$. Moreover, due to statistical isotropy, the only parity-odd structure we can form using $\mathbf{k}_1$, $\mathbf{k}_2$, and $\mathbf{k}_3$ must be proportional to the triple product $\mathbf{k}_1\cdot(\mathbf{k}_2\cross\mathbf{k}_3)$. Therefore, as in \cite{coulton2024signatures}, we now define a shape function, $\tau_{-}$, for the parity-odd trispectrum via
\begin{equation}
\label{eqn:symmetrized_trispectrum}
T_{-}(\mathbf{k}_1,\mathbf{k}_2,\mathbf{k}_3,\mathbf{k}_4)\equiv\left[\mathbf{k}_1\cdot\left(\mathbf{k}_2\cross\mathbf{k}_3\right)\right]\tau_{-}(k_1,k_2,k_3,k_4).
\end{equation}
The trispectrum must be symmetric under the interchange of any pair of the four $\mathbf{k}$, and the triple product is antisymmetric by definition. Taken together, these symmetries mean that $\tau_{-}(k_1, k_2, k_3, k_4)$ must be antisymmetric under interchange of any pair of wave vectors. By approximating again that $n_s(k_p)\simeq1$ and following \cite{jamieson2024parity}, we have
\begin{equation}
\label{eqn:drew_template}
\tau_{-}(k_1,k_2,k_3,k_4)\equiv g_{-}[2\pi^2A_s(k_p)]^3\left(\frac{k_1^{n_a}k_2^{n_b}k_3^{n_c}k_4^0}{k_1^3k_2^3k_3^3k_4^0}\mp\text{$23$ perm.}\right),
\end{equation}
where $g_{-}$ controls the amplitude of the parity-odd trispectrum, and the $24$ terms in the parenthesis are the permutations of $\{k_1, k_2, k_3, k_4\}$, holding $n_a$, $n_b$ and $n_c$ fixed. As detailed under Eq.~(45) in \cite{jamieson2024parity}, here permutations with an even number of transpositions
get a positive sign, while those with an odd one get a negative sign in the sum. We notice that if $n_a+n_b+n_c=-3$, in Eq.~\eqref{eqn:primordial_trispectrum} we recover the expected dependence on the wavenumbers of the trispectrum, i.e. $k^{-9}$. We next rewrite the triple product using the isotropic basis functions \cite{cahn2023isotropic} as 
\begin{equation}
\label{eqn:triple_product}
\begin{split}
\mathbf{k}_1\cdot\left(\mathbf{k}_2\cross\mathbf{k}_3\right)=-\frac{i\sqrt{2}}{3}(4\pi)^{3/2}\,k_1k_2k_3\sum_{m_1^{\prime}=-1}^{1}\sum_{m_2^{\prime}=-1}^1&\sum_{m_3^{\prime}=-1}^{1}\begin{pmatrix}
1 & 1 & 1 \\
m_1^{\prime} & m_2^{\prime} & m_3^{\prime}
\end{pmatrix}\\
&\times Y_{1m_1^{\prime}}(\versor{k}_2)Y_{1m_2^{\prime}}(\versor{k}_3)Y_{1m_3^{\prime}}(\versor{k}_3),
\end{split}
\end{equation}
so that, by setting the parity-even component equal to zero and comparing the expression above with Eq.~\eqref{eqn:isotropic_basis}, we find 
\begin{equation}
\label{eqn:primordial_trispectrum}
\begin{split}
T^{\mathcal{R}}_{\ell_1^{\prime}\ell_2^{\prime}\ell_3^{\prime}}\left(k_1,k_2,k_3,k_4\right)\equiv \lambda\,[2\pi^2A_s(k_p)]^3\delta^{\mathrm{K}}_{1\ell_1^{\prime}}\delta^{\mathrm{K}}_{1\ell_2^{\prime}}\delta^{\mathrm{K}}_{1\ell_3^{\prime}}\,k_1&k_2k_3\\
&\times\left(\frac{k_1^{n_a}k_2^{n_b}k_3^{n_c}k_4^0}{k_1^3k_2^3k_3^3k_4^0}\mp\text{$23$ perm.}\right),
\end{split}
\end{equation}
where
\begin{equation}
\label{eqn:definition_of_lambda}
\lambda\equiv\left(\frac{5}{3}\right)^4\frac{\sqrt{2}}{3}(4\pi)^{3/2}g_{-}.
\end{equation}
We recall that only the $k$'s in the parentheses in Eq.~\eqref{eqn:drew_template} are permuted. Indeed, we can always isolate the parity-odd part of the trispectrum by factorising out the triple product, while the rest of the angular dependence can be absorbed into the $\tau_{-}$ template. Therefore, expressing Eq.~\eqref{eqn:primordial_trispectrum} with only $\ell^{\prime}_1=\ell^{\prime}_2=\ell^{\prime}_3=1$ is the most general approach. We can now insert the expressions for $W(\chi)$, $\gamma(\chi)$ and $T^{\mathcal{R}}_{\ell_1^{\prime}\ell_2^{\prime}\ell_3^{\prime}}$ in Eq.~\eqref{eqn:integral_trispectrum_general}, so that, after performing the change of variable $y\equiv\chi_{\ast}/\chi$, we find
\begin{equation}
\label{eqn:integral_trispectrum}
\begin{split}
\mathcal{I}^{\,\ell_1^{\prime}\ell_2^{\prime}\ell_3^{\prime}}_{\ell_1\ell_2\ell_3\ell_4}&\equiv\lambda\,[2\pi^2A_s(k_p)]^3\left(\frac{6}{5}D_0\right)^4\delta^{\mathrm{K}}_{1\ell_1^{\prime}}\delta^{\mathrm{K}}_{1\ell_2^{\prime}}\delta^{\mathrm{K}}_{1\ell_3^{\prime}}\left(\ell_1^{\,n_a-3}\ell_2^{\,n_b-3}\ell_3^{\,n_c-3}\ell_4^{\,0}\mp\text{$23$ perm.}\right)\\
&\qquad\times(\ell_1\ell_2\ell_3)\int_1^{\infty}\frac{\mathrm{d}y}{y^{2-p}}\,\frac{(1-y)^4}{\chi_{\ast}^{3+p}}\,\mathcal{T}_{\delta}\left(\frac{\ell_1y}{\chi_\ast}\right)\mathcal{T}_{\delta}\left(\frac{\ell_2y}{\chi_\ast}\right)\mathcal{T}_{\delta}\left(\frac{\ell_3y}{\chi_\ast}\right)\mathcal{T}_{\delta}\left(\frac{\ell_4y}{\chi_\ast}\right),
\end{split}
\end{equation}
where we have defined $p\equiv n_a+n_b+n_c$, and used $D\sim D_0a$. Indeed, the advantage of this template for the primordial trispectrum is that it simplifies the summation over $\ell^{\prime}$, allowing us to separate the resulting integral from the summation\footnote{This was possible because of our power-law toy model of the primordial trispectrum. Some alternative models, such as axion inflation, predict more complicated trispectra \cite{Niu:2022fki, Cabass:2022rhr, Fujita:2023inz, Creque-Sarbinowski:2023wmb, Moretti:2024fzb}.} in Eq.~\eqref{eqn:reduced_trispectrum_compact}.

\subsection{Specifying the Matter Transfer Function}
\begin{figure}
\centering
\includegraphics[width =\linewidth]{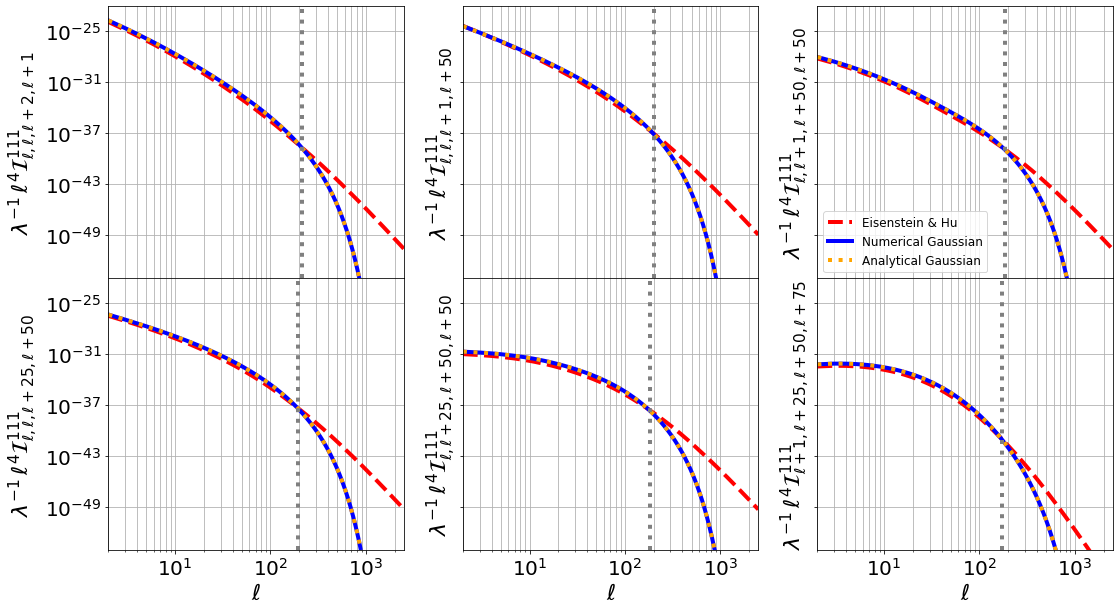}
\caption{We plot $\mathcal{I}_{\ell_1\ell_2\ell_3\ell_4}^{111}$ given Eq.~\eqref{eqn:integral_trispectrum}, for different $\ell_1$, $\ell_2$, $\ell_3$, $\ell_4$, specified by the combinations shown in the subscripts on the vertical axis of each panel. We always use powers of $n_a,n_b,n_c=-2,-1,0$, given Eq.~\eqref{eqn:drew_template}. No relation between the multipoles is enforced here apart from $\ell_1+\ell_2+\ell_3+\ell_4=$ odd. The red dashed curve is using the EH matter transfer function. The blue solid line follows from the numerical integration of Eq.~\eqref{eqn:integral_trispectrum_final} for the Gaussian transfer function, while the orange dotted line is its analytical solution, given \eqref{eqn:trispectrum_exponentials}. The almost perfect match between the orange and blue lines nicely verifies our numerical pipeline. The vertical dotted grey lines indicate when the mismatch between $\mathcal{I}_{\ell_1\ell_2\ell_3\ell_4}^{111}$ evaluated using the Gaussian transfer function and using the EH transfer function starts to exceed 10\%. The weighting by $\ell^4$ is chosen for legibility.}
\label{fig:trispectrum_1D_configs}
\end{figure}

\begin{figure}
\centering
\includegraphics[width =\linewidth]{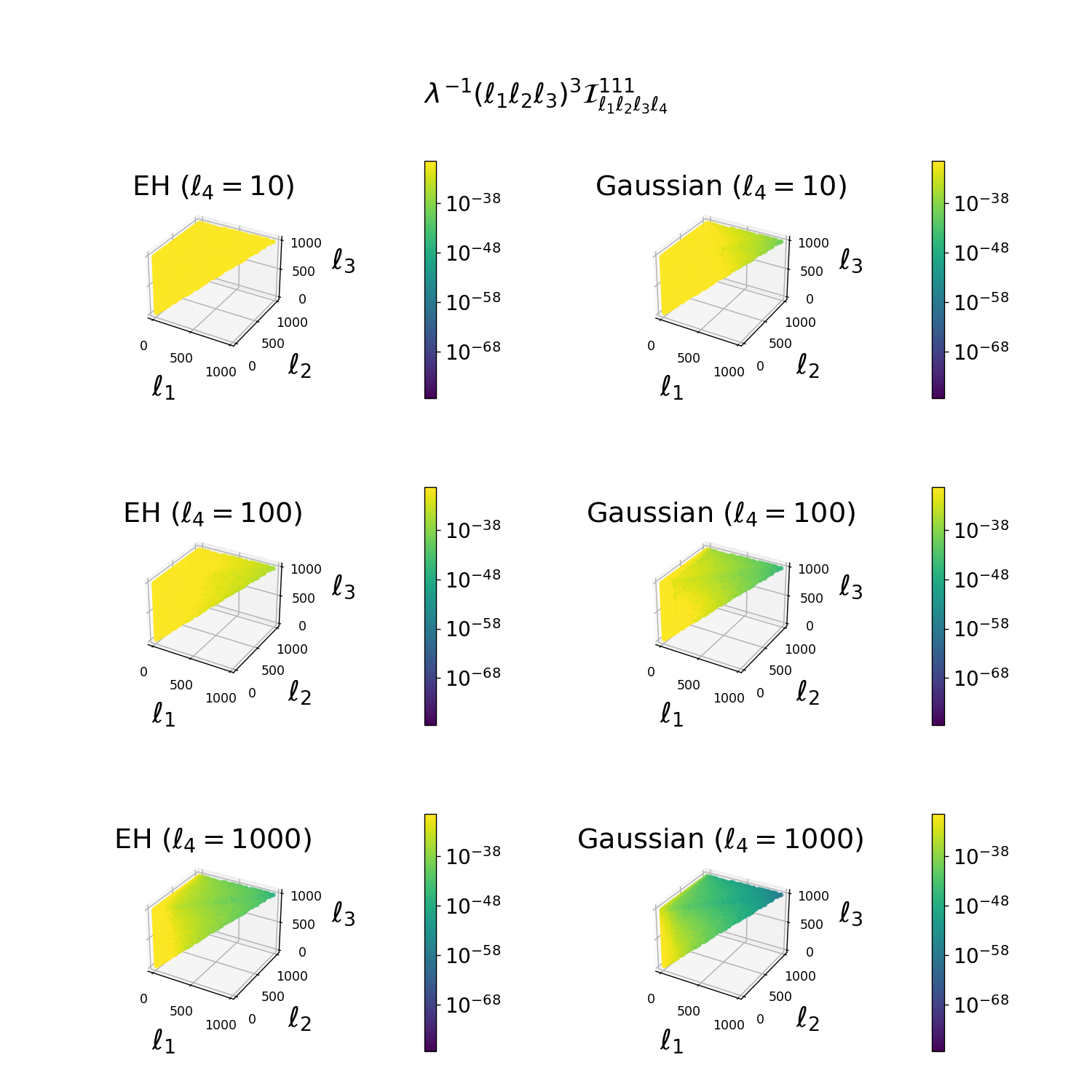}
\caption{$\mathcal{I}_{\ell_1\ell_2\ell_3\ell_4}^{111}$ given Eq.~\eqref{eqn:integral_trispectrum} as a function of $\ell_1$, $\ell_2$, $\ell_3$, after fixing $\ell_4$ to the value indicated in each panel title and setting $n_a,n_b,n_c=-2,-1,0$, given Eq.~\eqref{eqn:drew_template}. We have restricted our analysis to the $\ell_1<\ell_2<\ell_3$ domain to avoid redundant combinations, and we have considered only parity-odd configurations, i.e. $\ell_1+\ell_2+\ell_3+\ell_4=$ odd. The plots in the left column are using the EH transfer function, while those in the right column are using our Gaussian toy model for the matter transfer function, given Eq.~\eqref{eqn:integral_trispectrum_final}. Consistent with our previous analysis, at high multipoles, the use of the Gaussian toy model transfer function results in a significant mismatch with the results of using EH transfer function. The weighting by $(\ell_1\ell_2\ell_3)^3$ is chosen for legibility.}
\label{fig:trispectrum_grid}
\end{figure}

As in the previous section, we insert the Gaussian transfer function introduced in Eq.\eqref{eqn:gaussian_transfer_function} to obtain an analytical solution. 
\begin{equation}
\label{eqn:integral_trispectrum_final}
\begin{split}
\mathcal{I}^{\,\ell_1^{\prime}\ell_2^{\prime}\ell_3^{\prime}}_{\ell_1\ell_2\ell_3\ell_4}&\equiv\lambda\,[2\pi^2A_s(k_p)]^3\alpha^4\delta^{\mathrm{K}}_{1\ell_1^{\prime}}\delta^{\mathrm{K}}_{1\ell_2^{\prime}}\delta^{\mathrm{K}}_{1\ell_3^{\prime}}\left(\ell_1^{\,n_a-3}\ell_2^{\,n_b-3}\ell_3^{\,n_c-3}\ell_4^{\,0}\mp\text{$23$ perm.}\right)\\
&\qquad\qquad\qquad\qquad\times(\ell_1\ell_2\ell_3)\int_1^{\infty}\frac{\mathrm{d}y}{y^{2-p}}\,\frac{(1-y)^4}{\chi_{\ast}^{3+p}}\,e^{-\beta^2(\ell_1^2+\ell_2^2+\ell_3^2+\ell_4^2)y^2},
\end{split}
\end{equation}
where we have used Eq.~\eqref{eqn:alpha_and_beta}.
We recall that only the quantities in the parentheses in the last line of the right-hand side of Eq.~\eqref{eqn:integral_trispectrum_final} are permuted. Eq.~\eqref{eqn:integral_trispectrum_final} then reduces to
\begin{equation}
\label{eqn:trispectrum_exponentials}
\begin{split}
\mathcal{I}^{\,\ell_1^{\prime}\ell_2^{\prime}\ell_3^{\prime}}_{\ell_1\ell_2\ell_3\ell_4}&\equiv\frac{\alpha^4\lambda}{2\chi_{\ast}^{3+p}}[2\pi^2A_s(k_p)]^3(\ell_1\ell_2\ell_3)\\
&\qquad\times\Bigg[\mathrm{E}_{(3-p)/2}\left(\xi_4\right)-4\mathrm{E}_{1-p/2}\left(\xi_4\right)+6\mathrm{E}_{(1-p)/2}\left(\xi_4\right)-4\mathrm{E}_{-p/2}\left(\xi_4\right)\\
&\qquad\qquad+\mathrm{E}_{-(1+p)/2}\left(\xi_4\right)\Bigg]\delta^{\mathrm{K}}_{1\ell_1^{\prime}}\delta^{\mathrm{K}}_{1\ell_2^{\prime}}\delta^{\mathrm{K}}_{1\ell_3^{\prime}}\left(\ell_1^{\,n_a-3}\ell_2^{\,n_b-3}\ell_3^{\,n_c-3}\ell_4^{\,0}\mp\text{$23$ perm.}\right),
\end{split}
\end{equation}
where we have used Eq.~\eqref{eqn:generalized_exponential_integral} and defined $\xi_4\equiv\beta^2(\ell_1^2+\ell_2^2+\ell_3^2+\ell_4^2)$.
We have numerically evaluated the trispectrum for $n_a=-2$, $n_b=-1$, $n_c=0$ by using the Gaussian and EH matter transfer functions\footnote{We also verified that using the BBKS transfer function instead of the EH transfer function has no impact on the results, indicating that the effect of Baryon Acoustic Oscillations (BAO) is negligible in this case.} in Figs.~\ref{fig:trispectrum_1D_configs}--\ref{fig:trispectrum_grid}. As we saw for the lower-order polyspectra, here too the results with the two different transfer functions differ as the multipoles increase.

Moving forward, we will not make further use of the Gaussian transfer function defined in Eq.~\eqref{eqn:gaussian_transfer_function}, but rather, only the EH one. Nevertheless, the former has been useful for checking our numerical work. Indeed, since we were able to perform the integrals in closed form and also numerically, getting agreement (as we did) shows that our numerical pipeline is working well in terms of the integration scheme and other details. Furthermore, one of the reasons for using the Gaussian transfer function was to have a closed-form toy model, which also performed well up to $\ell\lesssim100$.

\section{\label{sec:phenomenology} Interpretation}

We now present a qualitative explanation as to why CMB lensing is sensitive to parity. One might naively expect it not to be, as parity requires a 3D statistic (as explained in, e.g., \cite{cahn2023test}), while CMB measurements are taken on a 2D surface. However, as discussed in Appendix~\ref{app:parity_on_sphere}, parity is not equivalent to a rotation when the surface is curved. Nevertheless, if the four points are strictly co-planar, the trispectrum becomes completely insensitive to parity. Deviations from strict co-planarity can occur, which allow for a small, but generally suppressed, parity-odd signal.

This highlights the key difference between primary CMB anisotropies and CMB lensing: the physical scales probed. In the primary CMB, configurations involving only large multipoles correspond to very large physical scales, comparable to the distance to the last scattering surface ($\sim 10,000\,\mathrm{Mpc}/h$), where parity-violating signals are strongly suppressed. In contrast, CMB lensing is sensitive to smaller physical scales, closer to those where 3D parity violation has been observed in late-time large-scale structure, typically on scales $\lesssim 400\,\mathrm{Mpc}/h$. This makes the separation vectors of the four LSS points effectively coplanar. Hence, the triple product that sources the parity-odd component of the angular trispectrum of CMB temperature anisotropies is suppressed, as detailed in App.~\ref{app:triple_product}. However, on a 2D plane, parity is equivalent to a rotation, so statistical isotropy implies that parity violation should be negligible.

\begin{figure}
    \centering
    \includegraphics[width=0.85\linewidth]{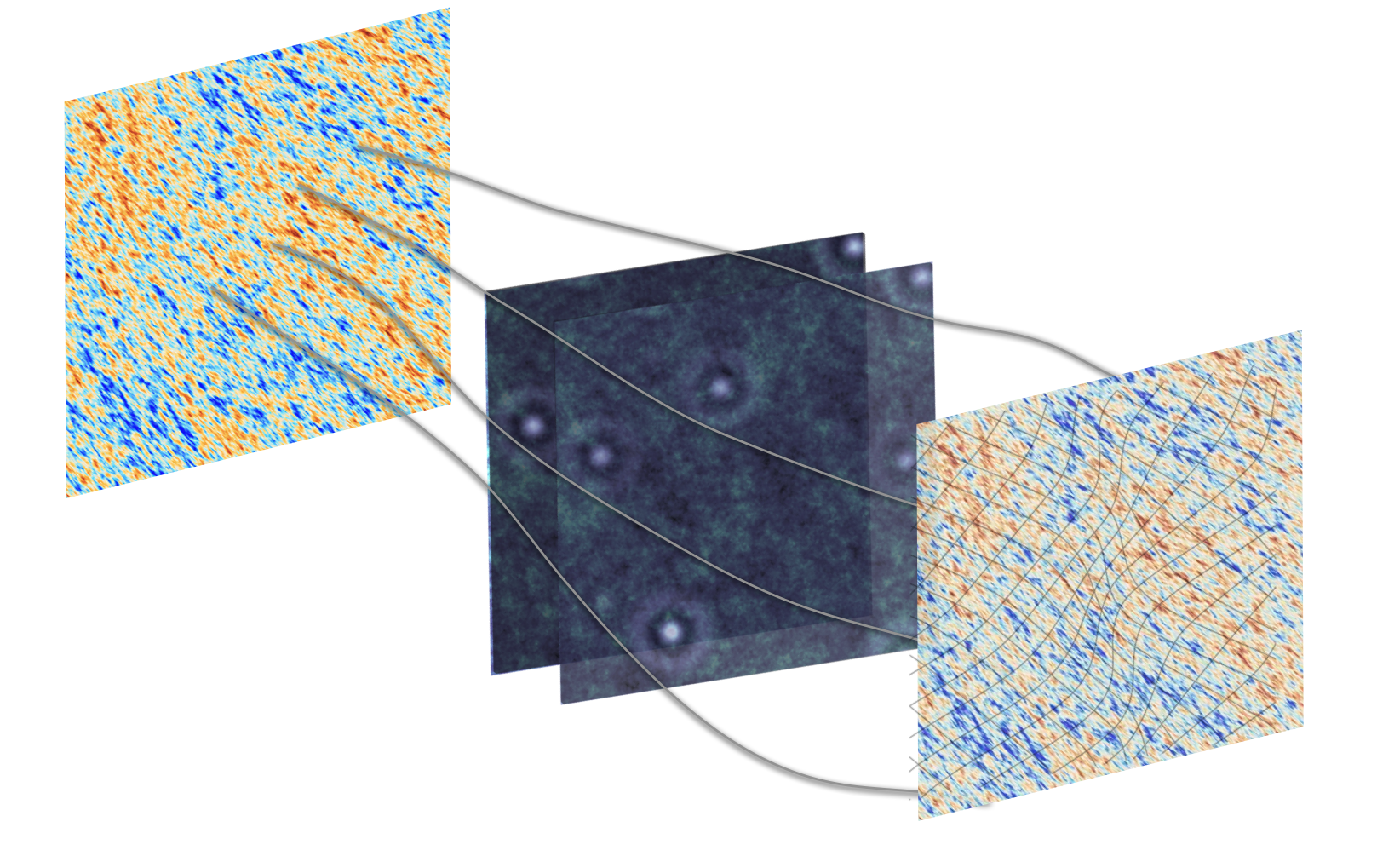}
    \caption{Illustration of CMB lensing: photons emitted from the last scattering surface (left) are deflected by the foreground large-scale structure (middle). This gravitational lensing distorts the observed photon distribution, resulting in subtle shifts in the CMB map observed (right).}
    \label{fig:illustration}
\end{figure}

\subsection{Parity and Projection}
To explain why CMB lensing is more sensitive to parity violation than CMB temperature anisotropies, we have to analyze the trispectrum's geometrical properties.
We first demonstrate that the projection of a tetrahedron onto a plane cannot probe parity. This scenario is considered because, if the lensing potential behaved like primary CMB anisotropies, the four trispectrum points would tend to lie in a single plane, as previously noted. In the perfect flat-sky limit, the 4PCF would thus lose sensitivity to parity.

\begin{figure}
\centering
\includegraphics[width = \textwidth, height = 0.65\textwidth]{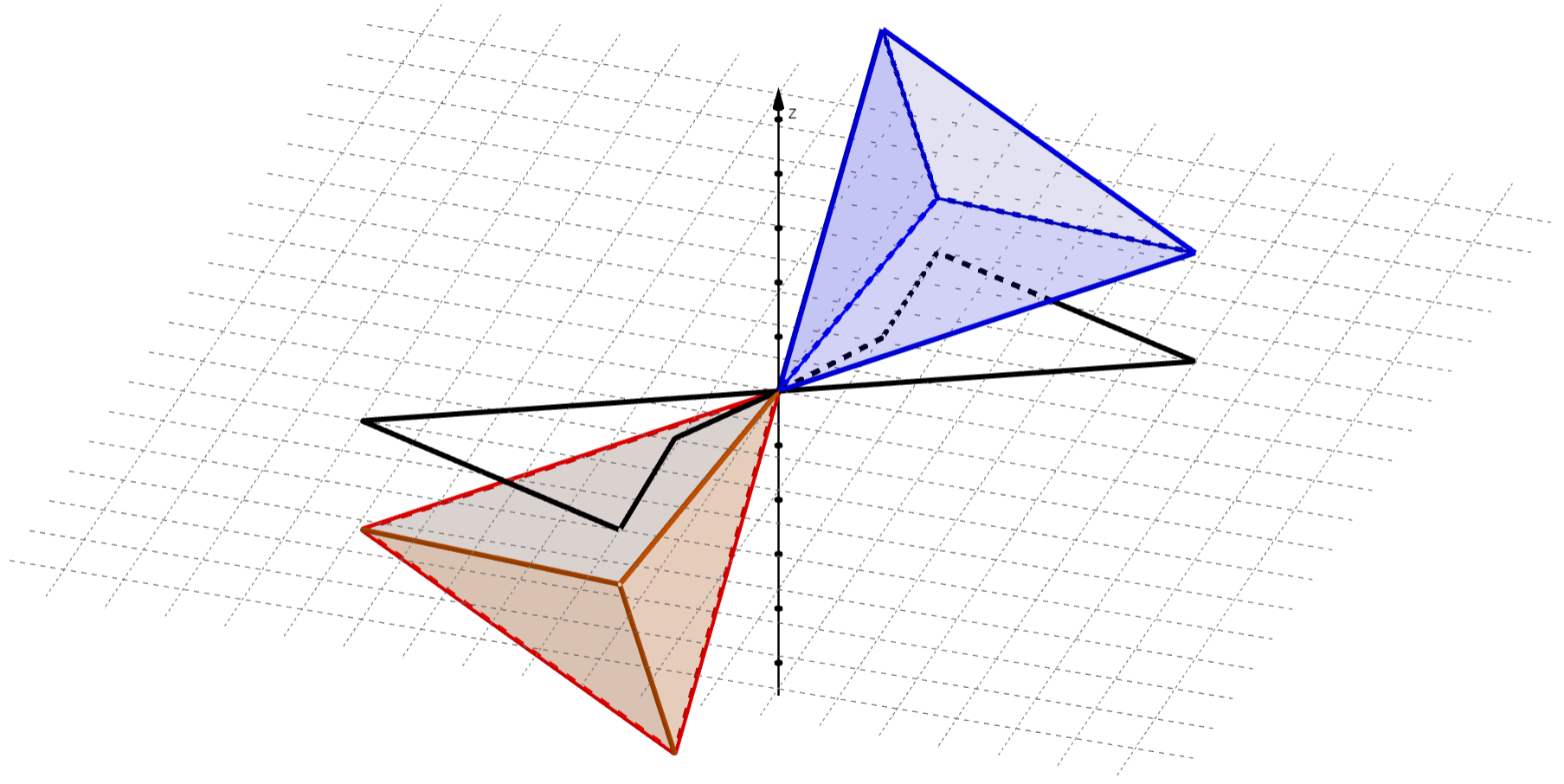}
\caption{Here we display two distinct tetrahedra (blue and red) with a shared vertex at the origin of the reference frame. The red tetrahedron is obtained from the blue one by performing a parity transformation, i.e. $(x,y,z)\mapsto(-x,-y,-z)$. This process reflects each point of the blue tetrahedron across the origin, resulting in the red tetrahedron. The black quadrilaterals in the plot represent the projections of each tetrahedron onto the plane represented by the grid. The projection of one tetrahedron can be obtained by rotating the projection of the other by $180^{\circ}$ around the axis orthogonal to the plane that is shown in the picture. This rotation effectively mirrors the projection of the blue tetrahedron, aligning it with that of the red one.}
\label{fig:tethraedra}
\end{figure}

Indeed, a pure projection renders any information about 3D parity violation degenerate with a rotation in the 2D plane of projection. Fig.~\ref{fig:tethraedra} shows two tetrahedra, where the red one is obtained by performing a spatial inversion of the coordinates of the blue tetrahedron with respect to the origin, i.e., reflecting all points through the origin. This parity transformation reverses the orientation of the tetrahedron while preserving its structure. The black polygons represent the projections of the tetrahedra onto the $(x,y)$ plane, where the projection of one tetrahedron can be derived from the projection of the other by a $180^{\circ}$ rotation around the $z$-axis (perpendicular to the plane). 

We now provide a mathematical proof of the statement we just made. Let $\{\mathbf{v}_i\}_{i=1}^{4}$ denote the set of four vectors representing the vertices of the blue tetrahedron in our reference frame. For our purposes, we define the vectors as follows:
\begin{equation}
\mathbf{v}_1 = (0, 0, 0)^{\mathsf{T}}, \quad \mathbf{v}_2 = (v_2^{\,x}, v_2^{\,y}, v_2^{\,z})^{\mathsf{T}}, \quad \mathbf{v}_3 = (v_3^{\,x}, v_3^{\,y}, v_3^{\,z})^{\mathsf{T}}, \quad \mathbf{v}_4 = (v_4^{\,x}, v_4^{\,y}, v_4^{\,z})^{\mathsf{T}},
\end{equation}
where $\mathbf{v}_1$ is the null vector, corresponding to the origin of the reference frame, and $^{\mathsf{T}}$ denotes transpose. The remaining vectors $\mathbf{v}_2$, $\mathbf{v}_3$, and $\mathbf{v}_4$ represent the other vertices of the tetrahedron, situated at specified coordinates in 3D space. Now, let $\mathbb{P}$ denote the parity operator. The action of $\mathbb{P}$ on the $i^{\,\mathrm{th}}$ 3D vector $\mathbf{v}_i = (v_i^{\,x}, v_i^{\,y}, v_i^{\,z})^{\mathsf{T}}$ is
\begin{equation}
\label{eqn:parity_transformation}
\mathbb{P}[\mathbf{v}_i] = (-v_i^{\,x}, -v_i^{\,y}, -v_i^{\,z})^{\mathsf{T}}.
\end{equation}
The projection of the $i^{\,\mathrm{th}}$ vector $\mathbf{v}^i$ onto the $(x,y)$ plane, denoted by $\mathsf{Proj}_{z=0}$, is  
\begin{equation}
\label{eqn:projection_on_a_plane}
\mathsf{Proj}_{z=0}[\mathbf{v}_i] = (\hat{x} \cdot \mathbf{v}_i) \hat{x} + (\hat{y} \cdot \mathbf{v}_i) \hat{y} = (v_i^{\,x}, v_i^{\,y}, 0)^{\mathsf{T}}.
\end{equation}
Finally, the rotation of the $i^{\,\mathrm{th}}$ vector $\mathbf{v}_i$ by an angle $\pi$ around the $z$-axis, denoted $\mathsf{R}_{\hat{z}}(\pi)$, is
\begin{equation}
\label{eqn:rotation_around_z}
\mathsf{R}_{\hat{z}}(\pi)[\mathbf{v}_i] = \begin{pmatrix}
\cos\pi & -\sin\pi & 0 \\
\sin\pi &  \cos\pi & 0 \\
0             &   0            & 1
\end{pmatrix}
\begin{pmatrix}
v_i^{\,x} \\
v_i^{\,y} \\
v_i^{\,z}
\end{pmatrix}
= \begin{pmatrix}
-v_i^{\,x} \\
-v_i^{\,y} \\
v_i^{\,z}
\end{pmatrix}.
\end{equation}
Therefore, rotating the projection of the parity-transformed tetrahedron on the $(x,y)$ plane by $180^{\circ}$ gives exactly the projection of the original tetrahedron on the same plane. 

Below we show a summary diagram of the transformations of the vector $\mathbf{v}_i$ under parity and projection: the vector $\mathbf{v}_i$ undergoes a parity transformation $\mathbb{P}$, mapping it to $-\mathbf{v}_i$; both $\mathbf{v}_i$ and $-\mathbf{v}_i$ are then projected onto the $z=0$ plane, resulting in $(v_i^{\,x},v_i^{\,y},0)^{\mathsf{T}}$ and $-(v_i^{\,x},v_i^{\,y},0)^{\mathsf{T}}$, respectively. A rotation $\mathsf{R}_{\hat{z}}$ by $180^\circ$ around the $z$-axis then relates these projections.
\begin{equation}
\begin{tikzcd}[row sep=6em, column sep=6em]
\mathbf{v}_i \arrow[r, "\mathbb{P}"{font=\large}, shorten >=2pt, shorten <=2pt] \arrow[d, "\mathsf{Proj}_{z=0}"{font=\large}, swap, shorten >=2pt, shorten <=2pt] 
& -\mathbf{v}_i \arrow[d, "\mathsf{Proj}_{z=0}"{font=\large}, shorten >=2pt, shorten <=2pt] \\
(v_i^{\,x},v_i^{\,y},0)^{\mathsf{T}} \arrow[leftarrow, r, "\mathsf{R}_{\hat{z}}(\pi)"'{font=\large}, shorten >=2pt, shorten <=2pt] 
& -(v_i^{\,x},v_i^{\,y},0)^{\mathsf{T}}.
\end{tikzcd}
\end{equation}

\subsection{Comparison with Primary CMB Anisotropies}
We now present a heuristic mathematical argument to explain why a parity-odd trispectrum is suppressed in the case of \textit{primary} CMB anisotropies, whereas this suppression does not apply to the CMB lensing potential. In the latter case, the difference arises from sensitivity to different (smaller or broader) physical scales. Let us rewrite here the general expression for the reduced trispectrum,
\begin{equation}
\label{eqn:reduced_angular_trispectrum}
\begin{split}
Q_{\ell_1\ell_2}^{\ell_3\ell_4}(L)=&(2L+1)\sum_{m=-L}^{L}(-1)^{M}\\
&\times\sum_{m_1m_2m_3m_4}\begin{pmatrix}
\ell_1 & \ell_2 & L \\
m_1 & m_2 & M
\end{pmatrix}\begin{pmatrix}
\ell_3 & \ell_4 & L \\
m_3 & m_4 & -M
\end{pmatrix}\expval{a_{\ell_1m_1}a_{\ell_2m_2}a_{\ell_3m_3}a_{\ell_4m_4}}.
\end{split}
\end{equation}
This expression, given in Eq.~\eqref{eqn:reduced_trispectrum}, holds for any observable associated with the harmonic coefficients $a_{\ell m}$, assuming statistical isotropy. We now focus on the case of a scalar field $A(\versor{n})$ projected on the 2D sphere, such as the CMB lensing potential or the CMB temperature anisotropies. We have
\begin{equation}
\label{eqn:alm}
a_{\ell m}=\int\mathrm{d}^2\versor{n}\,A(\versor{n})Y_{\ell m}^*(\versor{n}).
\end{equation}
We further assume that $A(\versor{n})$ can be expressed as the integral of a source function $\mathcal{S}(\chi\versor{n}, \eta_0 - \chi)$, weighted by a kernel $f(\chi)$, over the comoving distance:
\begin{equation}
A(\versor{n})=\int_0^{\chi_{\ast}}\mathrm{d}\chi\,f(\chi)\mathcal{S}(\chi\versor{n},\eta_0-\chi).
\end{equation}
This is the case in both the CMB lensing and CMB primary anisotropy cases, as summarized in Tab.~\ref{tab:cmb_summary}.
\begin{table}[ht]
\centering
\begin{tabular}{|c|c|c|c|}
\hline
\textbf{Observable} & \textbf{Kernel}  & \textbf{Source Function} \\
\hline
CMB Lensing Potential & $W(\chi)$ & $\psi(\chi\versor{n}, \eta_0 - \chi)$ \\
\hline
Primary CMB Anisotropies & $\mathcal{F}[\,g(\chi)\,]$ & $\mathcal{S}_{\text{cmb}}(\chi\versor{n}, \eta_0 - \chi)$ \\
\hline
\end{tabular}
\caption{Summary of the kernel and source functions for the CMB lensing potential and primary CMB anisotropies. We use Eq.~\eqref{eqn:lensing_potential_approximated} and factor out a functional of the visibility function, $\mathcal{F}[g(\chi)]$, from the source function for the CMB primary anisotropies, $\mathcal{S}_{\text{cmb}}(\chi\versor{n}, \eta_0 - \chi)$, as in \cite{shiraishi2011cmb}.}
\label{tab:cmb_summary}
\end{table}

\noindent We now write $S$ as an inverse Fourier transform,
\begin{equation}
\begin{split}
A(\versor{n})&=\int_0^{\chi_{\ast}}\mathrm{d}\chi\,f(\chi)\int\frac{\mathrm{d}^3\mathbf{k}}{(2\pi)^3}\,\tilde{\mathcal{S}}(\mathbf{k},\eta_0-\chi)e^{-i\mathbf{k}\cdot\chi\versor{n}}\\
&=4\pi\int_0^{\chi_{\ast}}\mathrm{d}\chi\,f(\chi)\int\frac{\mathrm{d}^3\mathbf{k}}{(2\pi)^3}\,\tilde{\mathcal{S}}(\mathbf{k},\eta_0-\chi)\sum_{LM}(-i)^Lj_{L}(k\chi)Y_{LM}(\versor{n})Y_{LM}^*(\versor{k}).
\end{split}
\end{equation}
We use the plane wave expansion from \cite{mehrem2011plane}, substitute the above expression into Eq.~\eqref{eqn:alm}, and integrate over $\versor{n}$. We fomd
\begin{equation}
a_{\ell m}=4\pi(-i)^{\ell}\int_0^{\chi_{\ast}}\mathrm{d}\chi\,f(\chi)\int\frac{\mathrm{d}^3\mathbf{k}}{(2\pi)^3}\tilde{\mathcal{S}}(\mathbf{k},\eta_0-\chi)j_{\ell}(k\chi)Y_{\ell m}^*(\versor{k}).
\end{equation}
If $f(\chi)$ is sharply peaked at $\chi = \chi_{\ast}$, as in the primary CMB anisotropy case, the integral reduces to
\begin{equation}
a_{\ell m}\simeq4\pi(-i)^{\ell}f(\chi_{\ast})\int\frac{\mathrm{d}^3\mathbf{k}}{(2\pi)^3}\,\tilde{\mathcal{S}}(\mathbf{k},\eta_0-\chi_{\ast})j_{\ell}(k\chi_{\ast})Y_{\ell m}^*(\versor{k}).
\end{equation}
The key point is that the argument of the spherical Bessel function is now
\begin{equation}
k\chi_{\ast}=k(\eta_{0}-\eta_{\ast})\simeq k\eta_0,
\end{equation}
since $\eta_0 \ll \eta_{\ast}$. For sufficiently small scales, $k\eta_0 \gg 1$, allowing us to consider the asymptotic behaviour of the spherical Bessel functions \cite{abramowitz1968handbook}. It is
\begin{equation}
j_{\ell}(k\eta_0\gg1)\sim\frac{1}{k\eta_0}\sin\left(k\eta_0-\frac{\ell\pi}{2}\right)+\mathcal{O}\left(\frac{1}{k^2\eta_0^2}\right).
\end{equation}
Therefore, since the spherical Bessel function oscillates rapidly, the main contributions to the integral over $k$ will come from those $k$ for which $\ell \sim (2/\pi) k \eta_0$. Thus, except for very large scales, we have $\ell \gg 1$. Let us now examine how this affects spherical harmonics for $\versor{k} = (\theta, \varphi)$. We have
\begin{equation}
Y_{\ell m}(\theta,\varphi)=\sqrt{\frac{(2\ell+1)}{4\pi}}\sqrt{\frac{(\ell-m)!}{(\ell+m)!}}e^{im\varphi}P^{m}_{\ell}(\cos\theta),
\end{equation}
where $P_{\ell}^m(\cos\theta)$ are the associated Legendre polynomials, for $\ell \gg 1$ these latter behave as \cite{olver1997asymptotics, Bernardeau:2010ac}
\begin{equation}
P_{\ell\to\infty}^{-m}(\cos\theta)\sim\frac{1}{\ell^{\,m}}\sqrt{\frac{\theta}{\sin\theta}}\left[J_m(\ell\theta)+\mathcal{O}(1/\ell)\right],\qquad\qquad m\ge0.
\end{equation}
Since the Bessel function $J_m(\ell \theta)$ peaks at $m \sim \ell \theta$, the main contribution to the integral over $\versor{k}$ comes from $\theta \sim m / \ell$. Given that $\ell \gg 1$, this implies $\theta \ll 1$, as expected from the rule that the typical angular size of a feature with a given multipole is $\theta \propto 1 / \ell$. 

So motivated, let us now examine the behaviour of spherical harmonics in the $\theta \to 0$ limit. As shown in Sec.~5.14 of \cite{varshalovich1988quantum}, we have
\begin{equation}
\begin{split}
Y_{\ell m}(\theta,\varphi)&\simeq Y_{\ell m}(0,\varphi)+\frac{\partial}{\partial\theta}Y_{\ell m}(\theta,\varphi)\Big|_{\theta=0}\theta+\mathcal{O}(\theta^2)\\
&=\sqrt{\frac{2\ell+1}{4\pi}}\left[\delta^{\mathrm{K}}_{m,0}+\frac{\sqrt{\ell(\ell+1)}}{2}\theta\left(\delta^{\mathrm{K}}_{m,-1}e^{-i\varphi}-\delta^{\mathrm{K}}_{m,1}e^{i\varphi}\right)\right]+\mathcal{O}(\theta^2),    
\end{split}
\end{equation}
so that, at leading order, Eq.~\eqref{eqn:reduced_angular_trispectrum} becomes
\begin{equation}
\begin{split}
Q_{\ell_1\ell_2}^{\ell_3\ell_4}(L)&\simeq(2L+1)\begin{pmatrix}
\ell_1 & \ell_2 & L \\
0 & 0 & 0
\end{pmatrix}\begin{pmatrix}
\ell_3 & \ell_4 & L \\
0 & 0 & 0
\end{pmatrix}\expval{a_{\ell_10}a_{\ell_20}a_{\ell_30}a_{\ell_40}},
\end{split}
\end{equation}
since the sum of the elements in the second row of the Wigner $3$-$j$ symbols must be zero. However, this implies
\begin{equation}
\begin{dcases}
&\text{$\ell_1+\ell_2+L=$ even}\\
&\text{$\ell_3+\ell_4+L=$ even}
\end{dcases}\qquad\implies\qquad\text{$\ell_1+\ell_2+\ell_3+\ell_4=$ even.}
\end{equation}
Therefore, at leading order, the parity-odd component of the angular trispectrum is suppressed, except on very large angular scales. For the primary CMB, these scales map to much larger physical scales than for CMB lensing, which also maps a given angular scale to a broad range of physical scales due to the broad lensing kernel.

This is not the case of CMB lensing, where the kernel is the lensing efficiency $W(\chi)$, as defined in Eq.~\eqref{eqn:lensing_efficiency}. Although CMB lensing anisotropies are still defined on a 2D spherical surface, integrating along the line of sight with a kernel depending on the comoving distance $\chi$ upweights contributions from spherical shells that are closer to the observer, and not just from the last scattering surface as with primary CMB anisotropies. Unlike the primary CMB, the angular trispectrum of the CMB lensing potential does not suffer from the same suppression of parity-odd signals. This difference arises from the physical scales involved. The broad lensing kernel enhances sensitivity to smaller, intermediate redshifts and to physical scales comparable to those where parity violation has been observed in large-scale structure. These scales are much smaller than those probed by the primary CMB, making parity-odd signals more accessible in the lensing case.

\section{\label{sec:SNR}Estimating the Signal-to-Noise Ratio}
In this section, we estimate the signal-to-noise ratio of the CMB lensing trispectrum. To compute the error bars that enter this ratio, we adopt a prescription similar to that of \cite{hu2001angular}. We define an unbiased estimator for the parity-odd component of the harmonic 4PCF:
\begin{equation}
\label{eqn:estimator}
\begin{split}
\hat{Q}_{\ell_3\ell_4}^{\ell_1\ell_2}(L)\equiv(2L+1)\sum_{m_a}\sum_{M}(-1)^M&\begin{pmatrix}
\ell_1 & \ell_2 & L \\
m_1 & m_2 & M
\end{pmatrix}\\
&\quad\times\begin{pmatrix}
\ell_3 & \ell_4 & L \\
m_3 & m_4 & -M
\end{pmatrix}\phi_{\ell_1m_1}\phi_{\ell_2m_3}\phi_{\ell_3m_3}\phi_{\ell_4m_4},
\end{split}
\end{equation}
for $a=1,2,3,4$, where $\ell_1+\ell_2+\ell_3+\ell_4=\text{odd}$, so that $\expval{\hat{Q}_{\ell_3\ell_4}^{\ell_1\ell_2}(L)}=Q_{\ell_3\ell_4}^{\ell_1\ell_2}(L)$. We now restrict our analysis to $\ell_1<\ell_2<\ell_3<\ell_4$ and $\ell_1^{\prime}<\ell_2^{\prime}<\ell_3^{\prime}<\ell_4^{\prime}$. One can show that \cite{hu2001angular}
\begin{equation}
\label{eqn:covariance}
\begin{split}
\frac{\expval{\hat{Q}_{\ell_3\ell_4}^{\ell_1\ell_2*}(L)\hat{Q}_{\ell_3^{\prime}\ell_4^{\prime}}^{\ell_1^{\prime}\ell_2^{\prime}}(L^{\prime})}}{(2L+1)C_{\ell_1}^{\phi\phi}C_{\ell_2}^{\phi\phi}C_{\ell_3}^{\phi\phi}C_{\ell_4}^{\phi\phi}}=\delta^{\mathrm{K}}_{LL^{\prime}}\delta^{12}_{34}+(2L^{\prime}+1)\Bigg[&(-1)^{\ell_2+\ell_3}\left\{\begin{matrix}\ell_1 & \ell_2 & L \\
\ell_4 & \ell_3 & L^{\prime}
\end{matrix}\right\}\delta_{24}^{13}\\
&\qquad+(-1)^{L+L^{\prime}}\left\{\begin{matrix}\ell_1 & \ell_2 & L \\
\ell_3 & \ell_4 & L^{\prime}
\end{matrix}\right\}\delta_{32}^{14}\Bigg],
\end{split}
\end{equation}
where we have used the expression for the Wigner $6$-$j$ symbol \cite{varshalovich1988quantum}
\begin{equation}
\label{eqn:wigner_6j}
\begin{split}
\left\{\begin{matrix}
j_1 & j_2 & j_3 \\
j_4 & j_5 & j_6
\end{matrix}\right\}\equiv(-1)^{j_4+j_5+j_6}\sum_{s_1s_2s_3}&\sum_{s_4s_5s_6}(-1)^{s_4+s_5+s_6}\begin{pmatrix}
j_1 & j_2 & j_3 \\
s_1 & s_2 & s_3
\end{pmatrix}\\
&\quad\times\begin{pmatrix}
j_5 & j_1 & j_6 \\
s_5 & -s_1 & -s_6
\end{pmatrix}\begin{pmatrix}
j_6 & j_2 & j_4 \\
s_6 & -s_2 & -s_4
\end{pmatrix}\begin{pmatrix}
j_4 & j_3 & j_5 \\
s_4 & -s_3 & -s_5
\end{pmatrix},
\end{split}
\end{equation}
together with Eq.~\eqref{eqn:wigner_orthogonality}, and defined
\begin{equation}
\label{eqn:delta_covariance}
\begin{split}
\delta_{cd}^{ab}&\equiv\left[\delta^{\mathrm{K}}_{\ell_1\ell_a}\delta^{\mathrm{K}}_{\ell_2\ell_b}+(-1)^{\ell_1+\ell_2+L}\delta^{\mathrm{K}}_{\ell_1\ell_b}\delta^{\mathrm{K}}_{\ell_2\ell_a}\right]\left[\delta^{\mathrm{K}}_{\ell_3\ell_c}\delta^{\mathrm{K}}_{\ell_4\ell_d}+(-1)^{\ell_3+\ell_4+L}\delta^{\mathrm{K}}_{\ell_3\ell_d}\delta^{\mathrm{K}}_{\ell_4\ell_c}\right]\\
&\qquad+\left[\delta^{\mathrm{K}}_{\ell_1\ell_c}\delta^{\mathrm{K}}_{\ell_2\ell_d}+(-1)^{\ell_1+\ell_2+L}\delta^{\mathrm{K}}_{\ell_1\ell_d}\delta^{\mathrm{K}}_{\ell_2\ell_c}\right]\left[\delta^{\mathrm{K}}_{\ell_3\ell_a}\delta^{\mathrm{K}}_{\ell_4\ell_b}+(-1)^{\ell_3+\ell_4+L}\delta^{\mathrm{K}}_{\ell_3\ell_b}\delta^{\mathrm{K}}_{\ell_4\ell_a}\right].
\end{split}
\end{equation}
We restrict to $\ell_1 < \ell_2 < \ell_3 < \ell_4$ and $\ell_1^{\prime} < \ell_2^{\prime} < \ell_3^{\prime} < \ell_4^{\prime}$ to ignore terms in the covariance where one of more pairs is from the same family (unprimed or primed). By definition, the squared signal-to-noise ratio is then
\begin{equation}
\label{eqn:SNR_general}
\left(\mathrm{SNR}\right)^2\equiv\sum_{\ell_1\ell_2\ell_3\ell_4}\sum_{\ell_1^{\prime}\ell_2^{\prime}\ell_3^{\prime}\ell_4^{\prime}}\sum_{LL^{\prime}}\frac{\expval{\hat{Q}_{\ell_3\ell_4}^{\ell_1\ell_2*}(L)}\expval{\hat{Q}_{\ell_3^{\prime}\ell_4^{\prime}}^{\ell_1^{\prime}\ell_2^{\prime}}(L^{\prime})}}{\expval{\hat{Q}_{\ell_3\ell_4}^{\ell_1\ell_2*}(L)\hat{Q}_{\ell_3^{\prime}\ell_4^{\prime}}^{\ell_1^{\prime}\ell_2^{\prime}}(L^{\prime})}-\expval{\hat{Q}_{\ell_3\ell_4}^{\ell_1\ell_2*}(L)}\expval{\hat{Q}_{\ell_3^{\prime}\ell_4^{\prime}}^{\ell_1^{\prime}\ell_2^{\prime}}(L^{\prime})}}.
\end{equation}
If we now adopt the approximation of weak non-Gaussianity for the covariance,\footnote{The observed CMB anisotropies are very close to Gaussian, consistent with the simplest models of inflation \cite{bartolo2004non, akrami2020planck_nongaussianity, akrami2020planck_inflation, komatsu2001acoustic, Kogo:2006kh}, justifying our use of power spectra only to form the covariance matrix employed in estimating the SNR.} we can reasonably assume that the second term in the denominator is subdominant to the first one, and so, by substituting Eq.~\eqref{eqn:covariance}, we obtain \cite{hu2001angular, Kogo:2006kh}
\begin{equation}
\label{eqn:SNR_weak}
\mathrm{SNR}\simeq\sqrt{\sum_{L=2}^{\ell_{\mathrm{max}}}(2L+1)^{-1}\sum^{\ell_{\mathrm{max}}}_{\ell_1<\ell_2<\ell_3<\ell_4}\frac{\lvert Q_{\ell_3\ell_4}^{\ell_1\ell_2}(L)\rvert^2}{C_{\ell_1}^{\phi\phi}C_{\ell_2}^{\phi\phi}C_{\ell_3}^{\phi\phi}C_{\ell_4}^{\phi\phi}}},
\end{equation}
since the Kronecker deltas in Eq.~\eqref{eqn:delta_covariance} simplify the summations over the $\ell^{\prime}$. By restricting the summation to $\ell_1<\ell_2<\ell_3<\ell_4$, we eliminate redundant permutations, though we are neglecting the signal-to-noise contributed when the $\ell$’s are not all different from each other. Notice that the summation starts at $L=2$, as lower multipoles are unobservable in CMB measurements: $L=0$ corresponds to the average CMB temperature, while $L=1$ is indistinguishable from the observer's motion relative to the CMB.

To estimate the SNR for the CMB lensing trispectrum, we extended the bispectrum numerics from Section~\ref{sec:bispectrum} to calculate the angular trispectrum. We validated our bispectrum calculation by comparing our results with \cite{bohm2016bias}, as shown in Appendix~\ref{app:recovering_bohm}, confirming the accuracy and reliability of our implementation.

The integral defined in Eq.~\eqref{eqn:integral_trispectrum} depends solely on the multipoles $\ell_1$, $\ell_2$, $\ell_3$, and $\ell_4$. To speed up the trispectrum calculation, we pre-compute $I_{\ell_1 \ell_2 \ell_3 \ell_4}^{111}$ for all valid multipole combinations, where the sum of the multipoles is odd and the order satisfies $\ell_1 < \ell_2 < \ell_3 < \ell_4$. This pre-computation allows for quick evaluation of the reduced trispectrum and SNR across these combinations, avoiding redundant calculations.

We further enhance the efficiency by vectorizing and parallelizing the computations. Vectorization reduces overhead by minimizing loop iterations, while parallelization, implemented using a process pool executor, distributes the workload across multiple CPUs. Together, these techniques enable efficient and scalable SNR calculations. In summary, we extend the trispectrum algorithm, precompute the necessary integrals for valid multipole combinations, and use vectorization and parallelization to efficiently compute the SNR.

\begin{figure}
\centering
\includegraphics[width = \linewidth]{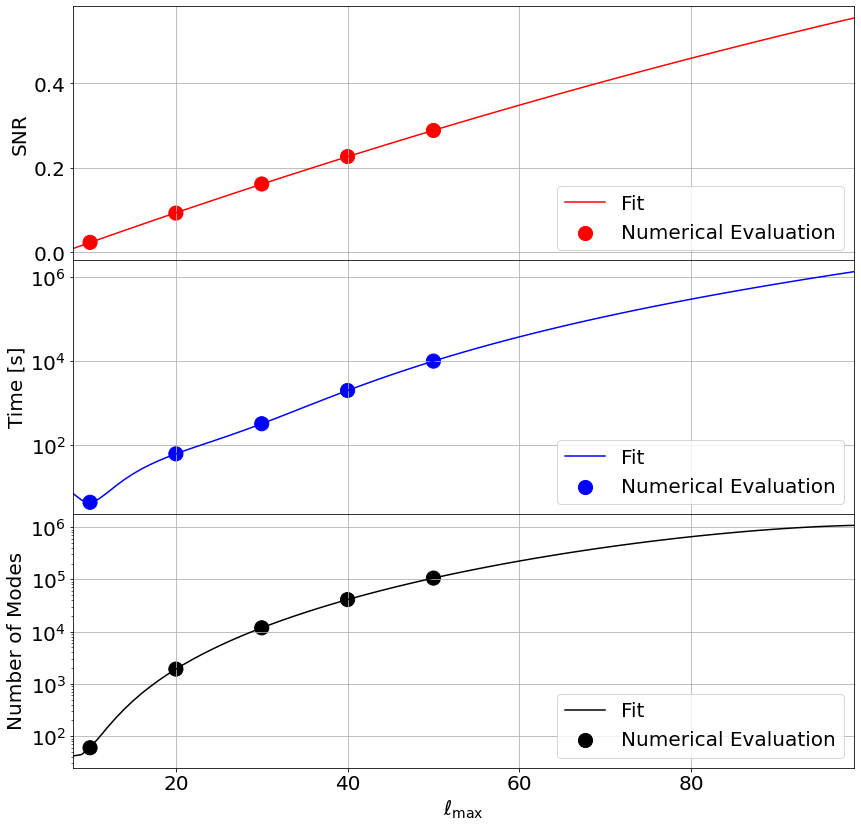}
\caption{The signal-to-noise ratio for the parity-odd CMB lensing trispectrum, the total evaluation time, and the number of modes (i.e. the different combination of $\ell_1,\ell_2,\ell_3,\ell_4, L$), as functions of $\ell_{\mathrm{max}}$, the multipole at which we truncate the summations. We have computed $Q_{\ell_3\ell_4}^{\ell_1\ell_2}(L)$ for the parity-odd toy model trispectrum defined in Eq.~\eqref{eqn:primordial_trispectrum} and we have used the Eisenstein \& Hu matter transfer function. We have performed the numerical evaluation for the toy model considered in this paper by setting $|g_{-}|=2\times10^7$, as done in \cite{Hou:2024udn}, and without including noise.}
\label{fig:SNR}
\end{figure}

We show the SNR as a function of $\ell_{\mathrm{max}}$ in the upper panel of Fig.~\ref{fig:SNR} in the absence of noise. Indeed, it is quite reasonable to assume a low-noise limit for future experiments\footnote{Achieving these low noise levels would require going beyond the standard quadratic estimator by \cite{okamoto2002angular} to optimal maximum likelihood estimators, from \cite{Hirata:2003ka, Smith:2010gu}.}, as suggested by Fig.~46 in \cite{CMB-S4:2016ple}. Due to the computational cost of evaluating the SNR for large multipoles, we fitted the numerically obtained data with a polynomial of degree $N$ to extrapolate the values for higher $\ell$'s.
We find that the signal-to-noise ratio increases roughly monotonically and is well fit by a quadratic polynomial in $\ell_{\mathrm{max}}$. As expected, the SNR does not saturate, since it is free from instrumental noise. Moreover, the execution time required by our code to evaluate the SNR seems to have approximately the same scaling as the number of modes: this is reasonable, since if we increase the latter, we also increase the complexity of the calculation. A straightforward extrapolation suggests that an SNR exceeding $1$ could be achieved for a larger $\ell_{\mathrm{max}}$. However, the signal will inevitably drop off at some $\ell_{\mathrm{max}}$ as we approach the flat-sky limit.

\begin{table}
\centering
\label{tab:poly_summary}
\begin{tabular}{crrr}
\toprule
 & \textbf{SNR} & \textbf{Time} & \textbf{\# of Modes} \\
\midrule
\boldsymbol{$c_0$}      & $-4.995\times10^{-02}$ & $8.883\times10^{+01}$  & $1.734\times10^{+02}$  \\
\boldsymbol{$c_1$}      & $7.451\times10^{-03}$  & $-1.814\times10^{+01}$ & $-9.949\times10^{+00}$ \\
\boldsymbol{$c_2$}      & $-1.357\times10^{-05}$ & $1.026\times10^{+00}$  & $-2.696\times10^{+00}$ \\
\boldsymbol{$c_3$}      &                        & $5.122\times10^{-04}$  & $1.669\times10^{-01}$  \\
\boldsymbol{$c_4$}      &                        & $-4.871\times10^{-04}$ & $7.309\times10^{-03}$  \\
\boldsymbol{$c_5$}      &                        & $-6.939\times10^{-06}$ & $1.460\times10^{-04}$  \\
\boldsymbol{$c_6$}      &                        & $1.814\times10^{-07}$  & $1.329\times10^{-06}$  \\
\boldsymbol{$c_7$}      &                        & $1.349\times10^{-08}$  & $-2.571\times10^{-08}$ \\
\midrule
\boldsymbol{$\sigma_0$} & $1.593\times10^{-03}$  & $3.184\times10^{-07}$  & $2.816\times10^{-08}$   \\
\boldsymbol{$\sigma_1$} & $1.214\times10^{-04}$  & $7.388\times10^{-08}$  & $6.526\times10^{-09}$  \\
\boldsymbol{$\sigma_2$} & $1.985\times10^{-06}$  & $6.423\times10^{-09}$  & $5.664\times10^{-10}$  \\
\boldsymbol{$\sigma_3$} &                        & $2.793\times10^{-10}$  & $2.458\times10^{-11}$  \\
\boldsymbol{$\sigma_4$} &                        & $6.696\times10^{-12}$  & $5.880\times10^{-13}$   \\
\boldsymbol{$\sigma_5$} &                        & $8.982\times10^{-14}$  & $7.867\times10^{-15}$   \\
\boldsymbol{$\sigma_6$} &                        & $6.311\times10^{-16}$  & $5.512\times10^{-17}$   \\
\boldsymbol{$\sigma_7$} &                        & $1.807\times10^{-18}$  & $1.573\times10^{-19}$   \\
\bottomrule
\end{tabular}
\caption{Summary of the coefficients $c_n$ and their uncertainties $\sigma_n$ for the polynomial fits of degree $N$ performed on three distinct quantities: the SNR, execution time, and the number of modes, plotted in Fig.~\ref{fig:SNR}. The first column shows the coefficients and uncertainties for the polynomial fit to the SNR curve ($N=2$); the second column for the execution time ($N=7$), and the third column for the number of modes ($N=7$). While the execution time and number of modes are fitted using polynomials of the same degree, their coefficients and uncertainties differ, reflecting independent fits.}
\end{table}

\section{\label{sec:conclusions}Conclusions}
In this work, we have explored the sensitivity of the CMB lensing trispectrum to parity violation, providing a novel probe for investigating fundamental physics through observations of the cosmic microwave background. Our work builds on the existing theoretical framework of CMB lensing correlators and extends it to the four-point correlation function. We demonstrate that the CMB lensing trispectrum is a powerful tool for testing parity-breaking inflationary models. We began by defining the weak lensing potential and computing the CMB lensing power spectrum to ensure the robustness of our formalism. Our analysis then revisited the bispectrum calculation, following the approach outlined in \cite{bohm2016bias}, and adapted it to our specific context.

In Section~\ref{sec:trispectrum}, we present our main findings, detailing the trispectrum computation with a focus on parity violation. We implemented a parity-odd toy model inflationary trispectrum, inspired by \cite{jamieson2024parity, coulton2024signatures}, and derived the corresponding CMB lensing angular trispectrum.

Our results reveal that the CMB lensing trispectrum is sensitive to parity violation. Moreover, we estimated the signal-to-noise ratio for detecting this parity-breaking observable. Notably, as shown in Fig.~\ref{fig:SNR}, the SNR increases approximately quadratically with $\ell_{\mathrm{max}}$. However, we estimated the signal-to-noise ratio only in the idealized case, neglecting instrumental noise, foregrounds, and systematics \cite{Hu:2002vu}. We acknowledge these effects and leave their inclusion to future studies. This result demonstrates the feasibility of detecting parity-odd features in the CMB lensing trispectrum under ideal conditions.

Our findings have significant implications. We show that the CMB lensing trispectrum can test specific parity-breaking inflationary models, offering a novel indirect method for probing fundamental physics. This approach complements existing tests of parity violation in both the LSS and primary CMB while opening new avenues for exploring parity violation through cosmological observations. Future observational efforts should focus on extending these results to realistic scenarios that account for physical inflationary models and instrumental noise in the upcoming CMB lensing experiments, like Simons Observatory \cite{SimonsObservatory:2018koc} and CMB-S4 \cite{CMB-S4:2016ple}. Moreover, as discussed in \cite{jamieson2024parity}, the trispectrum template defined in Eq.~\eqref{eqn:primordial_trispectrum} peaks in the squeezed limit. This raises the possibility that using a template with a different peak could alter our results.

\appendix

\section{\label{app:recovering_bohm}Recovering B{\"o}hm's Bispectrum}
\begin{figure}
\centering
\includegraphics[width = \linewidth]{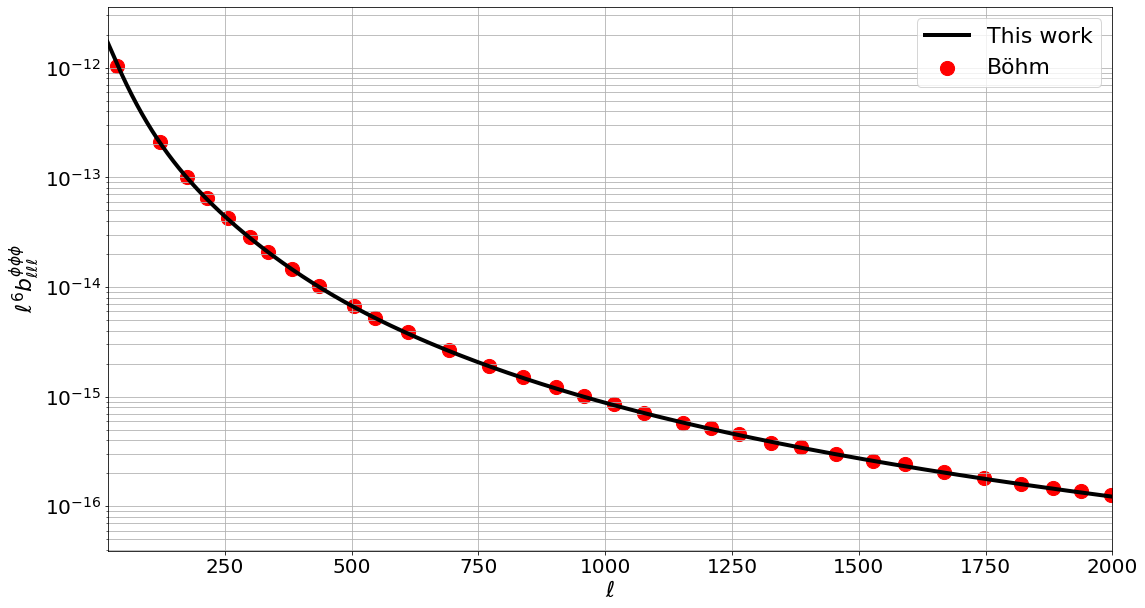}
\caption{The equilateral reduced lensing bispectrum, obtained using the model defined in Eq.~\eqref{eqn:bohm_matter_bispectrum}, is shown. The black curve represents the numerical result from this paper, while the red dots correspond to the values from \cite{bohm2016bias}. The numerical computation was performed using \texttt{CLASS} \cite{lesgourgues2011cosmic}, specifically the \texttt{HaloFit} method \cite{smith2003stable, takahashi2012revising} to obtain the non-linear matter power spectrum, with the \textit{Planck} Public Release 4 (2018) fiducial values for a $\Lambda$CDM cosmology \cite{aghanim2020planck}.}
\label{fig:recovering_bohm}
\end{figure}
In this appendix, we present the result of a consistency check we have performed to compare our results with those of \cite{bohm2016bias}. 

In that paper, the matter bispectrum is not sourced by primordial non-Gaussianity but is instead induced by the non-linearities in gravitational clustering. In this context, the matter bispectrum can be modelled using standard Eulerian perturbation theory, as described, for example, in \cite{bernardeau2002large}. Consequently, we have numerically evaluated the CMB lensing reduced bispectrum by replacing Eq.~\eqref{eqn:matter_bispectrum} with Eq.~(17) from \cite{bohm2016bias}:
\begin{equation}
\label{eqn:bohm_matter_bispectrum}
\begin{split}
&\expval{\tilde{\delta}_{\mathrm{m}}(\mathbf{k}_1,\eta_0-\chi_1)\tilde{\delta}_{\mathrm{m}}(\mathbf{k}_2,\eta_0-\chi_2)\tilde{\delta}_{\mathrm{m}}(\mathbf{k}_3,\eta_0-\chi_3)}=\\
&\qquad\qquad\qquad=(2\pi)^3\delta_{\mathrm{D}}^{[3]}\left(\mathbf{k}_1+\mathbf{k}_2+\mathbf{k}_3\right)\Big[2F_2(\mathbf{k}_1,\mathbf{k}_2)P_{\delta}^{\text{NL}}(k_1,\chi_1)P_{\delta}^{\text{NL}}(k_2,\chi_2)\\
&\qquad\qquad\qquad\qquad\qquad\qquad\qquad\qquad\qquad\qquad\qquad\qquad\qquad+\left(1\leftrightarrow3\right)+\left(2\leftrightarrow3\right)\Big],
\end{split}
\end{equation}
where $P_{\delta}^{\text{NL}}$ is the non-linear matter power spectrum, and
\begin{equation}
\label{eqn:F2}
F_2(\mathbf{k},\mathbf{k}^{\prime})\equiv\frac{5}{7}+\frac{1}{2}\left[\frac{k}{k^{\prime}}+\frac{k^{\prime}}{k}\right]\left(\versor{k}\cdot\versor{k}^{\prime}\right)+\frac{2}{7}\left(\versor{k}\cdot\versor{k}^{\prime}\right)^2.
\end{equation}
Moreover, because of the Dirac delta ensuring statistical homogeneity, $\mathbf{k}_1$, $\mathbf{k}_2$ and $\mathbf{k}_3$ form a closed triangle, and so we can use the law of cosines to eliminate the dependence on the direction of the wave-vectors, i.e.
\begin{equation}
\label{eqn:cosinus_law}
\versor{k}\cdot\versor{k}^{\prime}=\frac{{k^{\prime\prime}}^2-k^2-{k^{\prime}}^2}{2kk^{\prime}}\qquad\text{if}\qquad\mathbf{k}+\mathbf{k}^{\prime}+\mathbf{k}^{\prime\prime}=0.
\end{equation}
By substituting Eq.~\eqref{eqn:cosinus_law} into Eq.~\eqref{eqn:F2}, and then inserting the resulting expression into Eq.~\eqref{eqn:3PCF}, we obtain
\begin{equation}
\label{eqn:3PCF_bohm}
\begin{split}
&\expval{\phi_{\ell_1m_1}\phi_{\ell_2m_2}\phi_{\ell_3m_3}}=-\int\frac{\mathrm{d}^3\mathbf{k}_1}{(2\pi)^3}\int\frac{\mathrm{d}^3\mathbf{k}_2}{(2\pi)^3}\int\frac{\mathrm{d}^3\mathbf{k}_3}{(2\pi)^3}\int_{0}^{\chi_{\ast}}\frac{\mathrm{d}\chi_1}{k^2_1}\int_{0}^{\chi_{\ast}}\frac{\mathrm{d}\chi_2}{k^2_2}\int_{0}^{\chi_{\ast}}\frac{\mathrm{d}\chi_3}{k^2_3}\,\\
&\qquad\quad\times\prod_{n=1}^3\left[4\pi(-i)^{\ell_n}Y_{\ell_nm_n}^*(\versor{k}_n)W(\chi_n)\gamma(\chi_n)j_{\ell_n}(k_n\chi_n)\right](2\pi)^3\delta_{\mathrm{D}}^{[3]}\left(\mathbf{k}_1+\mathbf{k}_2+\mathbf{k}_3\right)\\
&\times \Bigg\{\left[\frac{10}{7}+\left(\frac{k_1}{k_2}+\frac{k_2}{k_1}\right)\left(\frac{{k_3}^2-k_1^2-{k_2^2}}{2k_1k_2}\right)+\frac{4}{7}\left(\frac{{k_3}^2-k_1^2-{k_2}^2}{2k_1k_2}\right)^2\right]\\
&\qquad\qquad\qquad\qquad\qquad\qquad\qquad\times P_{\delta}^{\text{NL}}(k_1,\chi_1)P_{\delta}^{\text{NL}}(k_2,\chi_2)+\left(1\leftrightarrow3\right)+\left(2\leftrightarrow3\right)\Bigg\}.
\end{split}
\end{equation}
We then follow the same steps outlined in Sec.~\ref{sec:bispectrum}, including rewriting the Dirac delta function in terms of plane waves and applying the Limber approximation. Ultimately, we derive the following expression for the reduced angular bispectrum of CMB lensing:
\begin{equation}
\label{eqn:bohm_bispectrum}
\begin{split}
b^{\phi\phi\phi}_{\ell_1\ell_2\ell_3}&=-\int_0^{\chi_{\ast}}\frac{\mathrm{d}\chi\,\chi^2}{(\ell_1\ell_2\ell_3)^2}W^3(\chi)\gamma(\chi)^3\\
&\times\Bigg\{\left[\frac{10}{7}+\left(\frac{\ell_1}{\ell_2}+\frac{\ell_2}{\ell_1}\right)\left(\frac{{\ell_3}^2-\ell_1^2-{\ell_2^2}}{2\ell_1\ell_2}\right)+\frac{4}{7}\left(\frac{{\ell_3}^2-\ell_1^2-{\ell_2}^2}{2\ell_1\ell_2}\right)^2\right]\\
&\qquad\qquad\qquad\qquad\qquad\quad\times P_{\delta}^{\text{NL}}\left(\frac{\ell_1}{\chi},\chi\right)P_{\delta}^{\text{NL}}\left(\frac{\ell_2}{\chi},\chi\right)+\left(1\leftrightarrow3\right)+\left(2\leftrightarrow3\right)\Bigg\},
\end{split}
\end{equation}
which matches Eq.~(16) in \cite{bohm2016bias}. In Fig.~\ref{fig:recovering_bohm} we compare the numerical evaluation of Eq.~\eqref{eqn:bohm_bispectrum} with the results of \cite{bohm2016bias}.

\section{\label{app:parity_on_sphere}Parity on a Spherical Surface}
In this appendix, we demonstrate that parity-transforming a scalar field projected on a spherical surface is not equivalent to rotating the projection. It is well known that three dimensions are required to distinguish a parity transformation from a simple rotation. On a flat 2D plane, a parity transformation is equivalent to a $180^{\circ}$ rotation around the axis perpendicular to the plane through the origin. However, this property does not hold on a curved 2D surface, such as a sphere. Here, we consider the case of a spherical surface, where any scalar field $A$ can be projected using spherical harmonics as \cite{varshalovich1988quantum}
\begin{equation}
A(\mathbf{r})=\sum_{\ell=0}^{\infty}\sum_{m=-\ell}^{\ell}a_{\ell m}(r)Y_{\ell m}(\versor{r}).
\end{equation}
Both parity and rotation preserve $r$ but affect $\versor{r}$. Indeed, the parity operator acts on spin-zero spherical harmonics as
\begin{equation}
\label{eqn:parity_Ylm}
\mathbb{P}\left[Y_{\ell m}(\versor{r})\right]= Y_{\ell m}(-\versor{r})=(-1)^{\ell}Y_{\ell m}(\versor{r}).
\end{equation}
There is no rotation able to do this. To see this explicitly, let us consider a rotation $\mathsf{R}=\mathsf{R}(\alpha,\beta,\gamma)$ about the origin, where $\alpha$, $\beta$ and $\gamma$ are the Euler angles. Under this operation,
\begin{equation}
\label{eqn:rotation_Ylm}
Y_{\ell m}(\mathsf{R}\versor{n})=\sum_{M=-\ell}^{\ell}\left[D^{(\ell)}_{mM}(\mathsf{R})\right]^*Y_{\ell M}(\versor{n}),
\end{equation}
with $D^{(\ell)}_{mM}(\mathsf{R})$ being an element of the Wigner $D$-matrix. The most general rotation can be expressed as the product of three rotations, one for each axis,
\begin{equation}
\label{eqn:rotation_matrix}
\begin{split}
\mathsf{R}(\alpha,\beta,\gamma)&=\mathsf{R}_x(\gamma)\mathsf{R}_y(\beta)\mathsf{R}_z(\alpha)\\
&=\begin{pmatrix}
\cos\gamma & -\sin\gamma & 0 \\
\sin\gamma & \cos\gamma & 0 \\
0 & 0 & 1
\end{pmatrix}\begin{pmatrix}
\cos\beta & 0 & \sin\beta \\
0 & 1 & 0 \\
-\sin\beta & 0 & \cos\beta
\end{pmatrix}\begin{pmatrix}
1 & 0 & 0 \\
0 & \cos\alpha & -\sin\alpha \\
0 & \sin\alpha & \cos\alpha
\end{pmatrix}\\
&=\begin{pmatrix}
\cos\beta\cos\gamma & \sin\alpha\sin\beta\cos\gamma-\cos\alpha\sin\gamma & \cos\alpha\sin\beta\cos\gamma+\sin\alpha\sin\gamma \\
\cos\beta\sin\gamma & \sin\alpha\sin\beta\sin\gamma + \cos\alpha\cos\gamma &  \cos\alpha\sin\beta\sin\gamma-\sin\alpha\cos\gamma \\
-\sin\beta & \sin\alpha\cos\beta & \cos\alpha\cos\beta
\end{pmatrix},
\end{split}
\end{equation}
i.e. a composite rotation whose yaw, pitch, and roll angles are the Euler angles. We can easily compute the determinant of the matrix above by using the Cauchy-Binet formula, which states that the determinant of a product of square matrices is equal to the product of their determinants:
\begin{equation}
|\mathsf{R}(\alpha,\beta,\gamma)|=|\mathsf{R}_x(\gamma)||\mathsf{R}_y(\beta)||\mathsf{R}_z(\alpha)|=1.
\end{equation}
On the contrary, the parity transformation of a 3D vector $\mathbf{r}$ is represented by a matrix whose determinant equals $-1$,
\begin{equation}
-\mathbf{r}=\begin{pmatrix}
-x\\
-y\\
-z\\
\end{pmatrix}=\mathbb{P}\left[\mathbf{r}\right]=\begin{pmatrix}
-1 & 0 & 0 \\
0 & -1 & 0 \\
0 & 0 & -1
\end{pmatrix}\begin{pmatrix}
x \\
y \\
z
\end{pmatrix}.
\end{equation}
The reason is that parity is a discrete transformation, while rotations are continuous ones, so their matrix representations belong to different groups. Let us now show that there is no combination of Euler angles able to induce a parity transformation. If we now equate the two matrix representations, we obtain a system of 9 equations with 3 unknowns, i.e. the Euler angles. By starting from the last row, we find
\begin{equation}
\begin{dcases}
\sin\beta = 0 \\
\sin\alpha\cos\beta = 0\\
\cos\alpha\cos\beta = -1
\end{dcases}\quad\implies\quad\begin{dcases}
\beta=b\pi\\
\alpha=a\pi\\
a+b=\text{odd}
\end{dcases}
\end{equation}
for $a,b\in\mathbb{Z}$. From the second row, we find
\begin{equation}
\begin{dcases}
\cos\beta\sin\gamma=0\\
\sin\alpha\sin\beta+\cos\alpha\cos\gamma=-1\\
\cos\alpha\sin\beta\sin\gamma-\sin\alpha\cos\gamma=0
\end{dcases}\quad\implies\quad\begin{dcases}
\gamma=c\pi\\
a+c=\text{odd}
\end{dcases}
\end{equation}
for $c\in\mathbb{Z}$. Finally, from the first row, we obtain
\begin{equation}
\begin{dcases}
\cos\beta\cos\gamma=-1\\
\sin\alpha\sin\beta\cos\gamma-\cos\alpha\sin\gamma=0\\
\cos\alpha\sin\beta\cos\gamma+\sin\alpha\sin\gamma=0
\end{dcases}\quad\implies\quad b+c=\text{odd}.
\end{equation}
By summing together the conditions on $a,b,c$, we find
\begin{equation}
2(a+b+c)=\text{odd}.
\end{equation}
However, the left-hand side of the equation above is even, so there is no combination of Euler angles able to satisfy this equality. This proves that parity-transforming a scalar field on a 2D curved surface is not equivalent to rotating it.

\section{Maximum Value of Triple Product Sourcing Primary CMB Trispectrum}
\label{app:triple_product}

Here, we explore a simple argument to illustrate why the primary CMB temperature anisotropies are not particularly sensitive probes of parity violation.

The primary CMB temperature perturbations originate from the gravitational redshifting or blueshifting of photons decoupling from the electron-proton plasma at recombination, as they emerge from gravitational potential wells (due to overdensities) or potential peaks (due to underdensities). Thus, we aim to determine the configuration of four density points that maximizes the scalar triple product, as all parity-odd basis functions are proportional to this quantity \cite{cahn2023isotropic}. However, these four density points must also approximately lie on the same spherical shell at the redshift of recombination.\footnote{
For simplicity, in this approximate calculation, we neglect the finite width of recombination.} 

\begin{figure}
\centering
\includegraphics[width = 0.8\linewidth]{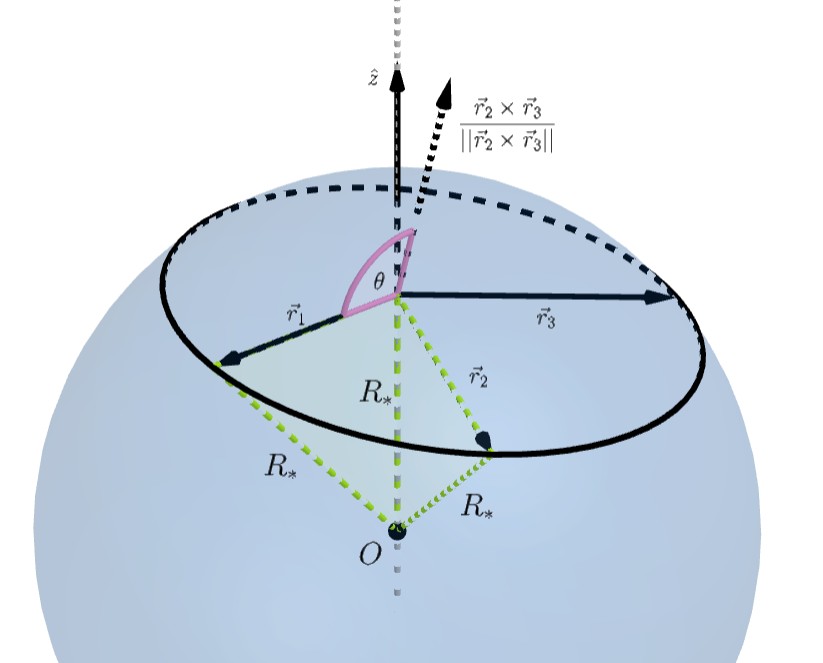}
\caption{Three density points located on the last-scattering surface, represented by the vectors $\mathbf{r}_1$, $\mathbf{r}_2$, and $\mathbf{r}_3$, which define the scalar triple product $\mathbf{r}_1 \cdot (\mathbf{r}_2 \times \mathbf{r}_3)$. The last-scattering surface is modeled as a sphere of radius $R_{\ast}$ centered at $O$, with all vector endpoints lying on this sphere. The configuration shown assumes $\mathbf{r}_3$ is orthogonal to $\mathbf{r}_2$ to maximize their cross product. The angle $\theta$ shown is defined as the angle between $\mathbf{r}_1$ and $\mathbf{r}_2 \times \mathbf{r}_3$.}
\label{fig:diag1}
\end{figure}

In Fig.~\ref{fig:diag1}, we illustrate three density points identified by the vectors $\mathbf{r}_1$, $\mathbf{r}_2$, and $\mathbf{r}_3$, which form the scalar triple product $\mathbf{r}_1 \cdot (\mathbf{r}_2 \times \mathbf{r}_3)$. The endpoints of these three vectors lie on the same spherical surface of radius $R_{\ast}$ centered at $O$, corresponding to the surface of last scattering. Our goal is to estimate the maximum value of this triple product, which requires determining the maximum possible value of $\cos\theta$.

\begin{figure}
\centering
\includegraphics[width = 0.5\linewidth]{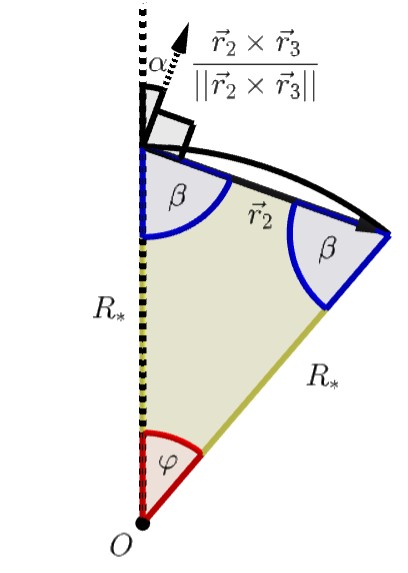}
\caption{Zoom-in on the isosceles triangle formed by the origin $O$ and the vectors $\mathbf{r}_2$ and $\mathbf{r}_3$, whose endpoints lie on the last-scattering surface of radius $R_{\ast}$. The triangle has two sides of length $R_{\ast}$ and base $\norm{\mathbf{r}_2}$. The internal angle $\varphi$ at the origin is determined via the law of cosines, and is used to compute the angle $\alpha$ between the $\hat{z}$-axis and the cross product $\mathbf{r}_2 \times \mathbf{r}_3$, which approximates to $\alpha \simeq \norm{\mathbf{r}_2}/(2R_{\ast})$ in the limit $\norm{\mathbf{r}_2} \ll R_{\ast}$.}
\label{fig:diag2}
\end{figure}

To do so, let us first consider the isosceles triangle with sides $R_{\ast}$, $R_{\ast}$, and $\norm{\mathbf{r}_2}$, which we zoom in on in Fig.~\ref{fig:diag2}. By applying the law of cosines, we find  
\begin{equation}
\norm{\mathbf{r}_2}^2 = R_{\ast}^2 + R_{\ast}^2 - 2R_{\ast}^2\cos(\varphi),\qquad\implies\qquad \cos\varphi = 1 - \frac{\norm{\mathbf{r}_2}^2}{2R_{\ast}^2} \simeq \frac{\norm{\mathbf{r}_2}}{R_{\ast}},
\end{equation}
assuming $\norm{\mathbf{r}_2} \ll R_{\ast}$.

\noindent We can now determine the angle $\alpha$ between the $\hat{z}$-axis and the cross product $\mathbf{r}_2 \times \mathbf{r}_3$ as 
\begin{equation}
\alpha = \pi - \frac{\pi}{2} - \beta.
\end{equation}
Similarly, since the internal angles of a triangle must sum to a flat angle, we have
\begin{equation}
\pi = \beta + \beta + \varphi,\qquad\implies\qquad \beta = \frac{1}{2}(\pi - \varphi) \simeq \frac{\pi}{2} - \frac{\norm{\mathbf{r}_2}}{2R_{\ast}}.
\end{equation}
Therefore, we find
\begin{equation}
\alpha \simeq \frac{\norm{\mathbf{r}_2}}{2R_{\ast}}.
\end{equation}

\begin{figure}
\centering
\includegraphics[width = 0.5\linewidth]{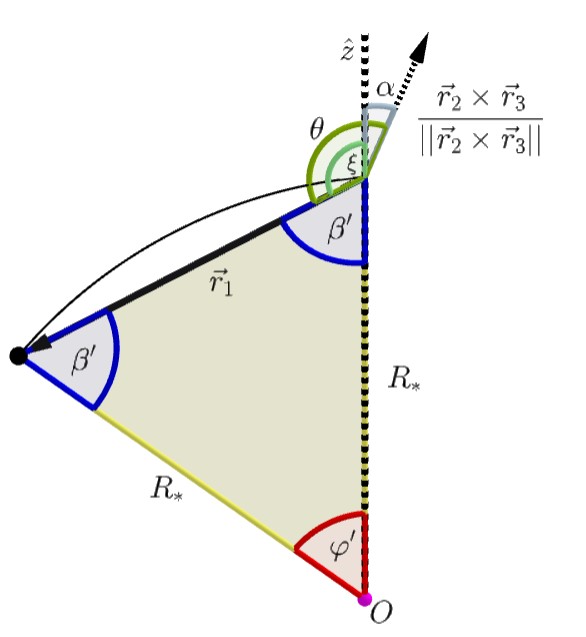}
\caption{Zoom-in on the isosceles triangle formed by the origin $O$ and the vectors $\mathbf{r}_1$ and $\mathbf{r}_3$, with two sides of length $R_{\ast}$ and base $\norm{\mathbf{r}_1}$, all endpoints lying on the last-scattering surface of radius $R_{\ast}$. The internal angle $\varphi^{\prime}$ at the origin is used to approximate the angle $\beta^{\prime}$, which in turn defines the angle $\xi$. Together with the previously defined angle $\alpha$, these determine the angle $\theta = \xi + \alpha$, representing the angle between $\mathbf{r}_1$ and $\mathbf{r}_2 \times \mathbf{r}_3$.}
\label{fig:diag3}
\end{figure}

Analogously, for the isosceles triangle with sides $R_{\ast}$, $R_{\ast}$, and $\norm{\mathbf{r}_1}$, zoomed in Fig.~\ref{fig:diag3}, we find
\begin{equation}
\beta^{\prime} = \frac{1}{2}(\pi - \varphi^{\prime}) \simeq \frac{\pi}{2} - \frac{\norm{\mathbf{r}_1}}{2R_{\ast}}.
\end{equation}
From the plot, we see that $\theta = \xi + \alpha$. Using the result for $\alpha$, we now derive an expression for $\xi$, given by
\begin{equation}
\xi = \pi - \beta^{\prime} \simeq \frac{\pi}{2} + \frac{\norm{\mathbf{r}_1}}{2R_{\ast}},
\end{equation}
so that 
\begin{equation}
\theta = \xi + \alpha \simeq \frac{\pi}{2} + \frac{\norm{\mathbf{r}_1} + \norm{\mathbf{r}_2}}{2R_{\ast}}.
\end{equation}

\noindent Assuming $\norm{\mathbf{r}_1} \simeq \norm{\mathbf{r}_2} \simeq 300$ Mpc$/h$, and $R_{\ast} \simeq 20{,}000$ Mpc$/h$, we obtain
\begin{equation}
\lvert\cos\theta\rvert|\simeq 10^{-2}.
\end{equation}
This very small number represents the maximum value of the cosine of the angle between $\mathbf{r}_1$ and $\mathbf{r}_2 \times \mathbf{r}_3$. Since the parity-odd component of the CMB temperature angular trispectrum is proportional to this geometric factor, it becomes clear why any parity-violating signal would be strongly suppressed.

\acknowledgments
AG and ZS acknowledge funding from NASA grant number 80NSSC24M0021. JH has received funding from the European Union’s Horizon 2020 research and innovation program under the Marie Sk\l{}odowska-Curie grant agreement No 101025187. AK was supported as a CITA National Fellow by the Natural Sciences and Engineering Research Council of Canada (NSERC), funding reference \#DIS-2022-568580.
ZS thanks Bob Cahn and JH thanks Eiichiro Komatsu for insightful discussions. All authors acknowledge the Slepian group for useful discussions.

\bibliographystyle{JHEP}
\bibliography{lens_tris}

\providecommand{\href}[2]{#2}\begingroup\raggedright\begin{thebibliography}{10}

\bibitem{schwinger1951theory}
J.S.~Schwinger, \emph{{The Theory of quantized fields. 1.}}, \href{https://doi.org/10.1103/PhysRev.82.914}{\emph{Phys. Rev.} {\bfseries 82} (1951) 914}.

\bibitem{lee1956question}
T.D.~Lee and C.-N.~Yang, \emph{{Question of Parity Conservation in Weak Interactions}}, \href{https://doi.org/10.1103/PhysRev.104.254}{\emph{Phys. Rev.} {\bfseries 104} (1956) 254}.

\bibitem{sakharov1998violation}
A.D.~Sakharov, \emph{{Violation of CP Invariance, C asymmetry, and baryon asymmetry of the universe}}, \href{https://doi.org/10.1070/PU1991v034n05ABEH002497}{\emph{Pisma Zh. Eksp. Teor. Fiz.} {\bfseries 5} (1967) 32}.

\bibitem{fukugita1986barygenesis}
M.~Fukugita and T.~Yanagida, \emph{{Baryogenesis Without Grand Unification}}, \href{https://doi.org/10.1016/0370-2693(86)91126-3}{\emph{Phys. Lett. B} {\bfseries 174} (1986) 45}.

\bibitem{Lue:1998mq}
A.~Lue, L.-M.~Wang and M.~Kamionkowski, \emph{{Cosmological signature of new parity violating interactions}}, \href{https://doi.org/10.1103/PhysRevLett.83.1506}{\emph{Phys. Rev. Lett.} {\bfseries 83} (1999) 1506} [\href{https://arxiv.org/abs/astro-ph/9812088}{{\ttfamily astro-ph/9812088}}].

\bibitem{Komatsu:2022nvu}
E.~Komatsu, \emph{{New physics from the polarized light of the cosmic microwave background}}, \href{https://doi.org/10.1038/s42254-022-00452-4}{\emph{Nature Rev. Phys.} {\bfseries 4} (2022) 452} [\href{https://arxiv.org/abs/2202.13919}{{\ttfamily 2202.13919}}].

\bibitem{Philcox:2023uor}
O.H.E.~Philcox, M.J.~K\"onig, S.~Alexander and D.N.~Spergel, \emph{{What can galaxy shapes tell us about physics beyond the standard model?}}, \href{https://doi.org/10.1103/PhysRevD.109.063541}{\emph{Phys. Rev. D} {\bfseries 109} (2024) 063541} [\href{https://arxiv.org/abs/2309.08653}{{\ttfamily 2309.08653}}].

\bibitem{Shim:2024tue}
J.~Shim, U.-L.~Pen, H.-R.~Yu and T.~Okumura, \emph{{Probing vector chirality in the early Universe}},  \href{https://arxiv.org/abs/2406.06080}{{\ttfamily 2406.06080}}.

\bibitem{Zhu:2024wme}
H.-M.~Zhu and U.-L.~Pen, \emph{{Systematic analysis of Parity-Violating modes}},  \href{https://arxiv.org/abs/2409.11400}{{\ttfamily 2409.11400}}.

\bibitem{Hou:2024udn}
J.~Hou, Z.~Slepian and D.~Jamieson, \emph{{Can Baryon Acoustic Oscillations Illuminate the Parity-Violating Galaxy 4PCF?}},  \href{https://arxiv.org/abs/2410.05230}{{\ttfamily 2410.05230}}.

\bibitem{cahn2023test}
R.N.~Cahn, Z.~Slepian and J.~Hou, \emph{{Test for Cosmological Parity Violation Using the 3D Distribution of Galaxies}}, \href{https://doi.org/10.1103/PhysRevLett.130.201002}{\emph{Phys. Rev. Lett.} {\bfseries 130} (2023) 201002} [\href{https://arxiv.org/abs/2110.12004}{{\ttfamily 2110.12004}}].

\bibitem{cahn2023isotropic}
R.N.~Cahn and Z.~Slepian, \emph{{Isotropic N-point basis functions and their properties}}, \href{https://doi.org/10.1088/1751-8121/acdfc4}{\emph{J. Phys. A} {\bfseries 56} (2023) 325204} [\href{https://arxiv.org/abs/2010.14418}{{\ttfamily 2010.14418}}].

\bibitem{Jeong:2012df}
D.~Jeong and M.~Kamionkowski, \emph{{Clustering Fossils from the Early Universe}}, \href{https://doi.org/10.1103/PhysRevLett.108.251301}{\emph{Phys. Rev. Lett.} {\bfseries 108} (2012) 251301} [\href{https://arxiv.org/abs/1203.0302}{{\ttfamily 1203.0302}}].

\bibitem{shiraishi2016parity}
M.~Shiraishi, \emph{{Parity violation in the CMB trispectrum from the scalar sector}}, \href{https://doi.org/10.1103/PhysRevD.94.083503}{\emph{Phys. Rev. D} {\bfseries 94} (2016) 083503} [\href{https://arxiv.org/abs/1608.00368}{{\ttfamily 1608.00368}}].

\bibitem{hou2023measurement}
J.~Hou, Z.~Slepian and R.N.~Cahn, \emph{{Measurement of parity-odd modes in the large-scale 4-point correlation function of Sloan Digital Sky Survey Baryon Oscillation Spectroscopic Survey twelfth data release CMASS and LOWZ galaxies}}, \href{https://doi.org/10.1093/mnras/stad1062}{\emph{Mon. Not. Roy. Astron. Soc.} {\bfseries 522} (2023) 5701} [\href{https://arxiv.org/abs/2206.03625}{{\ttfamily 2206.03625}}].

\bibitem{philcox2022probing}
O.H.E.~Philcox, \emph{{Probing parity violation with the four-point correlation function of BOSS galaxies}}, \href{https://doi.org/10.1103/PhysRevD.106.063501}{\emph{Phys. Rev. D} {\bfseries 106} (2022) 063501} [\href{https://arxiv.org/abs/2206.04227}{{\ttfamily 2206.04227}}].

\bibitem{krolewski2024evidenceparityviolationboss}
A.~Krolewski, S.~May, K.~Smith and H.~Hopkins, \emph{{No evidence for parity violation in BOSS}},  \href{https://arxiv.org/abs/2407.03397}{{\ttfamily 2407.03397}}.

\bibitem{adari2024searching}
P.~Adari and A.~Slosar, \emph{{Searching for Parity Violation in SDSS DR16 Lyman-\ensuremath{\alpha} Forest Data}},  \href{https://arxiv.org/abs/2405.04660}{{\ttfamily 2405.04660}}.

\bibitem{Cabass:2022oap}
G.~Cabass, M.M.~Ivanov and O.H.E.~Philcox, \emph{{Colliders and ghosts: Constraining inflation with the parity-odd galaxy four-point function}}, \href{https://doi.org/10.1103/PhysRevD.107.023523}{\emph{Phys. Rev. D} {\bfseries 107} (2023) 023523} [\href{https://arxiv.org/abs/2210.16320}{{\ttfamily 2210.16320}}].

\bibitem{Inomata:2024ald}
K.~Inomata, L.~Jenks and M.~Kamionkowski, \emph{{Parity-breaking galaxy 4-point function from lensing by chiral gravitational waves}},  \href{https://arxiv.org/abs/2408.03994}{{\ttfamily 2408.03994}}.

\bibitem{philcox2024could}
O.H.E.~Philcox and J.~Ereza, \emph{{Could Sample Variance be Responsible for the Parity-Violating Signal Seen in the BOSS Galaxy Survey?}},  \href{https://arxiv.org/abs/2401.09523}{{\ttfamily 2401.09523}}.

\bibitem{Ereza:2023zmz}
J.~Ereza, F.~Prada, A.~Klypin, T.~Ishiyama, A.~Smith, C.M.~Baugh et~al., \emph{{The Uchuu-glam BOSS and eBOSS LRG lightcones: exploring clustering and covariance errors}}, \href{https://doi.org/10.1093/mnras/stae1543}{\emph{Mon. Not. Roy. Astron. Soc.} {\bfseries 532} (2024) 1659} [\href{https://arxiv.org/abs/2311.14456}{{\ttfamily 2311.14456}}].

\bibitem{Taylor:2023deh}
P.L.~Taylor, M.~Craigie and Y.-S.~Ting, \emph{{Unsupervised searches for cosmological parity violation: An investigation with convolutional neural networks}}, \href{https://doi.org/10.1103/PhysRevD.109.083518}{\emph{Phys. Rev. D} {\bfseries 109} (2024) 083518} [\href{https://arxiv.org/abs/2312.09287}{{\ttfamily 2312.09287}}].

\bibitem{Hewson:2024rnb}
S.~Hewson, W.J.~Handley and C.G.~Lester, \emph{{On the spatial distribution of the Large-Scale structure: An Unsupervised search for Parity Violation}},  \href{https://arxiv.org/abs/2410.16030}{{\ttfamily 2410.16030}}.

\bibitem{Craigie:2024bhk}
M.~Craigie, P.L.~Taylor, Y.-S.~Ting, C.~Cuesta-Lazaro, R.~Ruggeri and T.M.~Davis, \emph{{Unsupervised Searches for Cosmological Parity Violation: Improving Detection Power with the Neural Field Scattering Transform}},  5, 2024.

\bibitem{philcox2023cmb}
O.H.E.~Philcox, \emph{{Do the CMB Temperature Fluctuations Conserve Parity?}}, \href{https://doi.org/10.1103/PhysRevLett.131.181001}{\emph{Phys. Rev. Lett.} {\bfseries 131} (2023) 181001} [\href{https://arxiv.org/abs/2303.12106}{{\ttfamily 2303.12106}}].

\bibitem{philcox2023testing}
O.H.E.~Philcox and M.~Shiraishi, \emph{{Testing parity symmetry with the polarized cosmic microwave background}}, \href{https://doi.org/10.1103/PhysRevD.109.083514}{\emph{Phys. Rev. D} {\bfseries 109} (2024) 083514} [\href{https://arxiv.org/abs/2308.03831}{{\ttfamily 2308.03831}}].

\bibitem{seljak1995gravitational}
U.~Seljak, \emph{{Gravitational lensing effect on cosmic microwave background anisotropies: A Power spectrum approach}}, \href{https://doi.org/10.1086/177218}{\emph{Astrophys. J.} {\bfseries 463} (1996) 1} [\href{https://arxiv.org/abs/astro-ph/9505109}{{\ttfamily astro-ph/9505109}}].

\bibitem{hu2000weak}
W.~Hu, \emph{{Weak lensing of the CMB: A harmonic approach}}, \href{https://doi.org/10.1103/PhysRevD.62.043007}{\emph{Phys. Rev. D} {\bfseries 62} (2000) 043007} [\href{https://arxiv.org/abs/astro-ph/0001303}{{\ttfamily astro-ph/0001303}}].

\bibitem{bartelmann2001weak}
M.~Bartelmann and P.~Schneider, \emph{{Weak gravitational lensing}}, \href{https://doi.org/10.1016/S0370-1573(00)00082-X}{\emph{Phys. Rept.} {\bfseries 340} (2001) 291} [\href{https://arxiv.org/abs/astro-ph/9912508}{{\ttfamily astro-ph/9912508}}].

\bibitem{challinor2005lensed}
A.~Challinor and A.~Lewis, \emph{{Lensed CMB power spectra from all-sky correlation functions}}, \href{https://doi.org/10.1103/PhysRevD.71.103010}{\emph{Phys. Rev. D} {\bfseries 71} (2005) 103010} [\href{https://arxiv.org/abs/astro-ph/0502425}{{\ttfamily astro-ph/0502425}}].

\bibitem{lewis2006weak}
A.~Lewis and A.~Challinor, \emph{{Weak gravitational lensing of the CMB}}, \href{https://doi.org/10.1016/j.physrep.2006.03.002}{\emph{Phys. Rept.} {\bfseries 429} (2006) 1} [\href{https://arxiv.org/abs/astro-ph/0601594}{{\ttfamily astro-ph/0601594}}].

\bibitem{bohm2016bias}
V.~B\"ohm, M.~Schmittfull and B.D.~Sherwin, \emph{{Bias to CMB lensing measurements from the bispectrum of large-scale structure}}, \href{https://doi.org/10.1103/PhysRevD.94.043519}{\emph{Phys. Rev. D} {\bfseries 94} (2016) 043519} [\href{https://arxiv.org/abs/1605.01392}{{\ttfamily 1605.01392}}].

\bibitem{Namikawa:2016jff}
T.~Namikawa, \emph{{CMB Lensing Bispectrum from Nonlinear Growth of the Large Scale Structure}}, \href{https://doi.org/10.1103/PhysRevD.93.121301}{\emph{Phys. Rev. D} {\bfseries 93} (2016) 121301} [\href{https://arxiv.org/abs/1604.08578}{{\ttfamily 1604.08578}}].

\bibitem{namikawa2019cmb}
T.~Namikawa, B.~Bose, F.R.~Bouchet, R.~Takahashi and A.~Taruya, \emph{{CMB lensing bispectrum: Assessing analytical predictions against full-sky lensing simulations}}, \href{https://doi.org/10.1103/PhysRevD.99.063511}{\emph{Phys. Rev. D} {\bfseries 99} (2019) 063511} [\href{https://arxiv.org/abs/1812.10635}{{\ttfamily 1812.10635}}].

\bibitem{kalaja2023reconstructed}
A.~Kalaja, G.~Orlando, A.~Bowkis, A.~Challinor, P.D.~Meerburg and T.~Namikawa, \emph{{The reconstructed CMB lensing bispectrum}}, \href{https://doi.org/10.1088/1475-7516/2023/04/041}{\emph{JCAP} {\bfseries 04} (2023) 041} [\href{https://arxiv.org/abs/2210.16203}{{\ttfamily 2210.16203}}].

\bibitem{Namikawa:2018erh}
T.~Namikawa, F.R.~Bouchet and A.~Taruya, \emph{{CMB lensing bispectrum as a probe of modified gravity theories}}, \href{https://doi.org/10.1103/PhysRevD.98.043530}{\emph{Phys. Rev. D} {\bfseries 98} (2018) 043530} [\href{https://arxiv.org/abs/1805.10567}{{\ttfamily 1805.10567}}].

\bibitem{hu2001angular}
W.~Hu, \emph{{Angular trispectrum of the CMB}}, \href{https://doi.org/10.1103/PhysRevD.64.083005}{\emph{Phys. Rev. D} {\bfseries 64} (2001) 083005} [\href{https://arxiv.org/abs/astro-ph/0105117}{{\ttfamily astro-ph/0105117}}].

\bibitem{okamoto2002angular}
T.~Okamoto and W.~Hu, \emph{{The angular trispectra of CMB temperature and polarization}}, \href{https://doi.org/10.1103/PhysRevD.66.063008}{\emph{Phys. Rev. D} {\bfseries 66} (2002) 063008} [\href{https://arxiv.org/abs/astro-ph/0206155}{{\ttfamily astro-ph/0206155}}].

\bibitem{kaiser1998weak}
N.~Kaiser, \emph{{Weak lensing and cosmology}}, \href{https://doi.org/10.1086/305515}{\emph{Astrophys. J.} {\bfseries 498} (1998) 26} [\href{https://arxiv.org/abs/astro-ph/9610120}{{\ttfamily astro-ph/9610120}}].

\bibitem{lesgourgues2011cosmic}
J.~Lesgourgues, \emph{{The Cosmic Linear Anisotropy Solving System (CLASS) I: Overview}},  \href{https://arxiv.org/abs/1104.2932}{{\ttfamily 1104.2932}}.

\bibitem{aghanim2020planck}
{\scshape Planck} collaboration, \emph{{Planck 2018 results. VI. Cosmological parameters}}, \href{https://doi.org/10.1051/0004-6361/201833910}{\emph{Astron. Astrophys.} {\bfseries 641} (2020) A6} [\href{https://arxiv.org/abs/1807.06209}{{\ttfamily 1807.06209}}].

\bibitem{ma1995cosmological}
C.-P.~Ma and E.~Bertschinger, \emph{{Cosmological perturbation theory in the synchronous and conformal Newtonian gauges}}, \href{https://doi.org/10.1086/176550}{\emph{Astrophys. J.} {\bfseries 455} (1995) 7} [\href{https://arxiv.org/abs/astro-ph/9506072}{{\ttfamily astro-ph/9506072}}].

\bibitem{mehrem2011plane}
R.~Mehrem, \emph{{The Plane Wave Expansion, Infinite Integrals and Identities involving Spherical Bessel Functions}},  \href{https://arxiv.org/abs/0909.0494}{{\ttfamily 0909.0494}}.

\bibitem{varshalovich1988quantum}
D.A.~Varshalovich, A.N.~Moskalev and V.K.~Khersonskii, \emph{{Quantum Theory of Angular Momentum}: {Irreducible Tensors, Spherical Harmonics, Vector Coupling Coefficients, 3nj Symbols}}, World Scientific Publishing Company (1988), \href{https://doi.org/10.1142/0270}{10.1142/0270}.

\bibitem{eisenstein1998baryonic}
D.J.~Eisenstein and W.~Hu, \emph{{Baryonic features in the matter transfer function}}, \href{https://doi.org/10.1086/305424}{\emph{Astrophys. J.} {\bfseries 496} (1998) 605} [\href{https://arxiv.org/abs/astro-ph/9709112}{{\ttfamily astro-ph/9709112}}].

\bibitem{eisenstein1999power}
D.J.~Eisenstein and W.~Hu, \emph{{Power spectra for cold dark matter and its variants}}, \href{https://doi.org/10.1086/306640}{\emph{Astrophys. J.} {\bfseries 511} (1997) 5} [\href{https://arxiv.org/abs/astro-ph/9710252}{{\ttfamily astro-ph/9710252}}].

\bibitem{borges2008evolution}
H.A.~Borges, S.~Carneiro, J.C.~Fabris and C.~Pigozzo, \emph{{Evolution of density perturbations in decaying vacuum cosmology}}, \href{https://doi.org/10.1103/PhysRevD.77.043513}{\emph{Phys. Rev. D} {\bfseries 77} (2008) 043513} [\href{https://arxiv.org/abs/0711.2689}{{\ttfamily 0711.2689}}].

\bibitem{slepian2016simple}
Z.~Slepian and D.J.~Eisenstein, \emph{{A Simple Analytic Treatment of Linear Growth of Structure with Baryon Acoustic Oscillations}}, \href{https://doi.org/10.1093/mnras/stv2889}{\emph{Mon. Not. Roy. Astron. Soc.} {\bfseries 457} (2016) 24} [\href{https://arxiv.org/abs/1509.08199}{{\ttfamily 1509.08199}}].

\bibitem{limber1953analysis}
D.N.~Limber, \emph{{The Analysis of Counts of the Extragalactic Nebulae in Terms of a Fluctuating Density Field. II}}, \href{https://doi.org/10.1086/145870}{\emph{Astrophys. J.} {\bfseries 119} (1954) 655}.

\bibitem{lemos2017effect}
P.~Lemos, A.~Challinor and G.~Efstathiou, \emph{{The effect of Limber and flat-sky approximations on galaxy weak lensing}}, \href{https://doi.org/10.1088/1475-7516/2017/05/014}{\emph{JCAP} {\bfseries 05} (2017) 014} [\href{https://arxiv.org/abs/1704.01054}{{\ttfamily 1704.01054}}].

\bibitem{akrami2020planck_inflation}
{\scshape Planck} collaboration, \emph{{Planck 2018 results. X. Constraints on inflation}}, \href{https://doi.org/10.1051/0004-6361/201833887}{\emph{Astron. Astrophys.} {\bfseries 641} (2020) A10} [\href{https://arxiv.org/abs/1807.06211}{{\ttfamily 1807.06211}}].

\bibitem{bardeen1986statistics}
J.M.~Bardeen, J.R.~Bond, N.~Kaiser and A.S.~Szalay, \emph{{The Statistics of Peaks of Gaussian Random Fields}}, \href{https://doi.org/10.1086/164143}{\emph{Astrophys. J.} {\bfseries 304} (1986) 15}.

\bibitem{kramer2020cosmolopy}
R.H.~Kramer, \emph{Cosmolopy: Cosmology package for python}, {\emph{Astrophysics Source Code Library} (2020) ascl}.

\bibitem{abramowitz1968handbook}
M.~Abramowitz and I.A.~Stegun, \emph{Handbook of mathematical functions with formulas, graphs, and mathematical tables}, vol.~56, US Government printing office (1988), \href{https://doi.org/doi:10.1119/1.15378}{doi:10.1119/1.15378}.

\bibitem{NIST:DLMF}
``{\it NIST Digital Library of Mathematical Functions}.'' \url{https://dlmf.nist.gov/}, Release 1.2.1 of 2024-06-15.

\bibitem{shiraishi2011cmb}
M.~Shiraishi, D.~Nitta, S.~Yokoyama, K.~Ichiki and K.~Takahashi, \emph{{CMB Bispectrum from Primordial Scalar, Vector and Tensor non-Gaussianities}}, \href{https://doi.org/10.1143/PTP.125.795}{\emph{Prog. Theor. Phys.} {\bfseries 125} (2011) 795} [\href{https://arxiv.org/abs/1012.1079}{{\ttfamily 1012.1079}}].

\bibitem{bartolo2004non}
N.~Bartolo, E.~Komatsu, S.~Matarrese and A.~Riotto, \emph{{Non-Gaussianity from inflation: Theory and observations}}, \href{https://doi.org/10.1016/j.physrep.2004.08.022}{\emph{Phys. Rept.} {\bfseries 402} (2004) 103} [\href{https://arxiv.org/abs/astro-ph/0406398}{{\ttfamily astro-ph/0406398}}].

\bibitem{akrami2020planck_nongaussianity}
{\scshape Planck} collaboration, \emph{{Planck 2018 results. IX. Constraints on primordial non-Gaussianity}}, \href{https://doi.org/10.1051/0004-6361/201935891}{\emph{Astron. Astrophys.} {\bfseries 641} (2020) A9} [\href{https://arxiv.org/abs/1905.05697}{{\ttfamily 1905.05697}}].

\bibitem{Spergel:1999xn}
D.N.~Spergel and D.M.~Goldberg, \emph{{Microwave background bispectrum. 1. Basic formalism}}, \href{https://doi.org/10.1103/PhysRevD.59.103001}{\emph{Phys. Rev. D} {\bfseries 59} (1999) 103001} [\href{https://arxiv.org/abs/astro-ph/9811252}{{\ttfamily astro-ph/9811252}}].

\bibitem{komatsu2001acoustic}
E.~Komatsu and D.N.~Spergel, \emph{{Acoustic signatures in the primary microwave background bispectrum}}, \href{https://doi.org/10.1103/PhysRevD.63.063002}{\emph{Phys. Rev. D} {\bfseries 63} (2001) 063002} [\href{https://arxiv.org/abs/astro-ph/0005036}{{\ttfamily astro-ph/0005036}}].

\bibitem{acquaviva2002second}
V.~Acquaviva, N.~Bartolo, S.~Matarrese and A.~Riotto, \emph{{Second order cosmological perturbations from inflation}}, \href{https://doi.org/10.1016/S0550-3213(03)00550-9}{\emph{Nucl. Phys. B} {\bfseries 667} (2003) 119} [\href{https://arxiv.org/abs/astro-ph/0209156}{{\ttfamily astro-ph/0209156}}].

\bibitem{maldacena2003non}
J.M.~Maldacena, \emph{{Non-Gaussian features of primordial fluctuations in single field inflationary models}}, \href{https://doi.org/10.1088/1126-6708/2003/05/013}{\emph{JHEP} {\bfseries 05} (2003) 013} [\href{https://arxiv.org/abs/astro-ph/0210603}{{\ttfamily astro-ph/0210603}}].

\bibitem{isserlis1918formula}
L.~Isserlis, \emph{{On a formula for the product-moment coefficient of any order of a normal frequency distribution in any number of variables}}, \href{https://doi.org/10.2307/2331932}{\emph{Biometrika} {\bfseries 12} (1918) 134}.

\bibitem{newman1966note}
E.T.~Newman and R.~Penrose, \emph{{Note on the Bondi-Metzner-Sachs group}}, \href{https://doi.org/10.1063/1.1931221}{\emph{J. Math. Phys.} {\bfseries 7} (1966) 863}.

\bibitem{wigner1927einige}
E.~Wigner, \emph{{Einige Folgerungen aus der Schr\"odingerschen Theorie f\"ur die Termstrukturen}}, \href{https://doi.org/10.1007/BF01397327}{\emph{Z. Phys.} {\bfseries 43} (1927) 624}.

\bibitem{eckart1930application}
C.~Eckart, \emph{{The Application of Group Theory to the Quantum Dynamics of Monatomic Systems}}, \href{https://doi.org/10.1103/RevModPhys.2.305}{\emph{Rev. Mod. Phys.} {\bfseries 2} (1930) 305}.

\bibitem{shiraishi2014signatures}
M.~Shiraishi, E.~Komatsu and M.~Peloso, \emph{{Signatures of anisotropic sources in the trispectrum of the cosmic microwave background}}, \href{https://doi.org/10.1088/1475-7516/2014/04/027}{\emph{JCAP} {\bfseries 04} (2014) 027} [\href{https://arxiv.org/abs/1312.5221}{{\ttfamily 1312.5221}}].

\bibitem{xiang2021program}
S.~Xiang, L.~Wang, Z.-C.~Yan and H.~Qiao, \emph{{A program for simplifying summation of Wigner 3j-symbols}}, \href{https://doi.org/10.1016/j.cpc.2021.107880}{\emph{Comput. Phys. Commun.} {\bfseries 264} (2021) 107880}.

\bibitem{jamieson2024parity}
D.~Jamieson, A.~Caravano, J.~Hou, Z.~Slepian and E.~Komatsu, \emph{{Parity-Odd Power Spectra: Concise Statistics for Cosmological Parity Violation}},  \href{https://arxiv.org/abs/2406.15683}{{\ttfamily 2406.15683}}.

\bibitem{coulton2024signatures}
W.R.~Coulton, O.H.E.~Philcox and F.~Villaescusa-Navarro, \emph{{Signatures of a parity-violating universe}}, \href{https://doi.org/10.1103/PhysRevD.109.023531}{\emph{Phys. Rev. D} {\bfseries 109} (2024) 023531} [\href{https://arxiv.org/abs/2306.11782}{{\ttfamily 2306.11782}}].

\bibitem{Niu:2022fki}
X.~Niu, M.H.~Rahat, K.~Srinivasan and W.~Xue, \emph{{Parity-odd and even trispectrum from axion inflation}}, \href{https://doi.org/10.1088/1475-7516/2023/05/018}{\emph{JCAP} {\bfseries 05} (2023) 018} [\href{https://arxiv.org/abs/2211.14324}{{\ttfamily 2211.14324}}].

\bibitem{Cabass:2022rhr}
G.~Cabass, S.~Jazayeri, E.~Pajer and D.~Stefanyszyn, \emph{{Parity violation in the scalar trispectrum: no-go theorems and yes-go examples}}, \href{https://doi.org/10.1007/JHEP02(2023)021}{\emph{JHEP} {\bfseries 02} (2023) 021} [\href{https://arxiv.org/abs/2210.02907}{{\ttfamily 2210.02907}}].

\bibitem{Fujita:2023inz}
T.~Fujita, T.~Murata, I.~Obata and M.~Shiraishi, \emph{{Parity-violating scalar trispectrum from a rolling axion during inflation}}, \href{https://doi.org/10.1088/1475-7516/2024/05/127}{\emph{JCAP} {\bfseries 05} (2024) 127} [\href{https://arxiv.org/abs/2310.03551}{{\ttfamily 2310.03551}}].

\bibitem{Creque-Sarbinowski:2023wmb}
C.~Creque-Sarbinowski, S.~Alexander, M.~Kamionkowski and O.~Philcox, \emph{{Parity-violating trispectrum from Chern-Simons gravity}}, \href{https://doi.org/10.1088/1475-7516/2023/11/029}{\emph{JCAP} {\bfseries 11} (2023) 029} [\href{https://arxiv.org/abs/2303.04815}{{\ttfamily 2303.04815}}].

\bibitem{Moretti:2024fzb}
T.~Moretti, N.~Bartolo and A.~Greco, \emph{{Breaking Parity: the case of the Trispectrum from Chiral Scalar-Tensor Theories of Gravity}},  \href{https://arxiv.org/abs/2410.11801}{{\ttfamily 2410.11801}}.

\bibitem{olver1997asymptotics}
F.~Olver, \emph{Asymptotics and special functions}, AK Peters/CRC Press (1997), \href{https://doi.org/10.1201/9781439864548}{10.1201/9781439864548}.

\bibitem{Bernardeau:2010ac}
F.~Bernardeau, C.~Pitrou and J.-P.~Uzan, \emph{{CMB spectra and bispectra calculations: making the flat-sky approximation rigorous}}, \href{https://doi.org/10.1088/1475-7516/2011/02/015}{\emph{JCAP} {\bfseries 02} (2011) 015} [\href{https://arxiv.org/abs/1012.2652}{{\ttfamily 1012.2652}}].

\bibitem{Kogo:2006kh}
N.~Kogo and E.~Komatsu, \emph{{Angular trispectrum of cmb temperature anisotropy from primordial non-gaussianity with the full radiation transfer function}}, \href{https://doi.org/10.1103/PhysRevD.73.083007}{\emph{Phys. Rev. D} {\bfseries 73} (2006) 083007} [\href{https://arxiv.org/abs/astro-ph/0602099}{{\ttfamily astro-ph/0602099}}].

\bibitem{Hirata:2003ka}
C.M.~Hirata and U.~Seljak, \emph{{Reconstruction of lensing from the cosmic microwave background polarization}}, \href{https://doi.org/10.1103/PhysRevD.68.083002}{\emph{Phys. Rev. D} {\bfseries 68} (2003) 083002} [\href{https://arxiv.org/abs/astro-ph/0306354}{{\ttfamily astro-ph/0306354}}].

\bibitem{Smith:2010gu}
K.M.~Smith, D.~Hanson, M.~LoVerde, C.M.~Hirata and O.~Zahn, \emph{{Delensing CMB Polarization with External Datasets}}, \href{https://doi.org/10.1088/1475-7516/2012/06/014}{\emph{JCAP} {\bfseries 06} (2012) 014} [\href{https://arxiv.org/abs/1010.0048}{{\ttfamily 1010.0048}}].

\bibitem{CMB-S4:2016ple}
{\scshape CMB-S4} collaboration, \emph{{CMB-S4 Science Book, First Edition}},  \href{https://arxiv.org/abs/1610.02743}{{\ttfamily 1610.02743}}.

\bibitem{Hu:2002vu}
W.~Hu, M.M.~Hedman and M.~Zaldarriaga, \emph{{Benchmark parameters for CMB polarization experiments}}, \href{https://doi.org/10.1103/PhysRevD.67.043004}{\emph{Phys. Rev. D} {\bfseries 67} (2003) 043004} [\href{https://arxiv.org/abs/astro-ph/0210096}{{\ttfamily astro-ph/0210096}}].

\bibitem{SimonsObservatory:2018koc}
{\scshape Simons Observatory} collaboration, \emph{{The Simons Observatory: Science goals and forecasts}}, \href{https://doi.org/10.1088/1475-7516/2019/02/056}{\emph{JCAP} {\bfseries 02} (2019) 056} [\href{https://arxiv.org/abs/1808.07445}{{\ttfamily 1808.07445}}].

\bibitem{smith2003stable}
{\scshape VIRGO Consortium} collaboration, \emph{{Stable clustering, the halo model and nonlinear cosmological power spectra}}, \href{https://doi.org/10.1046/j.1365-8711.2003.06503.x}{\emph{Mon. Not. Roy. Astron. Soc.} {\bfseries 341} (2003) 1311} [\href{https://arxiv.org/abs/astro-ph/0207664}{{\ttfamily astro-ph/0207664}}].

\bibitem{takahashi2012revising}
R.~Takahashi, M.~Sato, T.~Nishimichi, A.~Taruya and M.~Oguri, \emph{{Revising the Halofit Model for the Nonlinear Matter Power Spectrum}}, \href{https://doi.org/10.1088/0004-637X/761/2/152}{\emph{Astrophys. J.} {\bfseries 761} (2012) 152} [\href{https://arxiv.org/abs/1208.2701}{{\ttfamily 1208.2701}}].

\bibitem{bernardeau2002large}
F.~Bernardeau, S.~Colombi, E.~Gaztanaga and R.~Scoccimarro, \emph{{Large scale structure of the universe and cosmological perturbation theory}}, \href{https://doi.org/10.1016/S0370-1573(02)00135-7}{\emph{Phys. Rept.} {\bfseries 367} (2002) 1} [\href{https://arxiv.org/abs/astro-ph/0112551}{{\ttfamily astro-ph/0112551}}].

\end{thebibliography}\endgroup

\end{document}